%% file: main.tex
\documentclass[nofootinbib,reprint,superscriptaddress,pra,twocolumn,longbibliography,floatfix]{revtex4-2}

\input{preamble}

\begin{document}

\title{Phase separation of a repulsive two-component Fermi gas  at the two- to three-dimensional crossover}

\author{Piotr~T.~Grochowski\orcid{0000-0002-9654-4824}} 
\email{piotr.grochowski@uibk.ac.at}
\affiliation{Institute for Quantum Optics and Quantum Information of the Austrian Academy of Sciences, A-6020 Innsbruck, Austria}
\affiliation{Institute for Theoretical Physics, University of Innsbruck, A-6020 Innsbruck, Austria}
\affiliation{Center for Theoretical Physics, Polish Academy of Sciences, Aleja Lotnik\'ow 32/46, 02-668 Warsaw, Poland}

\author{Martin-Isbj\"orn~Trappe\orcid{0000-0002-2911-4162}}
\affiliation{Centre for Quantum Technologies, National University of Singapore, 3 Science Drive 2, Singapore 117543, Singapore}

\author{Tomasz~Karpiuk\orcid{0000-0001-7194-324X}}
\affiliation{Wydzia\l{} Fizyki, Uniwersytet w Bia\l{}ymstoku, ul. K. Cio\l{}kowskiego 1L, 15-245 Bia\l{}ystok, Poland}

\author{Kazimierz~Rz\k{a}\.zewski\orcid{0000-0002-6082-3565}}
\affiliation{Center for Theoretical Physics, Polish Academy of Sciences, Aleja Lotnik\'ow 32/46, 02-668 Warsaw, Poland}

\begin{abstract}
We present a theoretical analysis of phase separations between two repulsively interacting components in an ultracold fermionic gas, occurring at the dimensional crossover in a harmonic trap with varying aspect ratios.
A tailored kinetic energy functional is derived and combined with a density-potential functional approach to develop a framework that is benchmarked with the orbital-based method.
We investigate the changes in the density profile of the phase-separated gas under different interaction strengths and geometries.
The analysis reveals the existence of small, partially polarized domains in certain parameter regimes, which is similar to the purely two-dimensional limit.
However, the density profile is further enriched by a shell structure found in anisotropic traps.
We also track the transitions that can be driven by either a change in interaction strength or trap geometry.
The developed framework is noted to have applications for other systems with repulsive interactions that combine continuous and discrete degrees of freedom.
\end{abstract}

\maketitle 
\section{Introduction}
Reduced-dimensional systems are of great interest in condensed matter and statistical physics due to the enhanced influence of quantum fluctuations.
Such systems are crucial in experimental and technological applications, with examples spanning, e.g., high-temperature superconductors, layered semiconductors, and graphene.
Recent advancements in trapping ultracold atomic gases in quasi-two-dimensional geometries~\cite{Bloch2008,Lewenstein2012} have enabled the measurement of zero-~\cite{Boettcher2016,Holten2018,Toniolo2018} and finite-temperature effects~\cite{Tung2010,Plisson2011,Ries2015a,Fenech2016,Murthy2018}.
This is usually achieved through strongly anisotropic trapping potentials and one-dimensional optical lattices, which allow for the experimental realization of quasi-two-dimensional quantum gases and dimensional crossovers~\cite{
Dyke2011,Sommer2012,Cheng2016,Peppler2018,Gong2023,Guo2023,Guo2023a}.

Dimensional crossovers provide access to additional degrees of freedom, leading to the emergence of new quantum states with discrete energies.
In quasi-two-dimensional Fermi gases, where the transverse confinement energy is comparable to the Fermi energy, the occupation of new transverse states results in a shell structure~\cite{Dyke2011}.
This, in turn, causes steps in the density profile, chemical potential, and specific heat to appear as the system size increases due to Pauli exclusion principle~\cite{Schneider1998,Vignolo2003,Mueller2004,Kumagai2006a,Dyke2011}.

Most of the theoretical and experimental efforts have focused mostly on various many-body effects in two-dimensional Fermi gases, including crossover from Bose-Einstein condensate to Bardeen-Cooper-Schrieffer superfluid~\cite{Pitaevskii1997,Olshanii2010,Taylor2012,Hofmann2012,Gao2012,Chafin2013,Hu2019,Peppler2018,Murthy2019}, or pairing pseudogap~\cite{Tsuchiya2009,Gaebler2010,Murthy2018,Richie-Halford2020}.
As in realistic experimental scenarios, the anisotropy of confining potentials does not usually allow for purely two-dimensional regime, the analysis of dimensional crossovers has been of particular interest~\cite{Dyke2011,Sommer2012,Cheng2016,Ptok2017,Toniolo2017,Peppler2018,Toniolo2018,Adachi2018,Faigle-Cedzich2021,Gong2023,Zheng2023}.
Such transitions happen close to the ground state of the considered system and manifest the underlying bonding mechanism in a Fermi gas.

However, when excited, the spin components of a Fermi gas, i.e., given by two chosen hyperfine states, may exhibit repulsive correlations, being brought to the so-called repulsive branch.
In the case of a gas trapped in a harmonic potential, that repulsion may lead to a metastable phase separation between the components, an ultracold analog of celebrated Stoner instability in Coulomb-interacting electronic gas~\cite{Stoner1938}.
Experimentally elusive in both three and two dimensions due to the eventual decay to the ground state, this itinerant ferromagnetic state has been widely researched both theoretically~\cite{Sogo2002,Karpiuk2004,Duine2005,LeBlanc2009,Conduit2009,Cui2010,Pilati2010,Chang2011,Pekker2011,Massignan2011,Massignan2014,Levinsen2015,Trappe2016a,Miyakawa2017,Grochowski2017,Koutentakis2019,Grochowski2017,Ryszkiewicz2020,Karpiuk2020,Grochowski2020a,Koutentakis2020,Trappe2021a,Ryszkiewicz2022,Syrwid2022,Lebek2022} and experimentally~\cite{DeMarco2002,Du2008,Jo2009,Sommer2011,Sanner2012,Lee2012,Valtolina2017,Amico2018,Scazza2020,Adlong2020,Ji2022}, also in different mixtures~\cite{Lous2018,Huang2019,Grochowski2020}.

We focus on previously unexplored crossover of the Stoner instability between two and three dimensions, analyzing the purely repulsive system of two fermionic components trapped in an anisotropic trap with varying aspect ratio and interaction strength.
In the three-dimensional limit, the repulsive gas in a radially symmetric trap exhibits both partially and fully polarized phases, whose density profiles remain highly regular, either in a ring structure, with one of the components dominating in the center of the trap, or in two halves separated by a single domain wall~\cite{Trappe2016a}.
In contrast, in two dimensions, contact-interacting gas can form a plethora of partially polarized states manifesting many small domains, due to a lack of scaling of critical interaction strength with respect to the density of gas in the leading order~\cite{Trappe2021a}.
In each of these cases, refined treatment of the kinetic energy, going beyond the usual Thomas-Fermi approach, is necessary, as the competition between the kinetic, interaction, and correlation energy drives the exact shape of the density profile.

To take that into account and to describe systems of sizes reaching usual experimental scenarios of large Fermi clouds, we utilize orbital-free density-potential functional theory (DPFT) that was proven to accurately determine the complex phases of interacting Fermi gases~\cite{Englert1982,Englert1984,Englert1985,Englert1988,Englert1992,Trappe2016,Trappe2017,Chau2018,Englert2019}.
It achieves this by simplifying the many-body problem to two self-consistent equations: one for the single-particle density and another for an effective potential that accounts for the interactions.
DPFT is particularly useful for simulating trapped quantum gases, especially in two-dimensional configurations.
Conventional density-functional theory (DFT) techniques are unable to match DPFT's capabilities in this regard, as they are either limited to small particle numbers~\cite{Ancilotto2015,Das2018}, periodic confinement~\cite{Ma2012}, or rely on ad-hoc parameterizations of the kinetic energy~\cite{VanZyl2013,Gangwar2020}.
Although systematic gradient corrections in two dimensions are available for electronic systems~\cite{Vilhena2014}, DPFT offers a scalable approach that can be systematically expanded beyond the Thomas-Fermi approximation across one-, two-, and three-dimensional geometries.

In this work we examine phase transitions from para- to ferromagnetic states in a binary Fermi gas, involving two- to three-dimensional crossover using density-potential functional theory and Hartree-Fock methods.
To this end, we derive a tailored kinetic energy functional for a dimensional crossover of the Fermi gas and combine it with the density--potential functional approach to allow the description of large particle numbers. 
We use the Hartree-Fock method to benchmark our results in the low-atom-number limit.
Our findings indicate highly degenerate ground-state profiles during the transition, featuring various shapes like isotropic and anisotropic separations, ring-shaped polarization, and central splits within a shell structure. These profiles can be manipulated by adjusting particle number, interaction strength, and trapping aspect ratio, offering a versatile exploration of interacting quantum mixtures.

The paper is structured as follows. 
Section~\ref{methods} outlines the derivations of the kinetic and interaction energy functionals for the dimensional crossover, with the analysis of the accuracy placed in Appendix~\ref{app:nonas}, along with the presentation of the DPFT and Hartree-Fock methods.
Section~\ref{results} provides a description of the results we have obtained, including interaction- and aspect-ratio-driven phase transitions, comparison between two methods, and analysis of large-particle-number limit.
We conclude the paper in Section~\ref{conclusions}, providing a summary and outlook for the future.

\section{Methods}\label{methods}
\subsection{Density-functional approach for a Fermi gas at a two- to three-dimensional crossover}\label{methods:functional}
Let us consider a noninteracting, polarized Fermi gas of $N$ atoms with mass $m$ at zero temperature that is (i) confined in the $x-y$ plane in an area $S$ and (ii) harmonically trapped in the $z$-direction with frequency $\omega_z$.
The Hamiltonian of this system reads
\begin{equation}
    \hat{H} = \sum_{i=1}^{N} \hat{H}_i,
\end{equation}
where $\hat{H}_i$ is the single-body Hamiltonian that reads
\begin{equation}
    \hat{H}_i = \frac{\hat{\bf{p}}_{i}^2}{2 m} + \frac{1}{2} m \omega_z^2 \hat{z}_i^2 - \frac{1}{2} \hbar \omega_z,
\end{equation}
with two-dimensional momentum operator $\hat{\bf{p}}_{i}$ in the $x-y$ plane, and with the zero-point energy removed.
The energy of a single-particle eigenstate reads
\begin{equation}
E_{\state{j}}=\kin+j\hw,
\end{equation}
where state $\state{j}$ is described through two-dimensional wave vector $\bf{k}$ and oscillator state number $j \in \{0,1,\dots\}$.
The energy manifolds are degenerate due to both $k^2$ energy dependence and multiple possible oscillator states.
As such, the density of states $\rho(E)$ of the gas can be rewritten as a sum
\begin{equation}
\rho(E) = \sum_j \rho_j(E) = \sum_j \frac{m S}{2 \pi \hbar^2} \theta(E - j \hbar \omega_z),
\end{equation}
where $\theta(\cdot)$ is a Heaviside step function, summation is performed over available oscillator states, and we have used the expression for two-dimensional density of states $\rho_{\text{2D}} = m S / 2 \pi \hbar^2$.
We can then immediately write the total number of atoms as a function of the Fermi energy $\ef$,
\begin{equation} \label{nos}
N = \int_0^{\ef} \rho(E) \ \dd E = \frac{m S}{2 \pi \hbar^2} \left( l+1 \right) \left(  \ef - \frac{1}{2} \hw l  \right),
\end{equation}
where the highest occupied oscillator state number $l$ is introduced through
\begin{equation}
l = \floor{\ef / \hw},
\end{equation}
where $\floor{\cdot}$ is the floor function.
Similarly one gets the total energy,
\begin{align} \label{kineticq}
E_\text{T} = &\int_0^{\ef} \rho(E) E \ \dd E = \nonumber \\
&\frac{m S}{4 \pi \hbar^2} \left( l+1 \right) \left[  \ef^2 - \frac{1}{3} \hbar^2 \omega_z^2 l \left(l+\frac{1}{2}\right)  \right].
\end{align}
Now, combining~\eqref{nos} and~\eqref{kineticq}, one can arrive at the Thomas-Fermi energy functional,
\begin{align}\label{fullkA}
\epsilon\left[n_{\text{2D}}\right]=\frac{\pi \hbar^2}{(l+1)m}n_{\text{2D}}^2 + \frac{1}{2} l \hw n_{\text{2D}} - \frac{l (l+1) (l+2) m \omega_z^2}{48 \pi},
\end{align}
where we have introduced the two-dimensional density $n_{\text{2D}} = N / S$.
As we have gotten rid of $\ef$ from the expression, now $l$ needs to be computed as a solution to
\begin{align} \label{defl}
\left \lfloor{\frac{2 \pi \hbar^2}{(l+1)m \hw}n_{\text{2D}} +\frac{1}{2}l}\right \rfloor=l.
\end{align}
Let us now construct the following auxiliary density, 
\begin{align} 
n_z\left(z,n_{\text{2D}}\right) = \sum_j p_j\left(n_{\text{2D}}\right) n_j(z),
\end{align}
that we will use both for comparison of our framework with full three-dimensional density and for derivation of the mean-field energy functional.
Here, the densities of oscillator eigenstates,
\begin{align}
n_j (z) = \frac{1}{2^j j!} \sqrt{\frac{m \omega_z}{\pi \hbar}} e^{-\frac{m \omega_z z^2}{\hbar}} H_j^2\left( \sqrt{\frac{m \omega_z}{\hbar}} z \right),
\end{align}
are weighted by the total number of atoms in a given oscillator state,
\begin{align}
 p_j\left(n_{\text{2D}}\right) = \frac{\int_0^{\ef} \rho_j(E) \ \dd E}{\int_0^{\ef} \rho(E) \ \dd E} = \frac{1}{l+1} + \frac{m \omega_z}{4 \pi \hbar n_{\text{2D}}} \left(l-2j\right),
\end{align}
where $H_j(x)$ is the j$^\text{th}$ Hermite polynomial.
The density $n_z$ is normalized, $\int n_z\left(z,n_{\text{2D}}\right) \dd z=1 $.

Let us now consider an additional, slowly varying trapping in $x-y$ plane, $V(x,y)$, such that two-dimensional density becomes nonuniform, $n_{\text{2D}} = n_{\text{2D}}(x,y)$.
Then, we will approximate the real three-dimensional density $n(x,y,z)$ through 
\begin{align} \label{totdens}
n (x,y,z) = n_{\text{2D}} (x,y) n_z\left[z,n_{\text{2D}} (x,y)\right],
\end{align}
where $\int n(x,y,z) \ \dd x \ \dd y \ \dd z = N$.

Note that $l = l(x,y)$ now also becomes a position-dependent quantity.
We analyze the accuracy of the functional~\eqref{fullkA} in App.~\ref{app:nonas}.
We show that at the dimensional crossover in an anisotropic harmonic trap, it outperforms the usual three-dimensional Thomas-Fermi energy functional, both for total energy estimation and predicting the density profile.

Next, we will consider a binary mixture of two spin-polarized Fermi gases with densities $n_1 (x,y,z)$ and $n_2(x,y,z)$.
Let us introduce total contact interaction energy,
\begin{equation}\label{etotint3d}
    E_\text{int} = g \int n_1 (x,y,z) \ n_2 (x,y,z) \ \dd x \ \dd y \ \dd z, 
\end{equation}
where $g$ is a three-dimensional coupling constant.
Then, we can write 
\begin{equation}
    E_\text{int} =\int  \epsilon_\text{int} [n_1 (x,y), n_2 (x,y)] \ \dd x \ \dd y, 
\end{equation}
where $\epsilon_\text{int} [n_1 (x,y), n_2 (x,y)]$ is a two-dimensional interaction energy functional and we dropped subscript 2D for clarity.
Using~\eqref{totdens}, the functional can be written as
\begin{align}
    &\epsilon_\text{int} [n_1 (x,y), n_2 (x,y)] =  \nonumber \\ &g \sqrt{\frac{m \omega_z}{\hbar}} n_1 n_2 \eta_1(l_1,l_2) + g \sqrt{\frac{m \omega_z}{\hbar}} \frac{m \omega_z}{2 \pi \hbar} n_1 \eta_2(l_1,l_2) +  \nonumber \\ & g \sqrt{\frac{m \omega_z}{\hbar}} \frac{m \omega_z}{2 \pi \hbar} n_2 \eta_2(l_2,l_1) + g \sqrt{\frac{m \omega_z}{\hbar}} \frac{m^2 \omega_z^2}{4 \pi^2 \hbar^2}  \eta_3(l_1,l_2),
\end{align}
with
\begin{align}
     &\eta_1 (l_1,l_2) =\nonumber \\ & \frac{1}{(l_1+1)(l_2+1)} \sqrt{\frac{\hbar}{m \omega_z}} \sum_{j_1,j_2}^{l_1,l_2} \int n_{j_1}(z) n_{j_2}(z) \dd z, \nonumber \\
     &\eta_2 (l_a,l_b) =\nonumber \\ & \frac{1}{l_a+1} \sqrt{\frac{\hbar}{m \omega_z}} \sum_{j_1,j_2}^{l_1,l_2} \left( \frac{l_b}{2} - j_b \right) \int n_{j_a}(z) n_{j_b}(z) \dd z, \nonumber \\
      &\eta_3 (l_1,l_2) =\nonumber \\ & \sqrt{\frac{\hbar}{m \omega_z}}  \sum_{j_1,j_2}^{l_1,l_2} \left( \frac{l_1}{2} - j_1 \right) \left( \frac{l_2}{2} - j_2 \right) \int n_{j_1}(z) n_{j_2}(z) \dd z .
\end{align}
Here, analogously to~\eqref{defl}, $l_1$ and $l_2$ need to be self-consistently solved through
\begin{align}
\left \lfloor{\frac{2 \pi \hbar^2}{\left[l_s (x,y)+1 \right]m \hw}n_s(x,y) +\frac{1}{2}l_s(x,y)}\right \rfloor=l_s(x,y),
\end{align}
with $s \in \{1,2\}$.
With these formulas, we are equipped to construct a density--potential functional theory framework for the dimensional crossover of a Fermi gas that will allow us to include a semilocal kinetic energy description.

\subsection{Density--potential functional theory}
The exact DFT energy functional can be rephrased as a bifunctional of the two-dimensional densities $\vec n$ (here, ${\vec n=\{n_s\}=(n_1,n_2)}$ for fermion species ${s=1,2}$ and from now on we drop 2D subscript unless stated otherwise) and effective potential energies $\{V_s\}$ that combine the interaction effects with the external potential energies $V_s^{\mathrm{ext}}$~\cite{Trappe2016,Trappe2017,Trappe2019,Englert2019a,Trappe2021a,Trappe2023,Englert2023,Trappe2023a,Trappe2023b}.
We find the ground-state densities $n_1$ and $n_2$ among the stationary points of this bifunctional by self-consistently solving
\begin{align}
\label{n} n_s[V_s-\mu_s](\vec r)=\frac{\delta E_1[V_s-\mu_s]}{\delta V_s(\vec r)}
\intertext{and}
\label{V} V_s[\vec n](\vec r)=V_s^{\mathrm{ext}}(\vec r)+\frac{\delta E_{\mathrm{int}}[\vec n]}{\delta n_s(\vec r)}\, .
\end{align}
Here, $E_1[V_s-\mu_s]$ is the Legendre transform of the kinetic energy functional $E_{\mathrm{kin}}[n_s]$, $\mu_s$ is the chemical potential for species $s$, and the interaction energy $E_{\mathrm{int}}[\vec n]$ generally couples all densities.
For the general formalism and many applications of this \textit{density--potential functional theory} (DPFT) we refer to \cite{Trappe2016,Trappe2017,Trappe2019,Englert2019a,Trappe2021a,Trappe2023,Englert2023,Trappe2023a,Trappe2023b} and references therein.
In particular, we will deploy the exact same machinery for two-component interacting fermion gases as in \cite{Trappe2021a}, with the twist of transferring the kinetic energy contribution that stems from the transversal direction to the interaction energy.
This augmentation of the DPFT framework allows us to predict the properties of 3D systems at the cost of 2D calculations.

Specifically, by adding and subtracting the TF kinetic energy density ${\epsilon^{\mathrm{TF,2D}}_{\mathrm{kin}}=\pi \hbar^2 n^2/m}$ for spin-polarized fermions in 2D, we write the full energy density (\ref{fullkA}) as
\begin{align}\label{epsilon2D3D}
\epsilon[n]=\epsilon^{\mathrm{TF,2D}}_{\mathrm{kin}}+\epsilon^{\mathrm{2D3D}}_{\mathrm{int}}\,.
\end{align}
Here, the effective interaction energy density
\begin{align}\label{epsilon2D3Dint}
\epsilon^{\mathrm{2D3D}}_{\mathrm{int}}=-\frac{\pi \hbar^2}{m}\frac{l}{l+1}n^2+\frac{1}{2}\hbar\,l\,n\,\omega_z-\frac{m\omega_z^2}{48\pi}l(l+1)(l+2)
\end{align}
compensates for the introduction of the 2D TF kinetic energy in (\ref{epsilon2D3D}).
Since $l(n)$ is piecewise constant, the functional derivative of the energy ${E^{\mathrm{2D3D}}_{\mathrm{int}}[n]=\int\d\vec r\, \epsilon^{\mathrm{2D3D}}_{\mathrm{int}}}$ is 
\begin{align}
\frac{\delta E^{\mathrm{2D3D}}_{\mathrm{int}}[n]}{\delta n(\vec r)}=\frac{-2\pi\,\hbar^2\, n(\vec r)\,l}{m(l+1)}+\frac{\hbar\,\omega_z\,l}{2},
\end{align}
which we incorporate into the effective interaction potential in (\ref{V}), such that we can execute the self-consistent program of (\ref{n}) and (\ref{V}) in a pure 2D setting.

Finally, we may replace the quasi-classical TF kinetic energy $E^{\mathrm{TF,2D}}_{\mathrm{kin}}$ by semiclassical approximations of the (Legendre-transformed) kinetic energy functional, viz., approximations of $E_1[V-\mu]$; all details of the numerical procedures are discussed in \cite{Trappe2021a}.
Accordingly, we deploy the nonlocal quantum-corrected successor
\begin{align}\label{n3p}
n_{3'}(\vec r)=\int(\d\vec r')\left(\frac{k_{3'}}{2\pi r'}\right)^D J_D(2r'\,k_{3'})
\end{align}
of the local TF density for $D$ dimensions, see \cite{Chau2018,Trappe2021a}, with the Bessel function $J_D(\,)$ of order $D$ and the effective Fermi wave number
\begin{align}
k_{3'}=\frac{1}{\hbar}\big[2m\big(\mu-V(\vec r+\vec r')\big)\big]_+^{1/2},
\end{align}
where ${[z]_+=z\,\Theta(z)}$, and $\Theta(\,)$ is the Heaviside step function.

The approximate DPFT framework introduced here can be applied in situations where the qualitatively different treatment of $E_{\mathrm{int}}$ [with quasi-classical TF approximation, Eq.~\eqref{epsilon2D3Dint}] and $E_{\mathrm{kin}}$ [with semiclassical $n_{3'}$ density formula, Eq.~\eqref{n3p}] is acceptable. 
We will determine to which extent this holds true by benchmarking $n_{3'}$-based DPFT densities against Hartree--Fock (HF) results.

\subsection{Orbital approach}
To perform this benchmark, we use time-dependent HF equations
\begin{align}\label{hfeqweuse}
i\hbar \frac{\partial}{\partial t}  \varphi_i^{(s)} ({\vec r},t) 
 = \left[ -\frac{\hbar^2}{2 m} \nabla^2 + V^{\text{ext}}_s({\vec r},t) \right. \nonumber \\
 + \left. \frac{\delta E_{\text{int}}[\vec n]}{\delta n_{s} (\vec r, t)} \; \right] \; \varphi_i^{(s)} ({\vec r},t)
\end{align}
for the two-component spin mixture (${s\in\{1,2\}}$).
The derivation of these equations is shown in App.~\ref{app:tdhf}. Here, $\varphi_i^{(1)}(\vec r, t)$ and $\varphi_i^{(2)}(\vec r, t)$, with ${i=1,...,N/2}$, are spatial orbitals of the first and the second spin component, respectively.
The interaction terms $\frac{\delta E_{\text{int}}[\vec n]}{\delta n_{1/2} (\vec r, t)}$ are defined below through equation~(\ref{etotint3d}).
The one-particle densities
\begin{eqnarray}
 n_s(\vec r, t) & = & \sum_{i=1}^{N/2} |\varphi_i^{(s)} (\vec r, t)|^2
\end{eqnarray}
associated with the spin components $s$ sum to the total one-particle density $n(\vec r, t) = n_1(\vec r, t)+n_2(\vec r, t)$.

We are looking for the ground-state densities.
We are solving the set of equations~(\ref{hfeqweuse}) by the imaginary time propagation technique~\cite{Aichinger2005} where real time is replaced by imaginary time $t\rightarrow -i\tau$.
After that, the evolution operator is no longer a unitary operator.
Both the norm and the orthogonality are lost during the imaginary time propagation.
To keep the orthogonality and the norm of the spatial orbitals we use the Gram-Schmidt orthonormalization technique.
This way, propagating $N$ spatial orbitals we obtain the ground state of $N$ particles.
\begin{figure*}[h!t!]
    \centering
    \includegraphics[width=\linewidth]{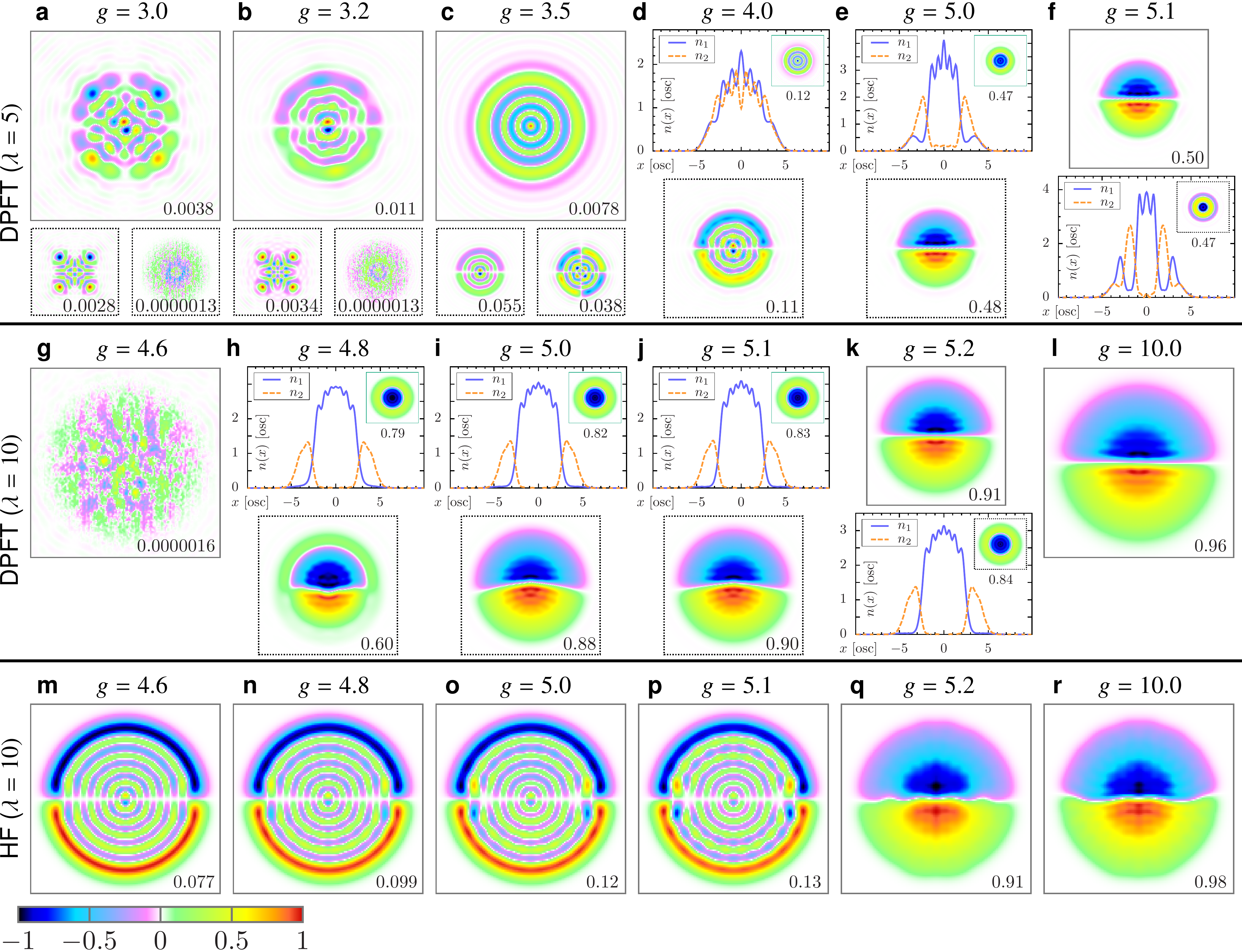}
    \caption{The density profiles for a two-component repulsive Fermi gas across interaction-driven phase transitions.
    The top (bottom) plots in each panel present the ground-state (metastable, closely matched in energy) profiles.
    The density profiles are plotted either as density differences $n_1-n_2$, scaled by $\mathrm{max}_{\vec r}|n_1(\vec r)-n_2(\vec r)|$, which are presented as contour plots in a square box with edge lengths of 6~[osc] (the color legend applies to all the contour plots); or as density cross-sections at ${y=0}$.
    (a-f) Ground-state density profiles for a fixed aspect ratio of the harmonic trap ($\lambda = 5$) and a constant total number of particles ($N_1 = N_2 = 55$).
    These profiles vary with the three-dimensional repulsive coupling constant between the two species, denoted as $g$, and are characterized through a global polarization $\mathcal{P}$.
    Notably, two distinct phase transitions are observable: one from a paramagnetic to an isotropic, partially polarized phase, followed by a transition to a two-hemisphere state.
    The findings are obtained using a density-potential functional approach.
    (g-l) Similarly presented are ground-state density profiles for a higher aspect ratio of the trap ($\lambda = 10$).
    (m-r) The results in this row are analogous to those in the second row; however, they are computed using the Hartree-Fock approach.
    Although the weak- and large-interaction limits yield comparable outcomes, disparities become evident within the phase transition regime. 
    See Table~\ref{Table1} for details of each panel, including interaction strength, scaling factor, energy, polarization, and transverse excitations.
    }
    \label{fig1}
\end{figure*}

\begin{figure*}[ht!]
    \centering
    \includegraphics[width=\linewidth]{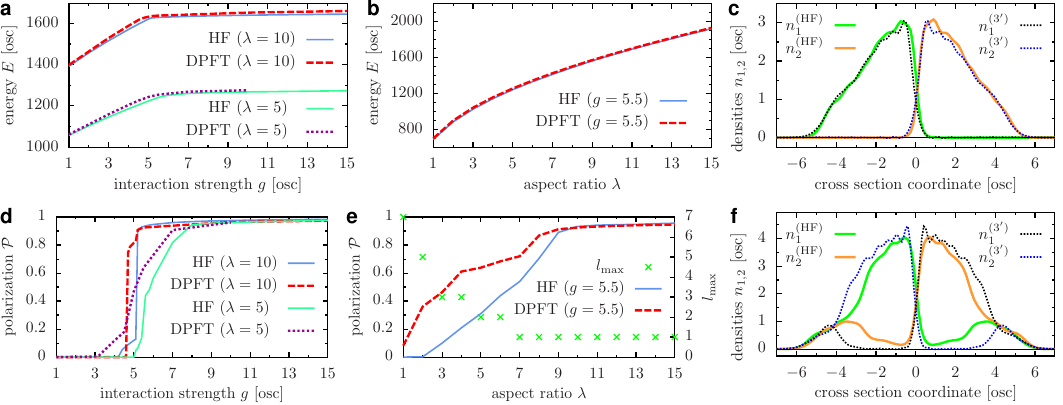}
    \caption{(a,b) Total energy of a two-component Fermi gas throughout interaction and aspect ratio sweeps for two methods---density--potential functional theory and Hartree-Fock.
    Both methods show similar predictions.
    (d,e) The same but for the total polarization. 
    In this case, the two methods match for strong interactions and high aspect ratios, but otherwise yield density profiles with quite different polarizations---a sign of near-degenerate states across the transitions w.r.t. $g$ and $\lambda$.
    (c,f) Comparison of density cross-section for both methods. The slopes of the densities match very well for strongly interacting clouds.}
    \label{fig2}
\end{figure*}

\section{Results}\label{results}
Our focus lies predominantly on analyzing phase transitions in a binary Fermi gas, dependent on three key physical parameters: the number of particles $N$, the aspect ratio $\lambda$, and the strength of three-dimensional interactions $g$.
Specifically, we consider a balanced two-component Fermi gas at zero temperature; each component consisting of $N_1 = N_2 = N/2$ atoms and trapped in the anisotropic harmonic potential with aspect ratio $\lambda = \omega_z / \omega$,
\begin{equation}
    V(x,y,z) = \frac{1}{2} m \omega^2 \left( x^2 + y^2 + \lambda^2 z^2  \right).
\end{equation}
Both species interact via contact repulsion given by three-dimensional pairwise delta interaction characterized by the coupling constant $g = 4 \pi a \hbar^2 / m $, where $a$ is the scattering length.
Here, we focus only on a mean-field contribution~\eqref{etotint3d} coming from this interaction, hence neglecting quantum contributions that stem from the exact, many-body ground state.
In the case of the repulsive binary Fermi mixture, such an omission shifts the critical values of interaction strengths at which phase transitions happen, both in the two-~\cite{Trappe2021a} and three-dimensional geometry~\cite{Grochowski2020}, however leaving qualitative features unchanged for nonuniform mixtures.
The refinement of the energy spectrum of repulsive Fermi gas beyond the mean field has been performed via many different approaches, including perturbation theory~\cite{Bloom1975,Duine2005,He2014}, quantum Monte Carlo simulations~\cite{Conduit2009,Conduit2010,Pilati2010,Chang2011,Conduit2013,Bertaina2013,Whitehead2016,Pilati2021}, lowest-order constraint variational calculation~\cite{Heiselberg2011}, nonperturbative ladder approximation ~\cite{He2012}, large-$N$ expansion, dimensional $\epsilon$-expansion~\cite{He2016}, and polaronic approach~\cite{Schmidt2012,Ngampruetikorn2012}.
However, most efforts focused on either pure two- or three-dimensional limit, while dimensional crossover has been analyzed only for attractive mixtures~\cite{Dyke2011,Sommer2012,Cheng2016,Ptok2017,Toniolo2017,Peppler2018,Toniolo2018,Adachi2018,Faigle-Cedzich2021,Gong2023,Zheng2023}.
As such, we focus on the mean-field description of interaction and leave further refinement as an outlook.

The transitions we consider manifest as distinct phase separations in the ground-state densities of the gas components.
Past research has highlighted that in radially symmetric harmonic trapping, transitions occur from uniform density profiles of the two fermion species to both isotropic and anisotropic separations.
Furthermore, within a purely two-dimensional configuration with bare contact interaction, an analogous transition arises~\cite{Trappe2021a}.
This transition shifts from a paramagnetic state at low repulsive interactions to ferromagnetic density profiles at higher interaction strengths.
However, intricate particle-number-dependent phases emerge between these limits.
Moreover, a range of metastable configurations with energy levels comparable to ground-state density profiles have been identified within the transitional regime. 
These configurations are likely to be observed in experimental settings.
Hence, our analysis seeks to uncover the crossover between these two scenarios, achievable by transitioning dimensionally through adjustments in the trap's aspect ratio.
For the analysis of spatial separation at the crossover, we will utilize a total polarization $\mathcal{P}$ of the trapped mixture:
\begin{align}\label{polarization}
\mathcal{P}= \frac{\int \d \mathbf{r} \left| n_1(\mathbf{r}) - n_2 (\mathbf{r}) \right|}{N}.
\end{align}

\subsection{Interaction-driven phase transitions}
We begin by examining a specific scenario where we vary interaction strengths while keeping the aspect ratio $\lambda = 5$ and particle count $N_1 = N_2 = 55$ constant.
We present it in Fig~\ref{fig1}(a-f).
At low interaction strengths, a paramagnetic phase is observed, where nearly identical density profiles with accordingly small energy differences make it difficult to distinguish the ground state from metastable states.
Although overall polarization remains low as we increase the interaction strength to around $g=3.5$, slight polarization modulations emerge near the center due to a relative decrease in interaction energy at the expense of kinetic energy.
At ${g=3.5}$ this trade-off between the energy components begins to favor ground-state profiles with isotropic separations.
This transition is characterized by visible domains without breaking radial symmetry.
We identify states in close energy proximity that break radial symmetry and possess nonzero polarization.
These states bear a resemblance to findings in purely 2D scenarios.
Further along, as the interaction strength reaches $g=5$, the gas segregates into two hemispheres, mirroring the analogous behavior observed in both 2D and 3D.
\begin{figure*}[h!t!]
    \centering
    \includegraphics[width=\linewidth]{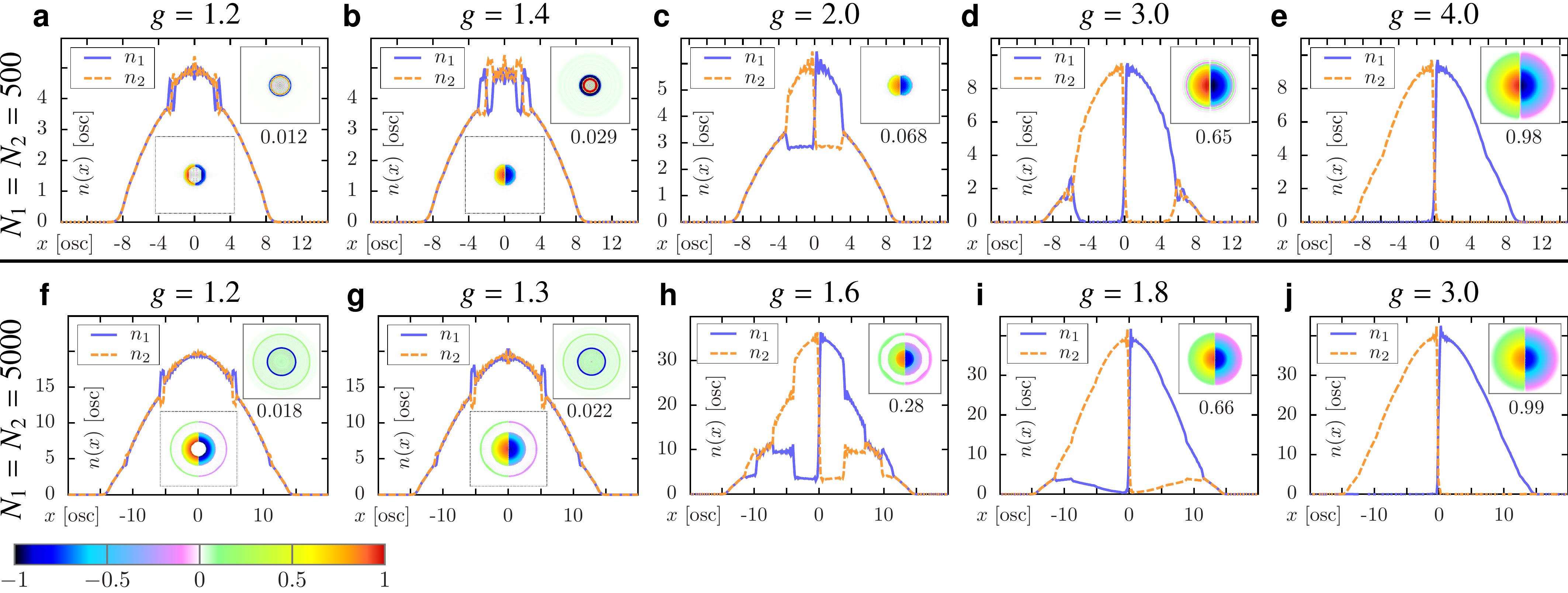}
    \caption{(a-e) Ground-state density profiles for a two-component repulsive Fermi gas, with a fixed aspect ratio of the harmonic trap ($\lambda = 25$) and a constant total number of particles ($N_1 = N_2 = 500$). These profiles vary with the three-dimensional repulsive coupling constant between the two species, denoted as $g$. Richer density profile structure across the para- to ferromagnetic phase transition is visible, as compared to the lower atom number case, exhibiting ring-shaped polarization and central-split patterns.
    (f-j) The same but for $\lambda = 30$ and $N_1 = N_2 = 5000$. The density profile gets even more intricate, showing the coexistence of ring-shaped polarization and central-split patterns for some values of interaction strength. The main plot and the top right inset in each panel present the ground-state profiles. In panels a, b, f, and g, we also show metastable profiles. The density differences $n_1-n_2$, scaled by ${\mathrm{max}}_{\vec r}|n_1(\vec r)-n_2(\vec r)|$, are presented as contour plots in a square box with edge lengths of 10~[osc] (15~[osc]) for $N_{1/2}=500$ ($N_{1/2}=5000$); the color legend applies to all the contour plots. See Table~\ref{Table2} for details.}
    \label{fig3}
\end{figure*}

Subsequently, we compare these results with those from an altered aspect ratio, $\lambda = 10$ [refer to Fig.~\ref{fig1}(g-l)].
We observe analogous behavior; however, the initial transition from no separation to isotropic separation is more pronounced, lacking an easily distinguishable transitional regime characterized by minor polarization modulations.
This transition occurs at $g=4.65$ and is followed by a mirror-symmetric separation at ${g=5.15}$ that grows into an almost complete split separation at ${g=10}$.
Again, we find metastable states across the transitions, which show competition between different types of splittings.
However, for this particular aspect ratio and atom number, the competition is mainly between isotropic and anisotropic separation states.

These findings are consistent with both two- and three-dimensional cases; however, they indicate that the fine structure of coexisting metastable states strongly depends on the perpendicular excitation structure.
Symmetry-breaking partially polarized profiles are unveiled for lower aspect ratios when the perpendicular degree of freedom is more intensely excited, in contrast to the higher aspect ratio case where such structures are not easily discernible, cf. Table~\ref{Table1}.
This showcases nontrivial behavior, similar to the scenario in the limit of pure two-dimensional geometry where these metastable states are present.

\subsection{Benchmarking DPFT against Hartree–Fock}
We now move on to validate the previously described low-atom-number outcomes using the orbital approach.
We compare density profiles throughout the interaction-induced phase transition for $\lambda = 10$ and $N_1 = N_2 = 55$, as displayed in Fig~\ref{fig1}(m-r).
While we observe that within the weak interaction limit, the paramagnetic, identical density profiles of both clouds are consistent across both methods, a slight discrepancy emerges at the onset of the phase transition.
The isotropic transition is not evident, but instead, the ground state just above the transition assumes a partially polarized, symmetry-breaking configuration.
It's worth noting that the density structure of this state bears a resemblance to certain profiles of metastable states identified using the DPFT method.
This observation suggests that the ground state at the transition is nearly degenerate, with various density profiles being realized by states with minute energy differences.
We hypothesize that the divergence between the methods regarding the true ground state arises from the proximity of these states.
Importantly, in Hartree-Fock calculations, the subsequent phase transition to a two-hemisphere state occurs at a similar value of the interaction strength as observed in the DPFT method.
In the scenario of a large interaction limit, density profiles between the two methods align remarkably well.

To conduct a more comprehensive analysis of this comparison, we present in Fig.~\ref{fig2} the comparison between the two methods across both interaction and aspect ratio variations.
We employ total energy and total polarization as metrics for assessment.
First, we find that energies match very well across both transitions, suggesting consistency between both methods.
As for the total polarization, the behavior across the transition is qualitatively captured with a good quantitative match in the strong coupling regime.
This slight quantitative mismatch suggests that the total polarization is a sensitive probe for specifying which state is realized experimentally.
Importantly, comparing the cuts through the density profiles, we find that the slopes of the density profiles match in both methods, implying a mutually consistent description of the interplay of kinetic and interaction energies in both methods. 
Such a behavior is of particular interest as domain-wall density profile determines, e.g., dimer formation rates in ultracold gases.

\begin{figure*}[h!t!]
    \centering
    \includegraphics[width=\linewidth]{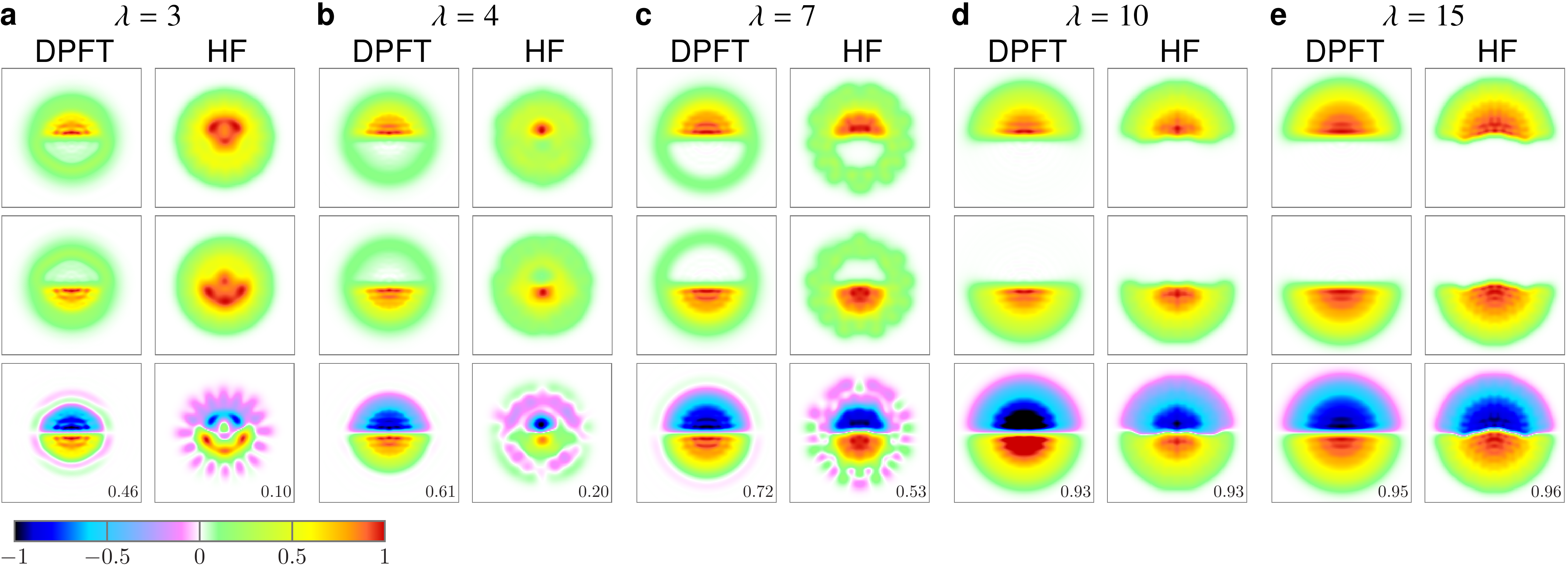}
    \caption{Ground-state density profiles for a two-component repulsive Fermi gas, with a fixed coupling constant between two fermionic components ($g = 5.5$) and a constant total number of particles ($N_1 = N_2 = 55$). These profiles vary with the aspect ratio of the harmonic trap $\lambda$, exhibiting a transition from partially to fully polarized state. The top (bottom) plots in each panel present the ground-state (metastable) profiles. The density differences $n_1-n_2$, scaled by ${\mathrm{max}}_{\vec r}|n_1(\vec r)-n_2(\vec r)|$, are presented as contour plots in a square box with edge lengths of 6~[osc]; the color legend applies to all the contour plots. See Table~\ref{Table3} for details.}
    \label{fig4}
\end{figure*}

\subsection{Large-atom-number limit}
We now proceed to analyze large atom number setups that go beyond the manageability of the Hartree-Fock approach due to numerical cost.
In Fig.~\ref{fig3} we present two cases, $N_1 = N_2 = 500$ and  $N_1 = N_2 = 5000$, both computed with DPFT approach.
In these cases, the shell structure of the density profiles, usual in low-dimensional experimental setups, i.e., sharp transitions in the densities due to discretized energy spectrum in transversal direction, becomes apparent.
First, we observe that the existence of these sharp density changes makes the competition between kinetic and interaction energy even more intricate.
As we go through the interaction-induced transition for $N_1 = N_2 = 500$, we find that partial polarization may be favored at these density changes, revealing a ring-shaped polarization pattern that might or might not preserve radial symmetry.
As the interaction strength increases above ${g \approx 1.4}$, the preferred density profile involves the anisotropic split at the center of the trap and a shell structure at the perimeter.
With the further growing interaction, this central split occupies more volume, becoming two fully separated hemispheres in the strong interaction regime.
Notably, in this case, no ground-state isotropic separation is observed. 
The $N_1 = N_2 = 5000$ case exhibits similar behavior.
Across the phase transition toward a ferromagnetic state, one observes the coexistence of both isotropic and anisotropic ring-shaped polarization structures and the central anisotropic split, shown in the lower atom number case.

\subsection{Geometry-driven transitions}
Up to now, we have analyzed a phase transition driven by the varying interaction.
Here, we would like to focus on the case in which the ground-state density profile is altered through the change of geometry via the aspect ratio of the harmonic trap.
In Fig.~\ref{fig4} we plot the transition for $g=5.5$ and $N_1 = N_2 = 55$, while the aspect ratio is changed from $\lambda = 3$ to $\lambda = 15$ [compare Fig.~\ref{fig2}(e), where total polarization is plotted for this transition].
Such a sweep realizes a transition from a partially polarized anisotropic split to the fully separated hemispheres.
In the limiting cases, the phase separation occurs at $g \approx 7$ for $\lambda = 1$ the~\cite{Trappe2016a} and at $g \approx 15.7 / \sqrt{\lambda}$ for the two-dimensional limit $\lambda \rightarrow \infty$~\cite{Trappe2021a} (which equals $g = 5.0$ and $g = 4.1$ for $\lambda = 10$ and $\lambda = 15$, respectively).
It shows how the geometry can be used to drive the phase transition, similarly to experiment with, e.g., confinement-induced resonances~\cite{Haller2010a}.

\section{Conclusions and outlook}\label{conclusions}

Summarizing, we have analyzed para- to ferromagnetic phase transition in a binary zero-temperature repulsive Fermi gas at a two- to three-dimensional crossover utilizing two distinctive methods---density--potential functional theory and Hartree-Fock methods.
We have found out that in a quasi-two-dimensional regime, the ground-state profile across the phase transition is nearly degenerate and exhibits a variety of shapes, including isotropic and anisotropic separations, ring-shaped polarization patterns, and central-split patterns in the usual shell structure.
These density profiles can be tuned via means of varying particle number, interaction strength, and aspect ratio of the external trapping, providing a versatile playground for the physics of interacting quantum mixtures.

As argued above, we have omitted the quantum corrections to the mean-field interaction that would shift the values of the interaction strength at which the transitions happen.
The inclusion of these corrections will be a subject of future work.
Another limitation of the considered model comes from the neglection of thermal excitations---in the context of itinerant ferromagnetism and phase separations in ultracold gases, finite-temperature effects have been considered for Fermi-Fermi~\cite{Massignan2013,Massignan2014,Ryszkiewicz2020,Ryszkiewicz2022} and Bose-Fermi~\cite{Grochowski2020} mixtures in two and three dimensions, repulsive polarons~\cite{Massignan2014,Tajima2018}, many-component mixtures~\cite{Huang2023}, and dipolar gases~\cite{Feng2020}.
Nonzero temperature results in the shift of the critical value of interaction strength towards higher values of the interaction~\cite{Massignan2014,Valtolina2017,Ryszkiewicz2020,Ryszkiewicz2022}, it can change the order of transition~\cite{Huang2023}, and if sufficiently large, thermal fluctuations make the phase separation disappear~\cite{Massignan2014}.
In the case we consider, exhibiting a plethora of fine polarization patterns, finite temperature would most likely additionally smear out small density fluctuations, similar to what we observed in two dimensions~\cite{Trappe2021a}.
However, such an analysis is outside of the scope of this work and is left out as an outlook for the future.

The experimental realizations of a two-component repulsive Fermi mixture in two and three dimensions, both single- and two-element, have been presented with many different species, including lithium~\cite{Jo2009,Dyke2011,Ong2015,Valtolina2017,Amico2018,Ji2022}, potassium~\cite{Frohlich2011,Kohstall2012,Koschorreck2012}, chromium~\cite{Ciamei2022}, ytterbium~\cite{DarkwahOppong2019}, and dysprosium~\cite{Ravensbergen2020}.
These realizations provide a natural experimental testbed for fermionic phase separation due to repulsion, also beyond the considerations of this work---involving different mass ratios and mixtures with different species---so that competing effects such as molecule formation could be suppressed and ferromagnetic correlations revealed.
Another avenue for fermionic phase separation involves repulsive Bose-Fermi mixtures that recently gained momentum, both theoretically~\cite{Huang2019,Grochowski2020,vonMilczewski2022,DAlberto2024} and experimentally~\cite{Lous2018,DeSalvo2019}.

As a concluding remark, it's worth noting that the multiparticle Hamiltonian governing the system in this paper exhibits axial symmetry.
This implies that, unless degenerate, the ground state must also possess axial symmetry.
However, the observed symmetry-breaking single-particle patterns, detailed in this study, are entirely physical as they accurately describe individual experimental shots.
This somewhat paradoxical aspect of the density functional method was recently highlighted in~\cite{Perdew2021a}.

\section{Acknowledgements}
The Center for Theoretical Physics of the Polish Academy of Sciences is a member of the National Laboratory of Atomic, Molecular and Optical Physics (KL FAMO).
This work has been supported by the National Research Foundation, Singapore and A*STAR under its CQT Bridging Grant and its Quantum Engineering Programme.
Part of the results were obtained using computers of the Computer Center of the University of Bia{\l}ystok.

\appendix

\section{Testing the energy functional in the noninteracting case}\label{app:nonas}
Let us consider a noninteracting spin-polarized Fermi gas trapped in an anisotropic harmonic trap with an aspect ratio $\lambda$,
\begin{equation}
    V(x,y,z) = \frac{1}{2} m \omega^2 \left( x^2 + y^2 + \lambda^2 z^2  \right).
\end{equation}
The ground state of the gas is given by a Slater determinant of $N$ lowest-energy harmonic single-particle eigensolutions,
\begin{align}
    \psi_{ijk}(x,y,z) &= \psi_i(x) \psi_j(y) \lambda^{1/4} \psi_k\left( \lambda^{1/2} z \right), \nonumber \\
    \psi_{i}(x) &= \frac{1}{\sqrt{2^i i!}} \left( \frac{m \omega}{\pi \hbar}\right)^{1/4} e^{- \frac{m \omega x^2}{2 \hbar}} H_i\left( \sqrt{\frac{m \omega}{\hbar}}\right),
\end{align}
with a single-particle energy $E_{ijk} = \hbar \omega (i + j + \lambda k + 1)$, where again we subtracted zero-point energy of $z-$mode.
The total energy of the gas then reads
\begin{align}\label{exEn}
    E_{\text{ex}} = \sum_{n=1}^{N} E_n,
\end{align}
where $E_n = E_{ijk}$ and associated single-particle wave function $\psi_n = \psi_{ijk}$ are ordered such that $E_{n+1} \geq E_n$.
The single-particle density $n_{\text{ex}}(x,y,z)$ then can be expressed as
\begin{align}
    n_{\text{ex}}(x,y,z) = \sum_{n=1}^{N} \left|\psi_{n}(x,y,z)\right|^2,
\end{align}
and we can additionally define a column density,
\begin{align}\label{n2dex}
    n^{\text{ex}}_\text{2D} (x,y) = \int \dd z \  n_{\text{ex}}(x,y,z).
\end{align}
To get a closed expression for each of these quantities in the large-$N$ limit, one can resolve to utilize a Thomas-Fermi approximation along with the local density approximation.
Here, we will compare the functional introduced in Sec.~\ref{methods:functional} and the usual approach involving three-dimensional Thomas-Fermi functional,
\begin{align} \label{TFeq1}
    E_\text{TF} \left[ n, \mu \right] = E_\text{kin} + E_\text{pot} + \mu \left( N - \int \dd x \dd y \dd z \ n(x,y,z) \right)
\end{align}
with 
\begin{align}
    E_\text{kin} \left[ n, \mu \right] &= A\int \dd x \dd y \dd z \ n^{5/3}(x,y,z)  \nonumber \\
    E_\text{pot} \left[ n, \mu \right] &= \int \dd x \dd y \dd z \ n(x,y,z)   V(x,y,z),
\end{align}
where $A = 6^{5/3} \hbar^2 \pi^{4/3} / 20 m$ and $\mu$ is a chemical potential.
The minimization of~\eqref{TFeq1} yields Thomas-Fermi equations
\begin{align}
    \frac{3}{5} A n^{2/3}(x,y,z) = \mu - V(x,y,z).
\end{align}
They can be readily solved to obtain
\begin{align} \label{n2dtf}
    n_\text{TF}(x,y,z) &= \left( \frac{3}{5} A^{-1}\right)^{3/2} \left[ \mu_\text{TF} - V(x,y,z) \right]^{3/2}, \nonumber \\
    \mu_\text{TF} &= 6^{1/3} \hbar \omega N^{1/3} \lambda^{1/3}, \nonumber \\
    n^{\text{TF}}_\text{2D} (r) &= \frac{m}{4 \pi \omega \lambda \hbar^3 } \left(\mu_\text{TF} - \frac{1}{2} m \omega^2 r^2  \right)^2,\nonumber \\
    E_{\text{TF}} &= 3^{4/3} 2^{-5/3} \hbar \omega N^{4/3} \lambda^{1/3} - E_0,
\end{align}
where $r = \sqrt{x^2 + y^2}$, $E^{\text{TF}}$ is a total energy with subtracted zero-point contribution of $z-$mode, $E_0 = \frac{1}{2} \hbar \omega N \lambda $.
Similar minimization for~\eqref{fullkA} gives
\begin{widetext}
\begin{align}\label{n2dus}
    n_\text{2D} (r) &= \frac{m}{2 \pi \hbar^2} 
    \begin{cases}
      \mu_\text{2D} - \frac{1}{2} m \omega^2 r^2, & r_1 < r < r_0, \\
      \cdots &  \\
      \left( j+1 \right)\left(\mu_\text{2D} - \frac{1}{2} m \omega^2 r^2  -\frac{1}{2} j \lambda \hbar \omega \right), & r_{j+1} < r < r_j, \\
      \cdots &  \\
      \left( l_{\text{m}}+1 \right)\left(\mu_\text{2D} - \frac{1}{2} m \omega^2 r^2  -\frac{1}{2} l_{\text{m}} \lambda \hbar \omega \right), & 0 < r < r_{l_{\text{m}}}, \\
    \end{cases} \nonumber \\
  \mu_\text{2D} &= \hbar \omega \frac{ 3 \lm \rounds{\lm + 1} \lambda  + \sqrt{72 \rounds{\lm+1} N - 3 \lm \rounds{\lm + 1}^2 \rounds{\lm + 2} \lambda^2 }}{6 \left(l_{\text{m}}+1 \right)}, \nonumber \\ 
  E_\text{2D} &= \hbar \omega \frac{\lm + 1}{24} \squares{ \lm^2 \rounds{\lm+1} \lambda^3 - \frac{1}{9} \rounds{ 3 \lm \lambda + \sqrt{ \frac{72 N}{ \lm + 1} - 3 \lm \rounds{\lm+2} \lambda^2  }  }^2 \rounds{ \frac{1}{2} \lm \lambda - \frac{1}{3} \sqrt{ \frac{72 N}{ \lm + 1} - 3 \lm \rounds{\lm+2} \lambda^2  }   }      },
\end{align}
\end{widetext}
where $\lm$ is given implicitly through
\begin{align}
 \lm =  \left \lfloor{  \frac{1}{2} \lm + \frac{1}{6 \lambda} \sqrt{ \frac{72 N}{ \lm + 1} - 3 \lm \rounds{\lm+2} \lambda^2  }  }\right \rfloor.
\end{align}
One can find an asymptotic behavior 
\begin{align}
 \lm  \xrightarrow{N/\lambda^2  \rightarrow \infty} 6^{1/3} N^{1/3} \lambda^{-2/3},
\end{align}
from which other asymptotics follow
\begin{align}
 \mu_\text{2D} & \xrightarrow{N/\lambda^2 \rightarrow \infty} \mu_\text{TF}, \nonumber \\
 E_\text{2D} & \xrightarrow{N/\lambda^2 \rightarrow \infty} E_\text{TF} + E_0, \nonumber \\
 n_\text{2D} (r_j) & \xrightarrow{N/\lambda^2 \rightarrow \infty} n^{\text{TF}}_\text{2D} (r_j),
\end{align}
showing that in the limit of a large number of atoms and low trap anisotropy, the two approaches tend to the same solution.
If $N$ is taken to be finite, then the two methods differ.
We present the comparison in Fig.~\ref{figa1}, showing that the usual three-dimensional Thomas-Fermi approach is less accurate for the energy estimation and does not reproduce the sharp features of the density profile appearing due to change of transverse states.
\renewcommand{\thefigure}{A\arabic{figure}}
\setcounter{figure}{0}
\begin{figure}[ht!]
    \centering
    \includegraphics[width=\linewidth]{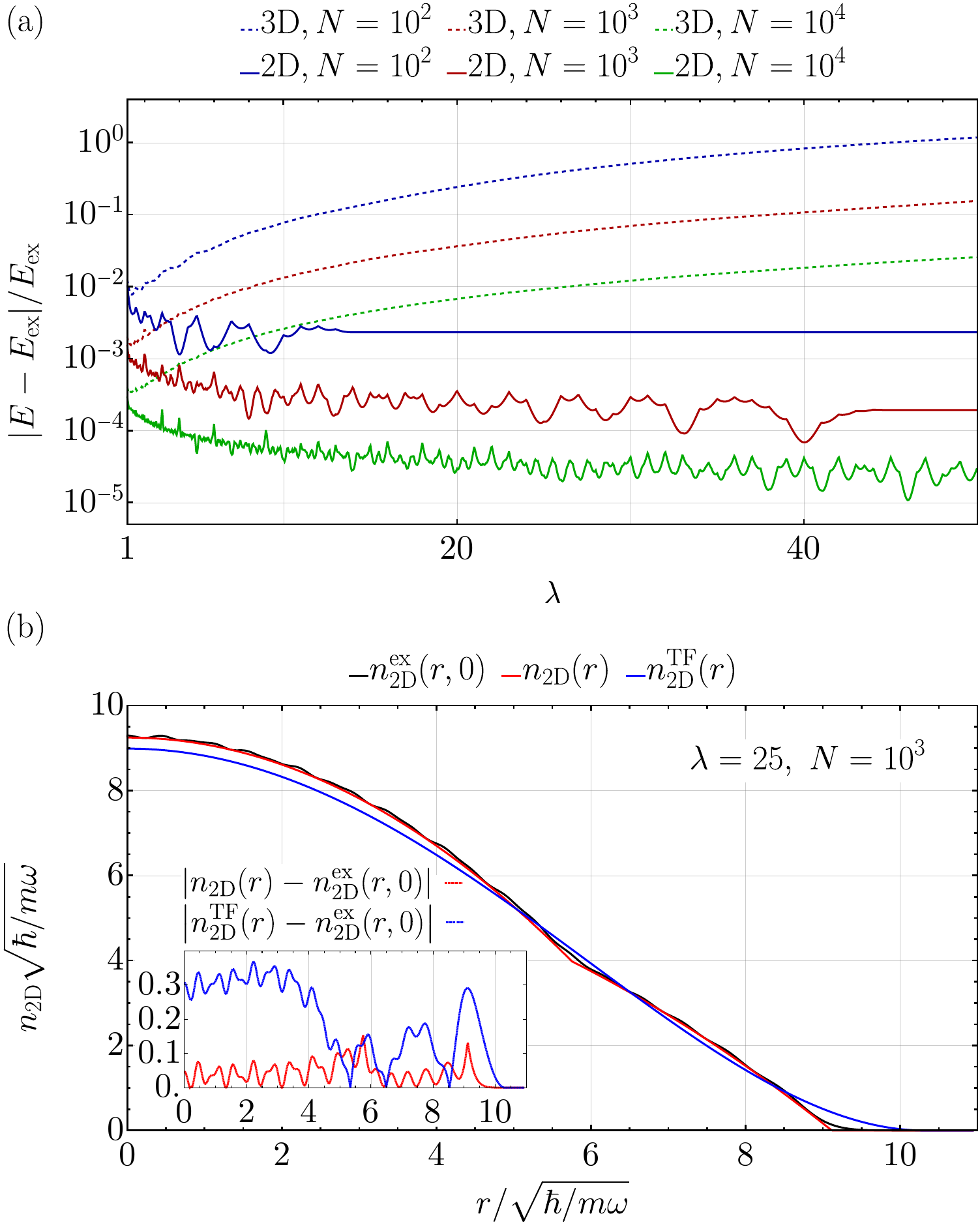}
    \caption{(a) Comparison of energies obtained through minimization of either functional~\eqref{fullkA} (label 2D, solid lines) or functional~\eqref{TFeq1} (label 3D, dashed lines) with respect to exact energy~\eqref{exEn} for different trap anisotropies $\lambda$ and atom numbers $N$.
    The former function provides more accurate results with the highest difference at large anisotropies.
    (b) Comparison of the density profiles obtained through two methods [given through Eqs.~\eqref{n2dus} and~\eqref{n2dtf}] and exact profile~\eqref{n2dex} for $\lambda = 25$ and $N = 1000$.
    The density profile exhibits a sharp transition around $r / \sqrt{\hbar / m \omega} = 5.5$ that is reproduced only with the former method.
    Inset: absolute density differences between two methods and exact density value.}
    \label{figa1}
\end{figure}

\section{Derivation of time-dependent Hartree-Fock equations}\label{app:tdhf}
In HF approximation one assumes that $N$ fermions are described by the wave function in the following form
\begin{eqnarray}
\Psi = \frac{1}{\sqrt{N!}} \left |
\begin{array}{llllll}
\psi_1(1) & \psi_1(2) & . & . & . & \psi_1(N) \\
\psi_2(1) & \psi_2(2) & . & . & . & \psi_2(N) \\
\phantom{aa}. & \phantom{aa}. & . &  &  & \phantom{aa}. \\
\phantom{aa}. & \phantom{aa}. &  & . &  & \phantom{aa}. \\
\phantom{aa}. & \phantom{aa}. &  &  & . & \phantom{aa}. \\
\psi_{N}(1) & \psi_{N}(2) & . & . & . & \psi_{N}(N)
\end{array}
\right |,
\label{slater}
\end{eqnarray}
where $\psi_i(j)$ are spin-orbitals.
In general, the spin-orbital can be written as
 \begin{equation}
  \psi_i(j) = \left[
  \begin{array}{c}
   \varphi_i(1,{\vec r}_j,t) \\
   \varphi_i(2,{\vec r}_j,t) \\
   \dots \\
   \varphi_i(s,{\vec r}_j,t)
  \end{array}
  \right],
 \end{equation}
where $\varphi_i(s_j,{\bf r}_j,t)$ are spatial orbitals. The spin-orbitals fulfill the orthonormality condition
 \begin{eqnarray}
  \langle\psi_i | \psi_k \rangle & = &
  \sum_{s_j=1}^{s_{max}}\int
   \varphi_i^{*}(s_j,{\vec r}_j,t)
   \varphi_k(s_j,{\vec r}_j,t)
   d{\vec r}_j \nonumber \\
   & = & \delta_{ik}\,,
 \end{eqnarray}
where $s_{max}$ is the number of spin components.
The wave function can be used to construct the lagrangian density
\begin{eqnarray}
\mathcal{L} & = & \frac{i\hbar}{2} \Psi^{\dagger} \frac{\partial \Psi}{\partial t}
- \frac{i\hbar}{2} \left( \frac{\partial \Psi^{\dagger}}{\partial t} \right) \Psi
- \frac{\hbar^2}{2 m} \sum_{i=1}^{N} \nabla_i \Psi^{\dagger} \nabla_i  \Psi
\nonumber \\
& - & \sum_{i=1}^{N} \Psi^{\dagger} V_{ext}({\vec r}_i) \Psi
 - \sum_{i<j} \Psi^{\dagger} V_{int}({\vec r}_i-{\vec r}_j)  \Psi\,.
\end{eqnarray}
Then one can build the Lagrangian
\begin{equation}
 L = \int \mathcal{L}\, d{\vec r}_1 \, d{\vec r}_2 \cdots d{\vec r}_N\,.
\end{equation}
And finally the action
\begin{equation}
 S = \int_{t_1}^{t_2} L\, dt\,.
\end{equation}
The principle of stationary action reads
\begin{eqnarray}
 0 = \delta S & = & \sum_{i}\left[ \frac{\partial S}{\partial \varphi_i^{*}(s_j,{\vec r}_j,t)}
 \delta \varphi_i^{*}(s_j,{\vec r}_j,t) \right. \nonumber \\
 & + & \left.
 \frac{\partial S}{\partial \varphi_i(s_j,{\vec r}_j,t)}
 \delta \varphi_i(s_j,{\vec r}_j,t) \right]
\end{eqnarray}
The variations over $\varphi_i^{*}(s_j,{\vec r}_j,t)$ and $\varphi_i(s_j,{\vec r}_j,t)$ are independent.
Taking the variation over $\varphi_i^{*}(s_j,{\vec r}_j,t)$ one gets the Euler-Lagrange equation for $\varphi_i(s_j,{\vec r}_j,t)$.
Here the Euler-Lagrange equations are called the Hartree-Fock equations and are the following
\begin{eqnarray}
&\phantom{.}& i\hbar \frac{\partial}{\partial t}  \varphi_i(s,{\vec r},t) =
 \left[ -\frac{\hbar^2}{2 m} \nabla^2 + V_{ext}({\bf r}) \right] \varphi_i(s,{\vec r},t) \nonumber \\
&\phantom{.}& + \sum_{k=1}^{N}\sum_{s^{\prime}=1}^{s_{max}}\int\,d^3r^{\prime}
\Big[ \varphi_i(s,{\vec r},t) V_{int}({\vec r}-{\vec r^{\prime}}) \times \nonumber \\  
&\phantom{.}& \varphi_k^{*}(s^{\prime},{\vec r^{\prime}},t)
\varphi_k(s^{\prime},{\vec r^{\prime}},t)  \nonumber \\
&\phantom{.}&  - \varphi_k(s,{\vec r},t) V_{int}({\vec r}-{\vec r^{\prime}}) \varphi_k^{*}(s^{\prime},{\vec r^{\prime}},t)
\varphi_i(s^{\prime},{\vec r^{\prime}},t) \Big]\,.
\end{eqnarray}
In our case
$s = 1, 2$ or $\uparrow, \downarrow$. Then we assume that 
\begin{equation}
 \left[
  \begin{array}{c}
   \varphi_i(1,{\vec r},t) \\
   \varphi_i(2,{\vec r},t)
  \end{array}
  \right]\equiv\left[
  \begin{array}{c}
   \varphi_i^{(1)}({\vec r},t) \\
   0
  \end{array}
  \right]
\end{equation}
  for $i=1,...,N/2$ and
\begin{equation}
  \left[
  \begin{array}{c}
   \varphi_i(1,{\vec r},t) \\
   \varphi_i(2,{\vec r},t)
  \end{array}
  \right]\equiv\left[
  \begin{array}{c}
  0\\
   \varphi_j^{(2)}({\vec r},t)
  \end{array}
  \right]
\end{equation}  
for $i=N/2+1,...,N$ and $j=1,...,N/2$.
We consider only low energy collisions in $\uparrow\downarrow$ channel
  \begin{equation}
   V_{int}^{\uparrow\downarrow}({\vec r}-{\vec r^{\prime}}) = g\,\delta({\vec r}-{\vec r^{\prime}})
  \end{equation}
Collisions in $\uparrow\uparrow$ and $\downarrow\downarrow$ channels are forbidden
  \begin{equation}
   V_{int}^{\uparrow\uparrow}({\vec r}-{\vec r^{\prime}}) =
   V_{int}^{\downarrow\downarrow}({\vec r}-{\vec r^{\prime}}) = 0
  \end{equation}
Finally one obtains the following equations of motion
\begin{align}
i\hbar \frac{\partial}{\partial t}  \varphi_i^{(s)} ({\vec r},t) 
 = \left[ -\frac{\hbar^2}{2 m} \nabla^2 + V_{ext}({\vec r},t) \right. \nonumber \\
 + \left. \frac{\delta E_{\text{int}}[\vec n]}{\delta n_{s} (\vec r, t)} \; \right] \; \varphi_i^{(s)} ({\vec r},t)\,.
\end{align}
\onecolumngrid
\newpage
\section{Details of Figures}

\aboverulesep=0ex
\belowrulesep=0ex
\renewcommand{\arraystretch}{0.85} 
\begin{table}[h!]
\caption{Details for the density plots in panels \textbf{a}--\textbf{r} of Fig.~\ref{fig1}.}
\label{Table1}
\begin{ruledtabular}
\setlength\extrarowheight{0.15em}
\begin{tabular}{ccccccc}
panel & $\lambda$ & $g$ & $\underset{\vec r}{\mathrm{max}}|n_1(\vec r)-n_2(\vec r)|$ & $E$ & $\mathcal P$ & $(l_1,l_2)$ \\[1.5ex]
\hline\\[-0.9em]
\multirow{3}{*}{\textbf{a}} & \multirow{3}{*}{5} & \multirow{3}{*}{3} & 0.0460 & 1159.34 & 0.0038 & (1,1) \\
 & & & 0.0389 & 1159.35 & 0.0028 & (1,1) \\
 & & & $1.7\times10^{-5}$ & 1159.41 & $1.3\times10^{-6}$ & (1,1) \\
\hline
\multirow{3}{*}{\textbf{b}} & \multirow{3}{*}{5} & \multirow{3}{*}{3.2} & 0.194 & 1168.62 & 0.0114 & (1,1) \\
 & & & 0.0455 & 1168.62 & 0.0034 & (1,1) \\
 & & & $1.6\times10^{-5}$ & 1168.73 & $1.3\times10^{-6}$ & (1,1) \\
\hline
\multirow{3}{*}{\textbf{c}} & \multirow{3}{*}{5} & \multirow{3}{*}{3.5} & 0.0652 & 1182.17 & 0.0078 & (1,1) \\
 & & & 1.1500 & 1182.20 & 0.0552 & (1,1) \\
 & & & 0.3417 & 1182.30 & 0.0380 & (1,1) \\
\hline
\multirow{2}{*}{\textbf{d}} & \multirow{2}{*}{5} & \multirow{2}{*}{4} & 1.3240 & 1204.50 & 0.1174 & (1,1) \\
 & & & 1.0176 & 1204.60 & 0.1086 & (1,1) \\
\hline
\multirow{2}{*}{\textbf{e}} & \multirow{2}{*}{5} & \multirow{2}{*}{5} & 3.9239 & 1242.91 & 0.4702 & (1,2) \\
 & & & 3.4668 & 1242.91 & 0.4773 & (2,2) \\
\hline
\multirow{2}{*}{\textbf{f}} & \multirow{2}{*}{5} & \multirow{2}{*}{5.1} & 3.6025 & 1245.63 & 0.5049 & (2,2) \\
 & & & 3.7670 & 1247.33 & 0.4700 & (2,2) \\
\hline
\textbf{g} & 10 & 4.6 & $1.8\times10^{-5}$ & 1621.74 & $1.6\times10^{-6}$ & (0,0) \\
\hline
\multirow{2}{*}{\textbf{h}} & \multirow{2}{*}{10} & \multirow{2}{*}{4.8} & 2.9592 & 1627.06 & 0.7875 & (0,1) \\
 & & & 3.0182 & 1629.09 & 0.6039 & (1,1) \\
 \hline
\multirow{2}{*}{\textbf{i}} & \multirow{2}{*}{10} & \multirow{2}{*}{5} & 3.0595 & 1631.64 & 0.8156 & (0,1) \\
 & & & 3.0144 & 1633.82 & 0.8762 & (1,1) \\
 \hline
\multirow{2}{*}{\textbf{j}} & \multirow{2}{*}{10} & \multirow{2}{*}{5.1} & 3.1148 & 1633.91 & 0.8254 & (0,1) \\
 & & & 3.0375 & 1634.71 & 0.8986 & (1,1) \\
 \hline
\multirow{2}{*}{\textbf{k}} & \multirow{2}{*}{10} & \multirow{2}{*}{5.2} & 3.0632 & 1635.45 & 0.9083 & (1,1) \\
 & & & 3.1545 & 1636.47 & 0.8357 & (0,1) \\
 \hline
\textbf{l} & 10 & 10 & 3.2809 & 1651.59 & 0.9644 & (1,1) \\
 \hline
\textbf{m} & 10 & 4.6 & 0.2606 & 1601.00 & 0.0774 & --- \\
 \hline
\textbf{n} & 10 & 4.8 & 0.3432 & 1610.79 & 0.0995 & --- \\
 \hline
\textbf{o} & 10 & 5 & 0.4113 & 1620.26 & 0.1193 & --- \\
 \hline
\textbf{p} & 10 & 5.1 & 0.4524 & 1624.87 & 0.1281 & --- \\
 \hline
\textbf{q} & 10 & 5.2 & 3.0423 & 1626.40 & 0.9086 & --- \\
 \hline
\textbf{r} & 10 & 10 & 3.1390 & 1639.25 & 0.9755 & --- \\
\end{tabular}
\end{ruledtabular}
\end{table}

\begin{table}
\caption{Details for the density plots in panels \textbf{a}--\textbf{j} of Fig.~\ref{fig3}.}
\label{Table2}
\begin{ruledtabular}
\setlength\extrarowheight{0.15em}
\begin{tabular}{cccccccc}
panel & $N_1=N_2$ & $\lambda$ & $g$ & $\underset{\vec r}{\mathrm{max}}|n_1(\vec r)-n_2(\vec r)|$ & $E$ & $\mathcal P$ & $(l_1,l_2)$ \\[1.5ex]
\hline\\[-0.9em]
\multirow{2}{*}{\textbf{a}} & \multirow{2}{*}{500} & \multirow{2}{*}{25} & \multirow{2}{*}{1.2} & 1.1999 & 37245.5 & 0.0117 & (1,1) \\
 & & & & 1.3325 & 37246.0 & 0.0163 & (1,1) \\
\hline 
\multirow{2}{*}{\textbf{b}} & \multirow{2}{*}{500} & \multirow{2}{*}{25} & \multirow{2}{*}{1.4} & 1.7187 & 37816.0 & 0.0290 & (1,1) \\
 & & & & 1.9441 & 38817.2 & 0.0345 & (1,1) \\
\hline 
\textbf{c} & 500 & 25 & 2 & 3.6595 & 39427.4 & 0.0678 & (1,1) \\ 
\hline
\textbf{d} & 500 & 25 & 3 & 9.3638 & 41337.5 & 0.6507 & (1,1) \\ 
\hline
\textbf{e} & 500 & 25 & 4 & 9.6879 & 41493.9 & 0.9835 & (1,1) \\ 
\hline
\multirow{2}{*}{\textbf{f}} & \multirow{2}{*}{5000} & \multirow{2}{*}{30} & \multirow{2}{*}{1.2} & 4.5722 & 863299 & 0.0179 & (2,2) \\
 & & & & 7.2395 & 863554 & 0.0608 & (2,2) \\
\hline 
\multirow{2}{*}{\textbf{g}} & \multirow{2}{*}{5000} & \multirow{2}{*}{30} & \multirow{2}{*}{1.3} & 5.8071 & 873039 & 0.0219 & (2,2) \\
 & & & & 10.191 & 873331 & 0.0996 & (2,2) \\
\hline 
\textbf{h} & 5000 & 30 & 1.6 & 36.125 & 900706 & 0.2781 & (3,3) \\ 
\hline
\textbf{i} & 5000 & 30 & 1.8 & 41.850 & 913258 & 0.6588 & (3,3) \\ 
\hline
\textbf{j} & 5000 & 30 & 3 & 46.096 & 923252 & 0.9944 & (3,3) \\ 
\end{tabular}
\end{ruledtabular}
\end{table}

\begin{table}
\caption{Details for the density plots in panels \textbf{a}--\textbf{e} of Fig.~\ref{fig4}.}
\label{Table3}
\begin{ruledtabular}
\setlength\extrarowheight{0.15em}
\begin{tabular}{ccccccc}
panel & $\lambda$ & method & $\underset{\vec r}{\mathrm{max}}|n_1(\vec r)-n_2(\vec r)|$ & $E$ & $\mathcal P$ & $(l_1,l_2)$ \\[1.5ex]
\hline\\[-0.9em]
\multirow{2}{*}{\textbf{a}} & \multirow{2}{*}{3} & DPFT & 4.3868 & 1039.752 & 0.4630 & (3,3) \\
 &  & HF & 1.0436 & 1022.84 & 0.0983 & --- \\
\hline
\multirow{2}{*}{\textbf{b}} & \multirow{2}{*}{4} & DPFT & 3.9943 & 1154.929 & 0.6109 & (3,3) \\
 &  & HF & 3.1287 & 1140.14 & 0.2008 & --- \\
\hline
\multirow{2}{*}{\textbf{c}} & \multirow{2}{*}{7} & DPFT & 3.2480 & 1424.63 & 0.7200 & (1,1) \\
 &  & HF & 3.0663 & 1411.58 & 0.5317 & --- \\
\hline
\multirow{2}{*}{\textbf{d}} & \multirow{2}{*}{10} & DPFT & 3.0962 & 1637.30 & 0.9255 & (1,1) \\
 &  & HF & 3.0997 & 1628.16 & 0.9332 & --- \\
\hline
\multirow{2}{*}{\textbf{e}} & \multirow{2}{*}{15} & DPFT & 2.4198 & 1931.07 & 0.9463 & (1,1) \\
 &  & HF & 2.4210 & 1920.94 & 0.9565 & --- \\ 
\end{tabular}
\end{ruledtabular}
\end{table}
 \FloatBarrier 

\renewcommand{\arraystretch}{1.0}

\normalem
\twocolumngrid

\bibliography{2d3dPaper}

\end{document}

%% file: preamble.tex
\usepackage{times}
\usepackage{amssymb}
\usepackage{color}
\usepackage{amsmath}
\usepackage{amsbsy}
\usepackage{amsthm}
\usepackage{booktabs}
\usepackage{graphicx}
\usepackage{bbm}
\usepackage{bm}
\usepackage{epsfig}
\usepackage{xfrac}
\usepackage{xcolor}
\usepackage{enumerate}
\usepackage{multirow}
\usepackage{physics}
\usepackage{placeins}
\usepackage{longtable}
\usepackage[T1]{fontenc}
\usepackage[english]{babel}
\usepackage{appendix}
\usepackage{dsfont}
\usepackage{mathtools}
\usepackage{float}
\usepackage{relsize}
\usepackage{xcolor}
\usepackage{scalerel}
\usepackage{tikz}
\usetikzlibrary{svg.path}
\usepackage{ulem}
\usepackage{amsfonts}
\usepackage{array}
\usepackage{bbding}
\usepackage{fixmath}

\definecolor{orcidlogocol}{HTML}{A6CE39}
\tikzset{
  orcidlogo/.pic={
    \fill[orcidlogocol] svg{M256,128c0,70.7-57.3,128-128,128C57.3,256,0,198.7,0,128C0,57.3,57.3,0,128,0C198.7,0,256,57.3,256,128z};
    \fill[white] svg{M86.3,186.2H70.9V79.1h15.4v48.4V186.2z}
                 svg{M108.9,79.1h41.6c39.6,0,57,28.3,57,53.6c0,27.5-21.5,53.6-56.8,53.6h-41.8V79.1z M124.3,172.4h24.5c34.9,0,42.9-26.5,42.9-39.7c0-21.5-13.7-39.7-43.7-39.7h-23.7V172.4z}
                 svg{M88.7,56.8c0,5.5-4.5,10.1-10.1,10.1c-5.6,0-10.1-4.6-10.1-10.1c0-5.6,4.5-10.1,10.1-10.1C84.2,46.7,88.7,51.3,88.7,56.8z};
  }
}

\newcommand\orcid[1]{\href{https://orcid.org/#1}{$\,$\mbox{\scalerel*{
\begin{tikzpicture}[yscale=-1,transform shape]
\pic{orcidlogo};
\end{tikzpicture}
}{|}}}}

\definecolor{myurlcolor}{rgb}{0.0,0.39,0.0}
\definecolor{myrefcolor}{rgb}{0.0,0.39,0.0}

\usepackage[colorlinks]{hyperref}
\hypersetup{unicode=true,
    bookmarksopen=false,
    breaklinks=false,
    pdfborder={0 0 0},
	bookmarksnumbered=false,
	pdfstartview={FitH},
	citecolor={myurlcolor},
	linkcolor={myrefcolor},
	urlcolor={myurlcolor}}

\definecolor{cyan(process)}{rgb}{0.0, 0.72, 0.92}

\newcommand{\hw}{\hbar \omega_z}
\newcommand{\kin}{\frac{\hbar^2 k^2}{2 m}}
\newcommand{\state}[1]{\left| \mathbf{k}, #1 \right\rangle }
\newcommand{\ef}{E_\text{F}}

\newcommand{\lm}{l_{\text{m}}}
\newcommand{\rounds}[1]{\left( #1 \right)}
\newcommand{\squares}[1]{\left[ #1 \right]}

\renewcommand{\vec}[1]{\mathbold{#1}}
\renewcommand{\d}{\mathrm{d}}

\DeclarePairedDelimiter\floor{\lfloor}{\rfloor}

%% file: main.bbl
\begin{thebibliography}{126}%
\makeatletter
\providecommand \@ifxundefined [1]{%
 \@ifx{#1\undefined}
}%
\providecommand \@ifnum [1]{%
 \ifnum #1\expandafter \@firstoftwo
 \else \expandafter \@secondoftwo
 \fi
}%
\providecommand \@ifx [1]{%
 \ifx #1\expandafter \@firstoftwo
 \else \expandafter \@secondoftwo
 \fi
}%
\providecommand \natexlab [1]{#1}%
\providecommand \enquote  [1]{``#1''}%
\providecommand \bibnamefont  [1]{#1}%
\providecommand \bibfnamefont [1]{#1}%
\providecommand \citenamefont [1]{#1}%
\providecommand \href@noop [0]{\@secondoftwo}%
\providecommand \href [0]{\begingroup \@sanitize@url \@href}%
\providecommand \@href[1]{\@@startlink{#1}\@@href}%
\providecommand \@@href[1]{\endgroup#1\@@endlink}%
\providecommand \@sanitize@url [0]{\catcode `\\12\catcode `\$12\catcode
  `\&12\catcode `\#12\catcode `\^12\catcode `\_12\catcode `\%12\relax}%
\providecommand \@@startlink[1]{}%
\providecommand \@@endlink[0]{}%
\providecommand \url  [0]{\begingroup\@sanitize@url \@url }%
\providecommand \@url [1]{\endgroup\@href {#1}{\urlprefix }}%
\providecommand \urlprefix  [0]{URL }%
\providecommand \Eprint [0]{\href }%
\providecommand \doibase [0]{https://doi.org/}%
\providecommand \selectlanguage [0]{\@gobble}%
\providecommand \bibinfo  [0]{\@secondoftwo}%
\providecommand \bibfield  [0]{\@secondoftwo}%
\providecommand \translation [1]{[#1]}%
\providecommand \BibitemOpen [0]{}%
\providecommand \bibitemStop [0]{}%
\providecommand \bibitemNoStop [0]{.\EOS\space}%
\providecommand \EOS [0]{\spacefactor3000\relax}%
\providecommand \BibitemShut  [1]{\csname bibitem#1\endcsname}%
\let\auto@bib@innerbib\@empty
\bibitem [{\citenamefont {Bloch}\ \emph {et~al.}(2008)\citenamefont {Bloch},
  \citenamefont {Dalibard},\ and\ \citenamefont {Zwerger}}]{Bloch2008}%
  \BibitemOpen
  \bibfield  {author} {\bibinfo {author} {\bibfnamefont {I.}~\bibnamefont
  {Bloch}}, \bibinfo {author} {\bibfnamefont {J.}~\bibnamefont {Dalibard}},\
  and\ \bibinfo {author} {\bibfnamefont {W.}~\bibnamefont {Zwerger}},\ }\href
  {https://doi.org/10.1103/RevModPhys.80.885} {\bibfield  {journal} {\bibinfo
  {journal} {Rev. Mod. Phys.}\ }\textbf {\bibinfo {volume} {80}},\ \bibinfo
  {pages} {885} (\bibinfo {year} {2008})}\BibitemShut {NoStop}%
\bibitem [{\citenamefont {Lewenstein}\ \emph {et~al.}(2012)\citenamefont
  {Lewenstein}, \citenamefont {Sanpera},\ and\ \citenamefont
  {Ahufinger}}]{Lewenstein2012}%
  \BibitemOpen
  \bibfield  {author} {\bibinfo {author} {\bibfnamefont {M.}~\bibnamefont
  {Lewenstein}}, \bibinfo {author} {\bibfnamefont {A.}~\bibnamefont
  {Sanpera}},\ and\ \bibinfo {author} {\bibfnamefont {V.}~\bibnamefont
  {Ahufinger}},\ }\href
  {https://doi.org/10.1093/acprof:oso/9780199573127.001.0001} {\emph {\bibinfo
  {title} {Ultracold {{Atoms}} in {{Optical Lattices}}}}}\ (\bibinfo
  {publisher} {Oxford University Press},\ \bibinfo {year} {2012})\BibitemShut
  {NoStop}%
\bibitem [{\citenamefont {Boettcher}\ \emph {et~al.}(2016)\citenamefont
  {Boettcher}, \citenamefont {Bayha}, \citenamefont {Kedar}, \citenamefont
  {Murthy}, \citenamefont {Neidig}, \citenamefont {Ries}, \citenamefont {Wenz},
  \citenamefont {Z{\"u}rn}, \citenamefont {Jochim},\ and\ \citenamefont
  {Enss}}]{Boettcher2016}%
  \BibitemOpen
  \bibfield  {author} {\bibinfo {author} {\bibfnamefont {I.}~\bibnamefont
  {Boettcher}}, \bibinfo {author} {\bibfnamefont {L.}~\bibnamefont {Bayha}},
  \bibinfo {author} {\bibfnamefont {D.}~\bibnamefont {Kedar}}, \bibinfo
  {author} {\bibfnamefont {P.~A.}\ \bibnamefont {Murthy}}, \bibinfo {author}
  {\bibfnamefont {M.}~\bibnamefont {Neidig}}, \bibinfo {author} {\bibfnamefont
  {M.~G.}\ \bibnamefont {Ries}}, \bibinfo {author} {\bibfnamefont {A.~N.}\
  \bibnamefont {Wenz}}, \bibinfo {author} {\bibfnamefont {G.}~\bibnamefont
  {Z{\"u}rn}}, \bibinfo {author} {\bibfnamefont {S.}~\bibnamefont {Jochim}},\
  and\ \bibinfo {author} {\bibfnamefont {T.}~\bibnamefont {Enss}},\ }\href
  {https://doi.org/10.1103/PhysRevLett.116.045303} {\bibfield  {journal}
  {\bibinfo  {journal} {Phys. Rev. Lett.}\ }\textbf {\bibinfo {volume} {116}},\
  \bibinfo {pages} {045303} (\bibinfo {year} {2016})}\BibitemShut {NoStop}%
\bibitem [{\citenamefont {Holten}\ \emph {et~al.}(2018)\citenamefont {Holten},
  \citenamefont {Bayha}, \citenamefont {Klein}, \citenamefont {Murthy},
  \citenamefont {Preiss},\ and\ \citenamefont {Jochim}}]{Holten2018}%
  \BibitemOpen
  \bibfield  {author} {\bibinfo {author} {\bibfnamefont {M.}~\bibnamefont
  {Holten}}, \bibinfo {author} {\bibfnamefont {L.}~\bibnamefont {Bayha}},
  \bibinfo {author} {\bibfnamefont {A.~C.}\ \bibnamefont {Klein}}, \bibinfo
  {author} {\bibfnamefont {P.~A.}\ \bibnamefont {Murthy}}, \bibinfo {author}
  {\bibfnamefont {P.~M.}\ \bibnamefont {Preiss}},\ and\ \bibinfo {author}
  {\bibfnamefont {S.}~\bibnamefont {Jochim}},\ }\href
  {https://doi.org/10.1103/PhysRevLett.121.120401} {\bibfield  {journal}
  {\bibinfo  {journal} {Phys. Rev. Lett.}\ }\textbf {\bibinfo {volume} {121}},\
  \bibinfo {pages} {120401} (\bibinfo {year} {2018})}\BibitemShut {NoStop}%
\bibitem [{\citenamefont {Toniolo}\ \emph {et~al.}(2018)\citenamefont
  {Toniolo}, \citenamefont {Mulkerin}, \citenamefont {Liu},\ and\ \citenamefont
  {Hu}}]{Toniolo2018}%
  \BibitemOpen
  \bibfield  {author} {\bibinfo {author} {\bibfnamefont {U.}~\bibnamefont
  {Toniolo}}, \bibinfo {author} {\bibfnamefont {B.~C.}\ \bibnamefont
  {Mulkerin}}, \bibinfo {author} {\bibfnamefont {X.-J.}\ \bibnamefont {Liu}},\
  and\ \bibinfo {author} {\bibfnamefont {H.}~\bibnamefont {Hu}},\ }\href
  {https://doi.org/10.1103/PhysRevA.97.063622} {\bibfield  {journal} {\bibinfo
  {journal} {Phys. Rev. A}\ }\textbf {\bibinfo {volume} {97}},\ \bibinfo
  {pages} {063622} (\bibinfo {year} {2018})}\BibitemShut {NoStop}%
\bibitem [{\citenamefont {Tung}\ \emph {et~al.}(2010)\citenamefont {Tung},
  \citenamefont {Lamporesi}, \citenamefont {Lobser}, \citenamefont {Xia},\ and\
  \citenamefont {Cornell}}]{Tung2010}%
  \BibitemOpen
  \bibfield  {author} {\bibinfo {author} {\bibfnamefont {S.}~\bibnamefont
  {Tung}}, \bibinfo {author} {\bibfnamefont {G.}~\bibnamefont {Lamporesi}},
  \bibinfo {author} {\bibfnamefont {D.}~\bibnamefont {Lobser}}, \bibinfo
  {author} {\bibfnamefont {L.}~\bibnamefont {Xia}},\ and\ \bibinfo {author}
  {\bibfnamefont {E.~A.}\ \bibnamefont {Cornell}},\ }\href
  {https://doi.org/10.1103/PhysRevLett.105.230408} {\bibfield  {journal}
  {\bibinfo  {journal} {Phys. Rev. Lett.}\ }\textbf {\bibinfo {volume} {105}},\
  \bibinfo {pages} {230408} (\bibinfo {year} {2010})}\BibitemShut {NoStop}%
\bibitem [{\citenamefont {Plisson}\ \emph {et~al.}(2011)\citenamefont
  {Plisson}, \citenamefont {Allard}, \citenamefont {Holzmann}, \citenamefont
  {Salomon}, \citenamefont {Aspect}, \citenamefont {Bouyer},\ and\
  \citenamefont {Bourdel}}]{Plisson2011}%
  \BibitemOpen
  \bibfield  {author} {\bibinfo {author} {\bibfnamefont {T.}~\bibnamefont
  {Plisson}}, \bibinfo {author} {\bibfnamefont {B.}~\bibnamefont {Allard}},
  \bibinfo {author} {\bibfnamefont {M.}~\bibnamefont {Holzmann}}, \bibinfo
  {author} {\bibfnamefont {G.}~\bibnamefont {Salomon}}, \bibinfo {author}
  {\bibfnamefont {A.}~\bibnamefont {Aspect}}, \bibinfo {author} {\bibfnamefont
  {P.}~\bibnamefont {Bouyer}},\ and\ \bibinfo {author} {\bibfnamefont
  {T.}~\bibnamefont {Bourdel}},\ }\href
  {https://doi.org/10.1103/PhysRevA.84.061606} {\bibfield  {journal} {\bibinfo
  {journal} {Phys. Rev. A}\ }\textbf {\bibinfo {volume} {84}},\ \bibinfo
  {pages} {061606} (\bibinfo {year} {2011})}\BibitemShut {NoStop}%
\bibitem [{\citenamefont {Ries}\ \emph {et~al.}(2015)\citenamefont {Ries},
  \citenamefont {Wenz}, \citenamefont {Z{\"u}rn}, \citenamefont {Bayha},
  \citenamefont {Boettcher}, \citenamefont {Kedar}, \citenamefont {Murthy},
  \citenamefont {Neidig}, \citenamefont {Lompe},\ and\ \citenamefont
  {Jochim}}]{Ries2015a}%
  \BibitemOpen
  \bibfield  {author} {\bibinfo {author} {\bibfnamefont {M.~G.}\ \bibnamefont
  {Ries}}, \bibinfo {author} {\bibfnamefont {A.~N.}\ \bibnamefont {Wenz}},
  \bibinfo {author} {\bibfnamefont {G.}~\bibnamefont {Z{\"u}rn}}, \bibinfo
  {author} {\bibfnamefont {L.}~\bibnamefont {Bayha}}, \bibinfo {author}
  {\bibfnamefont {I.}~\bibnamefont {Boettcher}}, \bibinfo {author}
  {\bibfnamefont {D.}~\bibnamefont {Kedar}}, \bibinfo {author} {\bibfnamefont
  {P.~A.}\ \bibnamefont {Murthy}}, \bibinfo {author} {\bibfnamefont
  {M.}~\bibnamefont {Neidig}}, \bibinfo {author} {\bibfnamefont
  {T.}~\bibnamefont {Lompe}},\ and\ \bibinfo {author} {\bibfnamefont
  {S.}~\bibnamefont {Jochim}},\ }\href
  {https://doi.org/10.1103/PhysRevLett.114.230401} {\bibfield  {journal}
  {\bibinfo  {journal} {Phys. Rev. Lett.}\ }\textbf {\bibinfo {volume} {114}},\
  \bibinfo {pages} {230401} (\bibinfo {year} {2015})}\BibitemShut {NoStop}%
\bibitem [{\citenamefont {Fenech}\ \emph {et~al.}(2016)\citenamefont {Fenech},
  \citenamefont {Dyke}, \citenamefont {Peppler}, \citenamefont {Lingham},
  \citenamefont {Hoinka}, \citenamefont {Hu},\ and\ \citenamefont
  {Vale}}]{Fenech2016}%
  \BibitemOpen
  \bibfield  {author} {\bibinfo {author} {\bibfnamefont {K.}~\bibnamefont
  {Fenech}}, \bibinfo {author} {\bibfnamefont {P.}~\bibnamefont {Dyke}},
  \bibinfo {author} {\bibfnamefont {T.}~\bibnamefont {Peppler}}, \bibinfo
  {author} {\bibfnamefont {M.~G.}\ \bibnamefont {Lingham}}, \bibinfo {author}
  {\bibfnamefont {S.}~\bibnamefont {Hoinka}}, \bibinfo {author} {\bibfnamefont
  {H.}~\bibnamefont {Hu}},\ and\ \bibinfo {author} {\bibfnamefont {C.~J.}\
  \bibnamefont {Vale}},\ }\href
  {https://doi.org/10.1103/PhysRevLett.116.045302} {\bibfield  {journal}
  {\bibinfo  {journal} {Phys. Rev. Lett.}\ }\textbf {\bibinfo {volume} {116}},\
  \bibinfo {pages} {045302} (\bibinfo {year} {2016})}\BibitemShut {NoStop}%
\bibitem [{\citenamefont {Murthy}\ \emph {et~al.}(2018)\citenamefont {Murthy},
  \citenamefont {Neidig}, \citenamefont {Klemt}, \citenamefont {Bayha},
  \citenamefont {Boettcher}, \citenamefont {Enss}, \citenamefont {Holten},
  \citenamefont {Z{\"u}rn}, \citenamefont {Preiss},\ and\ \citenamefont
  {Jochim}}]{Murthy2018}%
  \BibitemOpen
  \bibfield  {author} {\bibinfo {author} {\bibfnamefont {P.~A.}\ \bibnamefont
  {Murthy}}, \bibinfo {author} {\bibfnamefont {M.}~\bibnamefont {Neidig}},
  \bibinfo {author} {\bibfnamefont {R.}~\bibnamefont {Klemt}}, \bibinfo
  {author} {\bibfnamefont {L.}~\bibnamefont {Bayha}}, \bibinfo {author}
  {\bibfnamefont {I.}~\bibnamefont {Boettcher}}, \bibinfo {author}
  {\bibfnamefont {T.}~\bibnamefont {Enss}}, \bibinfo {author} {\bibfnamefont
  {M.}~\bibnamefont {Holten}}, \bibinfo {author} {\bibfnamefont
  {G.}~\bibnamefont {Z{\"u}rn}}, \bibinfo {author} {\bibfnamefont {P.~M.}\
  \bibnamefont {Preiss}},\ and\ \bibinfo {author} {\bibfnamefont
  {S.}~\bibnamefont {Jochim}},\ }\href
  {https://doi.org/10.1126/science.aan5950} {\bibfield  {journal} {\bibinfo
  {journal} {Science}\ }\textbf {\bibinfo {volume} {359}},\ \bibinfo {pages}
  {452} (\bibinfo {year} {2018})}\BibitemShut {NoStop}%
\bibitem [{\citenamefont {Dyke}\ \emph {et~al.}(2011)\citenamefont {Dyke},
  \citenamefont {Kuhnle}, \citenamefont {Whitlock}, \citenamefont {Hu},
  \citenamefont {Mark}, \citenamefont {Hoinka}, \citenamefont {Lingham},
  \citenamefont {Hannaford},\ and\ \citenamefont {Vale}}]{Dyke2011}%
  \BibitemOpen
  \bibfield  {author} {\bibinfo {author} {\bibfnamefont {P.}~\bibnamefont
  {Dyke}}, \bibinfo {author} {\bibfnamefont {E.~D.}\ \bibnamefont {Kuhnle}},
  \bibinfo {author} {\bibfnamefont {S.}~\bibnamefont {Whitlock}}, \bibinfo
  {author} {\bibfnamefont {H.}~\bibnamefont {Hu}}, \bibinfo {author}
  {\bibfnamefont {M.}~\bibnamefont {Mark}}, \bibinfo {author} {\bibfnamefont
  {S.}~\bibnamefont {Hoinka}}, \bibinfo {author} {\bibfnamefont
  {M.}~\bibnamefont {Lingham}}, \bibinfo {author} {\bibfnamefont
  {P.}~\bibnamefont {Hannaford}},\ and\ \bibinfo {author} {\bibfnamefont
  {C.~J.}\ \bibnamefont {Vale}},\ }\href
  {https://doi.org/10.1103/PhysRevLett.106.105304} {\bibfield  {journal}
  {\bibinfo  {journal} {Phys. Rev. Lett.}\ }\textbf {\bibinfo {volume} {106}},\
  \bibinfo {pages} {105304} (\bibinfo {year} {2011})}\BibitemShut {NoStop}%
\bibitem [{\citenamefont {Sommer}\ \emph {et~al.}(2012)\citenamefont {Sommer},
  \citenamefont {Cheuk}, \citenamefont {Ku}, \citenamefont {Bakr},\ and\
  \citenamefont {Zwierlein}}]{Sommer2012}%
  \BibitemOpen
  \bibfield  {author} {\bibinfo {author} {\bibfnamefont {A.~T.}\ \bibnamefont
  {Sommer}}, \bibinfo {author} {\bibfnamefont {L.~W.}\ \bibnamefont {Cheuk}},
  \bibinfo {author} {\bibfnamefont {M.~J.~H.}\ \bibnamefont {Ku}}, \bibinfo
  {author} {\bibfnamefont {W.~S.}\ \bibnamefont {Bakr}},\ and\ \bibinfo
  {author} {\bibfnamefont {M.~W.}\ \bibnamefont {Zwierlein}},\ }\href
  {https://doi.org/10.1103/PhysRevLett.108.045302} {\bibfield  {journal}
  {\bibinfo  {journal} {Phys. Rev. Lett.}\ }\textbf {\bibinfo {volume} {108}},\
  \bibinfo {pages} {045302} (\bibinfo {year} {2012})}\BibitemShut {NoStop}%
\bibitem [{\citenamefont {Cheng}\ \emph {et~al.}(2016)\citenamefont {Cheng},
  \citenamefont {Kangara}, \citenamefont {Arakelyan},\ and\ \citenamefont
  {Thomas}}]{Cheng2016}%
  \BibitemOpen
  \bibfield  {author} {\bibinfo {author} {\bibfnamefont {C.}~\bibnamefont
  {Cheng}}, \bibinfo {author} {\bibfnamefont {J.}~\bibnamefont {Kangara}},
  \bibinfo {author} {\bibfnamefont {I.}~\bibnamefont {Arakelyan}},\ and\
  \bibinfo {author} {\bibfnamefont {J.~E.}\ \bibnamefont {Thomas}},\ }\href
  {https://doi.org/10.1103/PhysRevA.94.031606} {\bibfield  {journal} {\bibinfo
  {journal} {Phys. Rev. A}\ }\textbf {\bibinfo {volume} {94}},\ \bibinfo
  {pages} {031606} (\bibinfo {year} {2016})}\BibitemShut {NoStop}%
\bibitem [{\citenamefont {Peppler}\ \emph {et~al.}(2018)\citenamefont
  {Peppler}, \citenamefont {Dyke}, \citenamefont {Zamorano}, \citenamefont
  {Herrera}, \citenamefont {Hoinka},\ and\ \citenamefont {Vale}}]{Peppler2018}%
  \BibitemOpen
  \bibfield  {author} {\bibinfo {author} {\bibfnamefont {T.}~\bibnamefont
  {Peppler}}, \bibinfo {author} {\bibfnamefont {P.}~\bibnamefont {Dyke}},
  \bibinfo {author} {\bibfnamefont {M.}~\bibnamefont {Zamorano}}, \bibinfo
  {author} {\bibfnamefont {I.}~\bibnamefont {Herrera}}, \bibinfo {author}
  {\bibfnamefont {S.}~\bibnamefont {Hoinka}},\ and\ \bibinfo {author}
  {\bibfnamefont {C.~J.}\ \bibnamefont {Vale}},\ }\href
  {https://doi.org/10.1103/PhysRevLett.121.120402} {\bibfield  {journal}
  {\bibinfo  {journal} {Phys. Rev. Lett.}\ }\textbf {\bibinfo {volume} {121}},\
  \bibinfo {pages} {120402} (\bibinfo {year} {2018})}\BibitemShut {NoStop}%
\bibitem [{\citenamefont {Gong}\ \emph {et~al.}(2023)\citenamefont {Gong},
  \citenamefont {Liu}, \citenamefont {Jiao}, \citenamefont {Zhang},
  \citenamefont {Yu}, \citenamefont {Peng}, \citenamefont {Peng}, \citenamefont
  {Shu}, \citenamefont {Zhu}, \citenamefont {Li},\ and\ \citenamefont
  {Luo}}]{Gong2023}%
  \BibitemOpen
  \bibfield  {author} {\bibinfo {author} {\bibfnamefont {H.}~\bibnamefont
  {Gong}}, \bibinfo {author} {\bibfnamefont {H.}~\bibnamefont {Liu}}, \bibinfo
  {author} {\bibfnamefont {B.}~\bibnamefont {Jiao}}, \bibinfo {author}
  {\bibfnamefont {H.}~\bibnamefont {Zhang}}, \bibinfo {author} {\bibfnamefont
  {H.}~\bibnamefont {Yu}}, \bibinfo {author} {\bibfnamefont {Q.}~\bibnamefont
  {Peng}}, \bibinfo {author} {\bibfnamefont {S.}~\bibnamefont {Peng}}, \bibinfo
  {author} {\bibfnamefont {T.}~\bibnamefont {Shu}}, \bibinfo {author}
  {\bibfnamefont {Y.}~\bibnamefont {Zhu}}, \bibinfo {author} {\bibfnamefont
  {J.}~\bibnamefont {Li}},\ and\ \bibinfo {author} {\bibfnamefont
  {L.}~\bibnamefont {Luo}},\ }\href
  {https://doi.org/10.1103/PhysRevA.107.053321} {\bibfield  {journal} {\bibinfo
   {journal} {Phys. Rev. A}\ }\textbf {\bibinfo {volume} {107}},\ \bibinfo
  {pages} {053321} (\bibinfo {year} {2023})}\BibitemShut {NoStop}%
\bibitem [{\citenamefont {Guo}\ \emph {et~al.}(2023{\natexlab{a}})\citenamefont
  {Guo}, \citenamefont {Yao}, \citenamefont {Dhar}, \citenamefont {Pizzino},
  \citenamefont {Horvath}, \citenamefont {Giamarchi}, \citenamefont {Landini},\
  and\ \citenamefont {N{\"a}gerl}}]{Guo2023}%
  \BibitemOpen
  \bibfield  {author} {\bibinfo {author} {\bibfnamefont {Y.}~\bibnamefont
  {Guo}}, \bibinfo {author} {\bibfnamefont {H.}~\bibnamefont {Yao}}, \bibinfo
  {author} {\bibfnamefont {S.}~\bibnamefont {Dhar}}, \bibinfo {author}
  {\bibfnamefont {L.}~\bibnamefont {Pizzino}}, \bibinfo {author} {\bibfnamefont
  {M.}~\bibnamefont {Horvath}}, \bibinfo {author} {\bibfnamefont
  {T.}~\bibnamefont {Giamarchi}}, \bibinfo {author} {\bibfnamefont
  {M.}~\bibnamefont {Landini}},\ and\ \bibinfo {author} {\bibfnamefont {H.-C.}\
  \bibnamefont {N{\"a}gerl}},\ }\href {http://arxiv.org/abs/2308.04144}
  {\bibfield  {journal} {\bibinfo  {journal} {arXiv.2308.04144}\ } (\bibinfo
  {year} {2023}{\natexlab{a}})}\BibitemShut {NoStop}%
\bibitem [{\citenamefont {Guo}\ \emph {et~al.}(2023{\natexlab{b}})\citenamefont
  {Guo}, \citenamefont {Yao}, \citenamefont {Ramanjanappa}, \citenamefont
  {Dhar}, \citenamefont {Horvath}, \citenamefont {Pizzino}, \citenamefont
  {Giamarchi}, \citenamefont {Landini},\ and\ \citenamefont
  {N{\"a}gerl}}]{Guo2023a}%
  \BibitemOpen
  \bibfield  {author} {\bibinfo {author} {\bibfnamefont {Y.}~\bibnamefont
  {Guo}}, \bibinfo {author} {\bibfnamefont {H.}~\bibnamefont {Yao}}, \bibinfo
  {author} {\bibfnamefont {S.}~\bibnamefont {Ramanjanappa}}, \bibinfo {author}
  {\bibfnamefont {S.}~\bibnamefont {Dhar}}, \bibinfo {author} {\bibfnamefont
  {M.}~\bibnamefont {Horvath}}, \bibinfo {author} {\bibfnamefont
  {L.}~\bibnamefont {Pizzino}}, \bibinfo {author} {\bibfnamefont
  {T.}~\bibnamefont {Giamarchi}}, \bibinfo {author} {\bibfnamefont
  {M.}~\bibnamefont {Landini}},\ and\ \bibinfo {author} {\bibfnamefont {H.-C.}\
  \bibnamefont {N{\"a}gerl}},\ }\href {http://arxiv.org/abs/2308.00411}
  {\bibfield  {journal} {\bibinfo  {journal} {arXiv.2308.00411}\ } (\bibinfo
  {year} {2023}{\natexlab{b}})}\BibitemShut {NoStop}%
\bibitem [{\citenamefont {Schneider}\ and\ \citenamefont
  {Wallis}(1998)}]{Schneider1998}%
  \BibitemOpen
  \bibfield  {author} {\bibinfo {author} {\bibfnamefont {J.}~\bibnamefont
  {Schneider}}\ and\ \bibinfo {author} {\bibfnamefont {H.}~\bibnamefont
  {Wallis}},\ }\href {https://doi.org/10.1103/PhysRevA.57.1253} {\bibfield
  {journal} {\bibinfo  {journal} {Phys. Rev. A}\ }\textbf {\bibinfo {volume}
  {57}},\ \bibinfo {pages} {1253} (\bibinfo {year} {1998})}\BibitemShut
  {NoStop}%
\bibitem [{\citenamefont {Vignolo}\ and\ \citenamefont
  {Minguzzi}(2003)}]{Vignolo2003}%
  \BibitemOpen
  \bibfield  {author} {\bibinfo {author} {\bibfnamefont {P.}~\bibnamefont
  {Vignolo}}\ and\ \bibinfo {author} {\bibfnamefont {A.}~\bibnamefont
  {Minguzzi}},\ }\href {https://doi.org/10.1103/PhysRevA.67.053601} {\bibfield
  {journal} {\bibinfo  {journal} {Phys. Rev. A}\ }\textbf {\bibinfo {volume}
  {67}},\ \bibinfo {pages} {053601} (\bibinfo {year} {2003})}\BibitemShut
  {NoStop}%
\bibitem [{\citenamefont {Mueller}(2004)}]{Mueller2004}%
  \BibitemOpen
  \bibfield  {author} {\bibinfo {author} {\bibfnamefont {E.~J.}\ \bibnamefont
  {Mueller}},\ }\href {https://doi.org/10.1103/PhysRevLett.93.190404}
  {\bibfield  {journal} {\bibinfo  {journal} {Phys. Rev. Lett.}\ }\textbf
  {\bibinfo {volume} {93}},\ \bibinfo {pages} {190404} (\bibinfo {year}
  {2004})}\BibitemShut {NoStop}%
\bibitem [{\citenamefont {Kumagai}\ \emph {et~al.}(2006)\citenamefont
  {Kumagai}, \citenamefont {Saitoh}, \citenamefont {Oyaizu}, \citenamefont
  {Furukawa}, \citenamefont {Takashima}, \citenamefont {Nohara}, \citenamefont
  {Takagi},\ and\ \citenamefont {Matsuda}}]{Kumagai2006a}%
  \BibitemOpen
  \bibfield  {author} {\bibinfo {author} {\bibfnamefont {K.}~\bibnamefont
  {Kumagai}}, \bibinfo {author} {\bibfnamefont {M.}~\bibnamefont {Saitoh}},
  \bibinfo {author} {\bibfnamefont {T.}~\bibnamefont {Oyaizu}}, \bibinfo
  {author} {\bibfnamefont {Y.}~\bibnamefont {Furukawa}}, \bibinfo {author}
  {\bibfnamefont {S.}~\bibnamefont {Takashima}}, \bibinfo {author}
  {\bibfnamefont {M.}~\bibnamefont {Nohara}}, \bibinfo {author} {\bibfnamefont
  {H.}~\bibnamefont {Takagi}},\ and\ \bibinfo {author} {\bibfnamefont
  {Y.}~\bibnamefont {Matsuda}},\ }\href
  {https://doi.org/10.1103/PhysRevLett.97.227002} {\bibfield  {journal}
  {\bibinfo  {journal} {Phys. Rev. Lett.}\ }\textbf {\bibinfo {volume} {97}},\
  \bibinfo {pages} {227002} (\bibinfo {year} {2006})}\BibitemShut {NoStop}%
\bibitem [{\citenamefont {Pitaevskii}\ and\ \citenamefont
  {Rosch}(1997)}]{Pitaevskii1997}%
  \BibitemOpen
  \bibfield  {author} {\bibinfo {author} {\bibfnamefont {L.~P.}\ \bibnamefont
  {Pitaevskii}}\ and\ \bibinfo {author} {\bibfnamefont {A.}~\bibnamefont
  {Rosch}},\ }\href {https://doi.org/10.1103/PhysRevA.55.R853} {\bibfield
  {journal} {\bibinfo  {journal} {Phys. Rev. A}\ }\textbf {\bibinfo {volume}
  {55}},\ \bibinfo {pages} {R853} (\bibinfo {year} {1997})}\BibitemShut
  {NoStop}%
\bibitem [{\citenamefont {Olshanii}\ \emph {et~al.}(2010)\citenamefont
  {Olshanii}, \citenamefont {Perrin},\ and\ \citenamefont
  {Lorent}}]{Olshanii2010}%
  \BibitemOpen
  \bibfield  {author} {\bibinfo {author} {\bibfnamefont {M.}~\bibnamefont
  {Olshanii}}, \bibinfo {author} {\bibfnamefont {H.}~\bibnamefont {Perrin}},\
  and\ \bibinfo {author} {\bibfnamefont {V.}~\bibnamefont {Lorent}},\ }\href
  {https://doi.org/10.1103/PhysRevLett.105.095302} {\bibfield  {journal}
  {\bibinfo  {journal} {Phys. Rev. Lett.}\ }\textbf {\bibinfo {volume} {105}},\
  \bibinfo {pages} {095302} (\bibinfo {year} {2010})}\BibitemShut {NoStop}%
\bibitem [{\citenamefont {Taylor}\ and\ \citenamefont
  {Randeria}(2012)}]{Taylor2012}%
  \BibitemOpen
  \bibfield  {author} {\bibinfo {author} {\bibfnamefont {E.}~\bibnamefont
  {Taylor}}\ and\ \bibinfo {author} {\bibfnamefont {M.}~\bibnamefont
  {Randeria}},\ }\href {https://doi.org/10.1103/PhysRevLett.109.135301}
  {\bibfield  {journal} {\bibinfo  {journal} {Phys. Rev. Lett.}\ }\textbf
  {\bibinfo {volume} {109}},\ \bibinfo {pages} {135301} (\bibinfo {year}
  {2012})}\BibitemShut {NoStop}%
\bibitem [{\citenamefont {Hofmann}(2012)}]{Hofmann2012}%
  \BibitemOpen
  \bibfield  {author} {\bibinfo {author} {\bibfnamefont {J.}~\bibnamefont
  {Hofmann}},\ }\href {https://doi.org/10.1103/PhysRevLett.108.185303}
  {\bibfield  {journal} {\bibinfo  {journal} {Phys. Rev. Lett.}\ }\textbf
  {\bibinfo {volume} {108}},\ \bibinfo {pages} {185303} (\bibinfo {year}
  {2012})}\BibitemShut {NoStop}%
\bibitem [{\citenamefont {Gao}\ and\ \citenamefont {Yu}(2012)}]{Gao2012}%
  \BibitemOpen
  \bibfield  {author} {\bibinfo {author} {\bibfnamefont {C.}~\bibnamefont
  {Gao}}\ and\ \bibinfo {author} {\bibfnamefont {Z.}~\bibnamefont {Yu}},\
  }\href {https://doi.org/10.1103/PhysRevA.86.043609} {\bibfield  {journal}
  {\bibinfo  {journal} {Phys. Rev. A}\ }\textbf {\bibinfo {volume} {86}},\
  \bibinfo {pages} {043609} (\bibinfo {year} {2012})}\BibitemShut {NoStop}%
\bibitem [{\citenamefont {Chafin}\ and\ \citenamefont
  {Sch{\"a}fer}(2013)}]{Chafin2013}%
  \BibitemOpen
  \bibfield  {author} {\bibinfo {author} {\bibfnamefont {C.}~\bibnamefont
  {Chafin}}\ and\ \bibinfo {author} {\bibfnamefont {T.}~\bibnamefont
  {Sch{\"a}fer}},\ }\href {https://doi.org/10.1103/PhysRevA.88.043636}
  {\bibfield  {journal} {\bibinfo  {journal} {Phys. Rev. A}\ }\textbf {\bibinfo
  {volume} {88}},\ \bibinfo {pages} {043636} (\bibinfo {year}
  {2013})}\BibitemShut {NoStop}%
\bibitem [{\citenamefont {Hu}\ \emph {et~al.}(2019)\citenamefont {Hu},
  \citenamefont {Mulkerin}, \citenamefont {Toniolo}, \citenamefont {He},\ and\
  \citenamefont {Liu}}]{Hu2019}%
  \BibitemOpen
  \bibfield  {author} {\bibinfo {author} {\bibfnamefont {H.}~\bibnamefont
  {Hu}}, \bibinfo {author} {\bibfnamefont {B.~C.}\ \bibnamefont {Mulkerin}},
  \bibinfo {author} {\bibfnamefont {U.}~\bibnamefont {Toniolo}}, \bibinfo
  {author} {\bibfnamefont {L.}~\bibnamefont {He}},\ and\ \bibinfo {author}
  {\bibfnamefont {X.-J.}\ \bibnamefont {Liu}},\ }\href
  {https://doi.org/10.1103/PhysRevLett.122.070401} {\bibfield  {journal}
  {\bibinfo  {journal} {Phys. Rev. Lett.}\ }\textbf {\bibinfo {volume} {122}},\
  \bibinfo {pages} {070401} (\bibinfo {year} {2019})}\BibitemShut {NoStop}%
\bibitem [{\citenamefont {Murthy}\ \emph {et~al.}(2019)\citenamefont {Murthy},
  \citenamefont {Defenu}, \citenamefont {Bayha}, \citenamefont {Holten},
  \citenamefont {Preiss}, \citenamefont {Enss},\ and\ \citenamefont
  {Jochim}}]{Murthy2019}%
  \BibitemOpen
  \bibfield  {author} {\bibinfo {author} {\bibfnamefont {P.~A.}\ \bibnamefont
  {Murthy}}, \bibinfo {author} {\bibfnamefont {N.}~\bibnamefont {Defenu}},
  \bibinfo {author} {\bibfnamefont {L.}~\bibnamefont {Bayha}}, \bibinfo
  {author} {\bibfnamefont {M.}~\bibnamefont {Holten}}, \bibinfo {author}
  {\bibfnamefont {P.~M.}\ \bibnamefont {Preiss}}, \bibinfo {author}
  {\bibfnamefont {T.}~\bibnamefont {Enss}},\ and\ \bibinfo {author}
  {\bibfnamefont {S.}~\bibnamefont {Jochim}},\ }\href
  {https://doi.org/10.1126/science.aau4402} {\bibfield  {journal} {\bibinfo
  {journal} {Science}\ }\textbf {\bibinfo {volume} {365}},\ \bibinfo {pages}
  {268} (\bibinfo {year} {2019})}\BibitemShut {NoStop}%
\bibitem [{\citenamefont {Tsuchiya}\ \emph {et~al.}(2009)\citenamefont
  {Tsuchiya}, \citenamefont {Watanabe},\ and\ \citenamefont
  {Ohashi}}]{Tsuchiya2009}%
  \BibitemOpen
  \bibfield  {author} {\bibinfo {author} {\bibfnamefont {S.}~\bibnamefont
  {Tsuchiya}}, \bibinfo {author} {\bibfnamefont {R.}~\bibnamefont {Watanabe}},\
  and\ \bibinfo {author} {\bibfnamefont {Y.}~\bibnamefont {Ohashi}},\ }\href
  {https://doi.org/10.1103/PhysRevA.80.033613} {\bibfield  {journal} {\bibinfo
  {journal} {Phys. Rev. A}\ }\textbf {\bibinfo {volume} {80}},\ \bibinfo
  {pages} {033613} (\bibinfo {year} {2009})}\BibitemShut {NoStop}%
\bibitem [{\citenamefont {Gaebler}\ \emph {et~al.}(2010)\citenamefont
  {Gaebler}, \citenamefont {Stewart}, \citenamefont {Drake}, \citenamefont
  {Jin}, \citenamefont {Perali}, \citenamefont {Pieri},\ and\ \citenamefont
  {Strinati}}]{Gaebler2010}%
  \BibitemOpen
  \bibfield  {author} {\bibinfo {author} {\bibfnamefont {J.~P.}\ \bibnamefont
  {Gaebler}}, \bibinfo {author} {\bibfnamefont {J.~T.}\ \bibnamefont
  {Stewart}}, \bibinfo {author} {\bibfnamefont {T.~E.}\ \bibnamefont {Drake}},
  \bibinfo {author} {\bibfnamefont {D.~S.}\ \bibnamefont {Jin}}, \bibinfo
  {author} {\bibfnamefont {A.}~\bibnamefont {Perali}}, \bibinfo {author}
  {\bibfnamefont {P.}~\bibnamefont {Pieri}},\ and\ \bibinfo {author}
  {\bibfnamefont {G.~C.}\ \bibnamefont {Strinati}},\ }\href
  {https://doi.org/10.1038/nphys1709} {\bibfield  {journal} {\bibinfo
  {journal} {Nat. Phys.}\ }\textbf {\bibinfo {volume} {6}},\ \bibinfo {pages}
  {569} (\bibinfo {year} {2010})}\BibitemShut {NoStop}%
\bibitem [{\citenamefont {{Richie-Halford}}\ \emph {et~al.}(2020)\citenamefont
  {{Richie-Halford}}, \citenamefont {Drut},\ and\ \citenamefont
  {Bulgac}}]{Richie-Halford2020}%
  \BibitemOpen
  \bibfield  {author} {\bibinfo {author} {\bibfnamefont {A.}~\bibnamefont
  {{Richie-Halford}}}, \bibinfo {author} {\bibfnamefont {J.~E.}\ \bibnamefont
  {Drut}},\ and\ \bibinfo {author} {\bibfnamefont {A.}~\bibnamefont {Bulgac}},\
  }\href {https://doi.org/10.1103/PhysRevLett.125.060403} {\bibfield  {journal}
  {\bibinfo  {journal} {Phys. Rev. Lett.}\ }\textbf {\bibinfo {volume} {125}},\
  \bibinfo {pages} {060403} (\bibinfo {year} {2020})}\BibitemShut {NoStop}%
\bibitem [{\citenamefont {Ptok}(2017)}]{Ptok2017}%
  \BibitemOpen
  \bibfield  {author} {\bibinfo {author} {\bibfnamefont {A.}~\bibnamefont
  {Ptok}},\ }\href {https://doi.org/10.1088/1361-648X/aa928d} {\bibfield
  {journal} {\bibinfo  {journal} {J. Phys.: Condens. Matter}\ }\textbf
  {\bibinfo {volume} {29}},\ \bibinfo {pages} {475901} (\bibinfo {year}
  {2017})}\BibitemShut {NoStop}%
\bibitem [{\citenamefont {Toniolo}\ \emph {et~al.}(2017)\citenamefont
  {Toniolo}, \citenamefont {Mulkerin}, \citenamefont {Vale}, \citenamefont
  {Liu},\ and\ \citenamefont {Hu}}]{Toniolo2017}%
  \BibitemOpen
  \bibfield  {author} {\bibinfo {author} {\bibfnamefont {U.}~\bibnamefont
  {Toniolo}}, \bibinfo {author} {\bibfnamefont {B.~C.}\ \bibnamefont
  {Mulkerin}}, \bibinfo {author} {\bibfnamefont {C.~J.}\ \bibnamefont {Vale}},
  \bibinfo {author} {\bibfnamefont {X.-J.}\ \bibnamefont {Liu}},\ and\ \bibinfo
  {author} {\bibfnamefont {H.}~\bibnamefont {Hu}},\ }\href
  {https://doi.org/10.1103/PhysRevA.96.041604} {\bibfield  {journal} {\bibinfo
  {journal} {Phys. Rev. A}\ }\textbf {\bibinfo {volume} {96}},\ \bibinfo
  {pages} {041604} (\bibinfo {year} {2017})}\BibitemShut {NoStop}%
\bibitem [{\citenamefont {Adachi}\ and\ \citenamefont
  {Ikeda}(2018)}]{Adachi2018}%
  \BibitemOpen
  \bibfield  {author} {\bibinfo {author} {\bibfnamefont {K.}~\bibnamefont
  {Adachi}}\ and\ \bibinfo {author} {\bibfnamefont {R.}~\bibnamefont {Ikeda}},\
  }\href {https://doi.org/10.1103/PhysRevB.98.184502} {\bibfield  {journal}
  {\bibinfo  {journal} {Phys. Rev. B}\ }\textbf {\bibinfo {volume} {98}},\
  \bibinfo {pages} {184502} (\bibinfo {year} {2018})}\BibitemShut {NoStop}%
\bibitem [{\citenamefont {{Faigle-Cedzich}}\ \emph {et~al.}(2021)\citenamefont
  {{Faigle-Cedzich}}, \citenamefont {Pawlowski},\ and\ \citenamefont
  {Wetterich}}]{Faigle-Cedzich2021}%
  \BibitemOpen
  \bibfield  {author} {\bibinfo {author} {\bibfnamefont {B.~M.}\ \bibnamefont
  {{Faigle-Cedzich}}}, \bibinfo {author} {\bibfnamefont {J.~M.}\ \bibnamefont
  {Pawlowski}},\ and\ \bibinfo {author} {\bibfnamefont {C.}~\bibnamefont
  {Wetterich}},\ }\href {https://doi.org/10.1103/PhysRevA.103.033320}
  {\bibfield  {journal} {\bibinfo  {journal} {Phys. Rev. A}\ }\textbf {\bibinfo
  {volume} {103}},\ \bibinfo {pages} {033320} (\bibinfo {year}
  {2021})}\BibitemShut {NoStop}%
\bibitem [{\citenamefont {Zheng}\ \emph {et~al.}(2023)\citenamefont {Zheng},
  \citenamefont {Wang}, \citenamefont {Liang}, \citenamefont {Huang},
  \citenamefont {Wang}, \citenamefont {Xiong}, \citenamefont {Zhou},
  \citenamefont {Chen}, \citenamefont {Chen},\ and\ \citenamefont
  {Hu}}]{Zheng2023}%
  \BibitemOpen
  \bibfield  {author} {\bibinfo {author} {\bibfnamefont {Q.}~\bibnamefont
  {Zheng}}, \bibinfo {author} {\bibfnamefont {Y.}~\bibnamefont {Wang}},
  \bibinfo {author} {\bibfnamefont {L.}~\bibnamefont {Liang}}, \bibinfo
  {author} {\bibfnamefont {Q.}~\bibnamefont {Huang}}, \bibinfo {author}
  {\bibfnamefont {S.}~\bibnamefont {Wang}}, \bibinfo {author} {\bibfnamefont
  {W.}~\bibnamefont {Xiong}}, \bibinfo {author} {\bibfnamefont
  {X.}~\bibnamefont {Zhou}}, \bibinfo {author} {\bibfnamefont {W.}~\bibnamefont
  {Chen}}, \bibinfo {author} {\bibfnamefont {X.}~\bibnamefont {Chen}},\ and\
  \bibinfo {author} {\bibfnamefont {J.}~\bibnamefont {Hu}},\ }\href
  {https://doi.org/10.1103/PhysRevResearch.5.013136} {\bibfield  {journal}
  {\bibinfo  {journal} {Phys. Rev. Res.}\ }\textbf {\bibinfo {volume} {5}},\
  \bibinfo {pages} {013136} (\bibinfo {year} {2023})}\BibitemShut {NoStop}%
\bibitem [{\citenamefont {Stoner}(1938)}]{Stoner1938}%
  \BibitemOpen
  \bibfield  {author} {\bibinfo {author} {\bibfnamefont {C.}~\bibnamefont
  {Stoner}},\ }\href {https://doi.org/10.1098/rspa.1938.0066} {\bibfield
  {journal} {\bibinfo  {journal} {Proc. R. Soc. Lond. A}\ }\textbf {\bibinfo
  {volume} {165}},\ \bibinfo {pages} {372} (\bibinfo {year}
  {1938})}\BibitemShut {NoStop}%
\bibitem [{\citenamefont {Sogo}\ and\ \citenamefont {Yabu}(2002)}]{Sogo2002}%
  \BibitemOpen
  \bibfield  {author} {\bibinfo {author} {\bibfnamefont {T.}~\bibnamefont
  {Sogo}}\ and\ \bibinfo {author} {\bibfnamefont {H.}~\bibnamefont {Yabu}},\
  }\href {https://doi.org/10.1103/PhysRevA.66.043611} {\bibfield  {journal}
  {\bibinfo  {journal} {Phys. Rev. A}\ }\textbf {\bibinfo {volume} {66}},\
  \bibinfo {pages} {043611} (\bibinfo {year} {2002})}\BibitemShut {NoStop}%
\bibitem [{\citenamefont {Karpiuk}\ \emph {et~al.}(2004)\citenamefont
  {Karpiuk}, \citenamefont {Brewczyk},\ and\ \citenamefont {Rz{\k
  a}{\.z}ewski}}]{Karpiuk2004}%
  \BibitemOpen
  \bibfield  {author} {\bibinfo {author} {\bibfnamefont {T.}~\bibnamefont
  {Karpiuk}}, \bibinfo {author} {\bibfnamefont {M.}~\bibnamefont {Brewczyk}},\
  and\ \bibinfo {author} {\bibfnamefont {K.}~\bibnamefont {Rz{\k
  a}{\.z}ewski}},\ }\href {https://doi.org/10.1103/PhysRevA.69.043603}
  {\bibfield  {journal} {\bibinfo  {journal} {Phys. Rev. A}\ }\textbf {\bibinfo
  {volume} {69}},\ \bibinfo {pages} {043603} (\bibinfo {year}
  {2004})}\BibitemShut {NoStop}%
\bibitem [{\citenamefont {Duine}\ and\ \citenamefont
  {MacDonald}(2005)}]{Duine2005}%
  \BibitemOpen
  \bibfield  {author} {\bibinfo {author} {\bibfnamefont {R.~A.}\ \bibnamefont
  {Duine}}\ and\ \bibinfo {author} {\bibfnamefont {A.~H.}\ \bibnamefont
  {MacDonald}},\ }\href {https://doi.org/10.1103/PhysRevLett.95.230403}
  {\bibfield  {journal} {\bibinfo  {journal} {Phys. Rev. Lett.}\ }\textbf
  {\bibinfo {volume} {95}},\ \bibinfo {pages} {230403} (\bibinfo {year}
  {2005})}\BibitemShut {NoStop}%
\bibitem [{\citenamefont {LeBlanc}\ \emph {et~al.}(2009)\citenamefont
  {LeBlanc}, \citenamefont {Thywissen}, \citenamefont {Burkov},\ and\
  \citenamefont {Paramekanti}}]{LeBlanc2009}%
  \BibitemOpen
  \bibfield  {author} {\bibinfo {author} {\bibfnamefont {L.~J.}\ \bibnamefont
  {LeBlanc}}, \bibinfo {author} {\bibfnamefont {J.~H.}\ \bibnamefont
  {Thywissen}}, \bibinfo {author} {\bibfnamefont {A.~A.}\ \bibnamefont
  {Burkov}},\ and\ \bibinfo {author} {\bibfnamefont {A.}~\bibnamefont
  {Paramekanti}},\ }\href {https://doi.org/10.1103/PhysRevA.80.013607}
  {\bibfield  {journal} {\bibinfo  {journal} {Phys. Rev. A}\ }\textbf {\bibinfo
  {volume} {80}},\ \bibinfo {pages} {013607} (\bibinfo {year}
  {2009})}\BibitemShut {NoStop}%
\bibitem [{\citenamefont {Conduit}\ \emph {et~al.}(2009)\citenamefont
  {Conduit}, \citenamefont {Green},\ and\ \citenamefont
  {Simons}}]{Conduit2009}%
  \BibitemOpen
  \bibfield  {author} {\bibinfo {author} {\bibfnamefont {G.~J.}\ \bibnamefont
  {Conduit}}, \bibinfo {author} {\bibfnamefont {A.~G.}\ \bibnamefont {Green}},\
  and\ \bibinfo {author} {\bibfnamefont {B.~D.}\ \bibnamefont {Simons}},\
  }\href {https://doi.org/10.1103/PhysRevLett.103.207201} {\bibfield  {journal}
  {\bibinfo  {journal} {Phys. Rev. Lett.}\ }\textbf {\bibinfo {volume} {103}},\
  \bibinfo {pages} {207201} (\bibinfo {year} {2009})}\BibitemShut {NoStop}%
\bibitem [{\citenamefont {Cui}\ and\ \citenamefont {Zhai}(2010)}]{Cui2010}%
  \BibitemOpen
  \bibfield  {author} {\bibinfo {author} {\bibfnamefont {X.}~\bibnamefont
  {Cui}}\ and\ \bibinfo {author} {\bibfnamefont {H.}~\bibnamefont {Zhai}},\
  }\href {https://doi.org/10.1103/PhysRevA.81.041602} {\bibfield  {journal}
  {\bibinfo  {journal} {Phys. Rev. A}\ }\textbf {\bibinfo {volume} {81}},\
  \bibinfo {pages} {041602(R)} (\bibinfo {year} {2010})}\BibitemShut {NoStop}%
\bibitem [{\citenamefont {Pilati}\ \emph {et~al.}(2010)\citenamefont {Pilati},
  \citenamefont {Bertaina}, \citenamefont {Giorgini},\ and\ \citenamefont
  {Troyer}}]{Pilati2010}%
  \BibitemOpen
  \bibfield  {author} {\bibinfo {author} {\bibfnamefont {S.}~\bibnamefont
  {Pilati}}, \bibinfo {author} {\bibfnamefont {G.}~\bibnamefont {Bertaina}},
  \bibinfo {author} {\bibfnamefont {S.}~\bibnamefont {Giorgini}},\ and\
  \bibinfo {author} {\bibfnamefont {M.}~\bibnamefont {Troyer}},\ }\href
  {https://doi.org/10.1103/PhysRevLett.105.030405} {\bibfield  {journal}
  {\bibinfo  {journal} {Phys. Rev. Lett.}\ }\textbf {\bibinfo {volume} {105}},\
  \bibinfo {pages} {030405} (\bibinfo {year} {2010})}\BibitemShut {NoStop}%
\bibitem [{\citenamefont {Chang}\ \emph {et~al.}(2011)\citenamefont {Chang},
  \citenamefont {Randeria},\ and\ \citenamefont {Trivedi}}]{Chang2011}%
  \BibitemOpen
  \bibfield  {author} {\bibinfo {author} {\bibfnamefont {S.-Y.}\ \bibnamefont
  {Chang}}, \bibinfo {author} {\bibfnamefont {M.}~\bibnamefont {Randeria}},\
  and\ \bibinfo {author} {\bibfnamefont {N.}~\bibnamefont {Trivedi}},\ }\href
  {https://doi.org/10.1073/pnas.1011990108} {\bibfield  {journal} {\bibinfo
  {journal} {Proc. Natl. Acad. Sci.}\ }\textbf {\bibinfo {volume} {108}},\
  \bibinfo {pages} {51} (\bibinfo {year} {2011})}\BibitemShut {NoStop}%
\bibitem [{\citenamefont {Pekker}\ \emph {et~al.}(2011)\citenamefont {Pekker},
  \citenamefont {Babadi}, \citenamefont {Sensarma}, \citenamefont {Zinner},
  \citenamefont {Pollet}, \citenamefont {Zwierlein},\ and\ \citenamefont
  {Demler}}]{Pekker2011}%
  \BibitemOpen
  \bibfield  {author} {\bibinfo {author} {\bibfnamefont {D.}~\bibnamefont
  {Pekker}}, \bibinfo {author} {\bibfnamefont {M.}~\bibnamefont {Babadi}},
  \bibinfo {author} {\bibfnamefont {R.}~\bibnamefont {Sensarma}}, \bibinfo
  {author} {\bibfnamefont {N.}~\bibnamefont {Zinner}}, \bibinfo {author}
  {\bibfnamefont {L.}~\bibnamefont {Pollet}}, \bibinfo {author} {\bibfnamefont
  {M.~W.}\ \bibnamefont {Zwierlein}},\ and\ \bibinfo {author} {\bibfnamefont
  {E.}~\bibnamefont {Demler}},\ }\href
  {https://doi.org/10.1103/PhysRevLett.106.050402} {\bibfield  {journal}
  {\bibinfo  {journal} {Phys. Rev. Lett.}\ }\textbf {\bibinfo {volume} {106}},\
  \bibinfo {pages} {050402} (\bibinfo {year} {2011})}\BibitemShut {NoStop}%
\bibitem [{\citenamefont {Massignan}\ and\ \citenamefont
  {Bruun}(2011)}]{Massignan2011}%
  \BibitemOpen
  \bibfield  {author} {\bibinfo {author} {\bibfnamefont {P.}~\bibnamefont
  {Massignan}}\ and\ \bibinfo {author} {\bibfnamefont {G.~M.}\ \bibnamefont
  {Bruun}},\ }\href {https://doi.org/10.1140/epjd/e2011-20084-5} {\bibfield
  {journal} {\bibinfo  {journal} {Eur. Phys. J. D}\ }\textbf {\bibinfo {volume}
  {65}},\ \bibinfo {pages} {83} (\bibinfo {year} {2011})}\BibitemShut {NoStop}%
\bibitem [{\citenamefont {Massignan}\ \emph {et~al.}(2014)\citenamefont
  {Massignan}, \citenamefont {Zaccanti},\ and\ \citenamefont
  {Bruun}}]{Massignan2014}%
  \BibitemOpen
  \bibfield  {author} {\bibinfo {author} {\bibfnamefont {P.}~\bibnamefont
  {Massignan}}, \bibinfo {author} {\bibfnamefont {M.}~\bibnamefont
  {Zaccanti}},\ and\ \bibinfo {author} {\bibfnamefont {G.~M.}\ \bibnamefont
  {Bruun}},\ }\href {https://doi.org/10.1088/0034-4885/77/3/034401} {\bibfield
  {journal} {\bibinfo  {journal} {Rep. Prog. Phys.}\ }\textbf {\bibinfo
  {volume} {77}},\ \bibinfo {pages} {034401} (\bibinfo {year}
  {2014})}\BibitemShut {NoStop}%
\bibitem [{\citenamefont {Levinsen}\ and\ \citenamefont
  {Parish}(2015)}]{Levinsen2015}%
  \BibitemOpen
  \bibfield  {author} {\bibinfo {author} {\bibfnamefont {J.}~\bibnamefont
  {Levinsen}}\ and\ \bibinfo {author} {\bibfnamefont {M.~M.}\ \bibnamefont
  {Parish}},\ }in\ \href {https://doi.org/10.1142/9789814667746_0001} {\emph
  {\bibinfo {booktitle} {Annual {{Review}} of {{Cold Atoms}} and
  {{Molecules}}}}}\ (\bibinfo  {publisher} {World Scientific},\ \bibinfo {year}
  {2015})\ pp.\ \bibinfo {pages} {1--75}\BibitemShut {NoStop}%
\bibitem [{\citenamefont {Trappe}\ \emph
  {et~al.}(2016{\natexlab{a}})\citenamefont {Trappe}, \citenamefont
  {Grochowski}, \citenamefont {Brewczyk},\ and\ \citenamefont {Rz{\k
  a}{\.z}ewski}}]{Trappe2016a}%
  \BibitemOpen
  \bibfield  {author} {\bibinfo {author} {\bibfnamefont {M.-I.}\ \bibnamefont
  {Trappe}}, \bibinfo {author} {\bibfnamefont {P.}~\bibnamefont {Grochowski}},
  \bibinfo {author} {\bibfnamefont {M.}~\bibnamefont {Brewczyk}},\ and\
  \bibinfo {author} {\bibfnamefont {K.}~\bibnamefont {Rz{\k a}{\.z}ewski}},\
  }\href {https://doi.org/10.1103/PhysRevA.93.023612} {\bibfield  {journal}
  {\bibinfo  {journal} {Phys. Rev. A}\ }\textbf {\bibinfo {volume} {93}},\
  \bibinfo {pages} {023612} (\bibinfo {year} {2016}{\natexlab{a}})}\BibitemShut
  {NoStop}%
\bibitem [{\citenamefont {Miyakawa}\ \emph {et~al.}(2017)\citenamefont
  {Miyakawa}, \citenamefont {Nakamura},\ and\ \citenamefont
  {Yabu}}]{Miyakawa2017}%
  \BibitemOpen
  \bibfield  {author} {\bibinfo {author} {\bibfnamefont {T.}~\bibnamefont
  {Miyakawa}}, \bibinfo {author} {\bibfnamefont {S.}~\bibnamefont {Nakamura}},\
  and\ \bibinfo {author} {\bibfnamefont {H.}~\bibnamefont {Yabu}},\ }\href
  {https://doi.org/10.7566/JPSJ.86.035004} {\bibfield  {journal} {\bibinfo
  {journal} {J. Phys. Soc. Jpn.}\ }\textbf {\bibinfo {volume} {86}},\ \bibinfo
  {pages} {035004} (\bibinfo {year} {2017})}\BibitemShut {NoStop}%
\bibitem [{\citenamefont {Grochowski}\ \emph {et~al.}(2017)\citenamefont
  {Grochowski}, \citenamefont {Karpiuk}, \citenamefont {Brewczyk},\ and\
  \citenamefont {Rz{\k a}{\.z}ewski}}]{Grochowski2017}%
  \BibitemOpen
  \bibfield  {author} {\bibinfo {author} {\bibfnamefont {P.~T.}\ \bibnamefont
  {Grochowski}}, \bibinfo {author} {\bibfnamefont {T.}~\bibnamefont {Karpiuk}},
  \bibinfo {author} {\bibfnamefont {M.}~\bibnamefont {Brewczyk}},\ and\
  \bibinfo {author} {\bibfnamefont {K.}~\bibnamefont {Rz{\k a}{\.z}ewski}},\
  }\href {https://doi.org/10.1103/PhysRevLett.119.215303} {\bibfield  {journal}
  {\bibinfo  {journal} {Phys. Rev. Lett.}\ }\textbf {\bibinfo {volume} {119}},\
  \bibinfo {pages} {215303} (\bibinfo {year} {2017})}\BibitemShut {NoStop}%
\bibitem [{\citenamefont {Koutentakis}\ \emph {et~al.}(2019)\citenamefont
  {Koutentakis}, \citenamefont {Mistakidis},\ and\ \citenamefont
  {Schmelcher}}]{Koutentakis2019}%
  \BibitemOpen
  \bibfield  {author} {\bibinfo {author} {\bibfnamefont {G.~M.}\ \bibnamefont
  {Koutentakis}}, \bibinfo {author} {\bibfnamefont {S.~I.}\ \bibnamefont
  {Mistakidis}},\ and\ \bibinfo {author} {\bibfnamefont {P.}~\bibnamefont
  {Schmelcher}},\ }\href {https://doi.org/10.1088/1367-2630/ab14ba} {\bibfield
  {journal} {\bibinfo  {journal} {New J. Phys.}\ }\textbf {\bibinfo {volume}
  {21}},\ \bibinfo {pages} {053005} (\bibinfo {year} {2019})}\BibitemShut
  {NoStop}%
\bibitem [{\citenamefont {Ryszkiewicz}\ \emph {et~al.}(2020)\citenamefont
  {Ryszkiewicz}, \citenamefont {Brewczyk},\ and\ \citenamefont
  {Karpiuk}}]{Ryszkiewicz2020}%
  \BibitemOpen
  \bibfield  {author} {\bibinfo {author} {\bibfnamefont {J.}~\bibnamefont
  {Ryszkiewicz}}, \bibinfo {author} {\bibfnamefont {M.}~\bibnamefont
  {Brewczyk}},\ and\ \bibinfo {author} {\bibfnamefont {T.}~\bibnamefont
  {Karpiuk}},\ }\href {https://doi.org/10.1103/PhysRevA.101.013618} {\bibfield
  {journal} {\bibinfo  {journal} {Phys. Rev. A}\ }\textbf {\bibinfo {volume}
  {101}},\ \bibinfo {pages} {013618} (\bibinfo {year} {2020})}\BibitemShut
  {NoStop}%
\bibitem [{\citenamefont {Karpiuk}\ \emph {et~al.}(2020)\citenamefont
  {Karpiuk}, \citenamefont {Grochowski}, \citenamefont {Brewczyk},\ and\
  \citenamefont {Rz{\k a}{\.z}ewski}}]{Karpiuk2020}%
  \BibitemOpen
  \bibfield  {author} {\bibinfo {author} {\bibfnamefont {T.}~\bibnamefont
  {Karpiuk}}, \bibinfo {author} {\bibfnamefont {P.~T.}\ \bibnamefont
  {Grochowski}}, \bibinfo {author} {\bibfnamefont {M.}~\bibnamefont
  {Brewczyk}},\ and\ \bibinfo {author} {\bibfnamefont {K.}~\bibnamefont {Rz{\k
  a}{\.z}ewski}},\ }\href {https://doi.org/10.21468/SciPostPhys.8.4.066}
  {\bibfield  {journal} {\bibinfo  {journal} {SciPost Phys.}\ }\textbf
  {\bibinfo {volume} {8}},\ \bibinfo {pages} {66} (\bibinfo {year}
  {2020})}\BibitemShut {NoStop}%
\bibitem [{\citenamefont {Grochowski}\ \emph
  {et~al.}(2020{\natexlab{a}})\citenamefont {Grochowski}, \citenamefont
  {Karpiuk}, \citenamefont {Brewczyk},\ and\ \citenamefont {Rz{\k
  a}{\.z}ewski}}]{Grochowski2020a}%
  \BibitemOpen
  \bibfield  {author} {\bibinfo {author} {\bibfnamefont {P.~T.}\ \bibnamefont
  {Grochowski}}, \bibinfo {author} {\bibfnamefont {T.}~\bibnamefont {Karpiuk}},
  \bibinfo {author} {\bibfnamefont {M.}~\bibnamefont {Brewczyk}},\ and\
  \bibinfo {author} {\bibfnamefont {K.}~\bibnamefont {Rz{\k a}{\.z}ewski}},\
  }\href {https://doi.org/10.1103/physrevresearch.2.013119} {\bibfield
  {journal} {\bibinfo  {journal} {Phys. Rev. Res.}\ }\textbf {\bibinfo {volume}
  {2}},\ \bibinfo {pages} {013119} (\bibinfo {year}
  {2020}{\natexlab{a}})}\BibitemShut {NoStop}%
\bibitem [{\citenamefont {Koutentakis}\ \emph {et~al.}(2020)\citenamefont
  {Koutentakis}, \citenamefont {Mistakidis},\ and\ \citenamefont
  {Schmelcher}}]{Koutentakis2020}%
  \BibitemOpen
  \bibfield  {author} {\bibinfo {author} {\bibfnamefont {G.~M.}\ \bibnamefont
  {Koutentakis}}, \bibinfo {author} {\bibfnamefont {S.~I.}\ \bibnamefont
  {Mistakidis}},\ and\ \bibinfo {author} {\bibfnamefont {P.}~\bibnamefont
  {Schmelcher}},\ }\href {https://doi.org/10.1088/1367-2630/ab90d6} {\bibfield
  {journal} {\bibinfo  {journal} {New J. Phys.}\ }\textbf {\bibinfo {volume}
  {22}},\ \bibinfo {pages} {63058} (\bibinfo {year} {2020})}\BibitemShut
  {NoStop}%
\bibitem [{\citenamefont {Trappe}\ \emph {et~al.}(2021)\citenamefont {Trappe},
  \citenamefont {Grochowski}, \citenamefont {Hue}, \citenamefont {Karpiuk},\
  and\ \citenamefont {Rz{\k a}{\.z}ewski}}]{Trappe2021a}%
  \BibitemOpen
  \bibfield  {author} {\bibinfo {author} {\bibfnamefont {M.-I.}\ \bibnamefont
  {Trappe}}, \bibinfo {author} {\bibfnamefont {P.~T.}\ \bibnamefont
  {Grochowski}}, \bibinfo {author} {\bibfnamefont {J.~H.}\ \bibnamefont {Hue}},
  \bibinfo {author} {\bibfnamefont {T.}~\bibnamefont {Karpiuk}},\ and\ \bibinfo
  {author} {\bibfnamefont {K.}~\bibnamefont {Rz{\k a}{\.z}ewski}},\ }\href
  {https://doi.org/10.1088/1367-2630/ac2b51} {\bibfield  {journal} {\bibinfo
  {journal} {New J. Phys.}\ }\textbf {\bibinfo {volume} {23}},\ \bibinfo
  {pages} {103042} (\bibinfo {year} {2021})}\BibitemShut {NoStop}%
\bibitem [{\citenamefont {Ryszkiewicz}\ \emph {et~al.}(2022)\citenamefont
  {Ryszkiewicz}, \citenamefont {Brewczyk},\ and\ \citenamefont
  {Karpiuk}}]{Ryszkiewicz2022}%
  \BibitemOpen
  \bibfield  {author} {\bibinfo {author} {\bibfnamefont {J.}~\bibnamefont
  {Ryszkiewicz}}, \bibinfo {author} {\bibfnamefont {M.}~\bibnamefont
  {Brewczyk}},\ and\ \bibinfo {author} {\bibfnamefont {T.}~\bibnamefont
  {Karpiuk}},\ }\href {https://doi.org/10.1103/PhysRevA.105.023315} {\bibfield
  {journal} {\bibinfo  {journal} {Phys. Rev. A}\ }\textbf {\bibinfo {volume}
  {105}},\ \bibinfo {pages} {023315} (\bibinfo {year} {2022})}\BibitemShut
  {NoStop}%
\bibitem [{\citenamefont {Syrwid}\ \emph {et~al.}(2022)\citenamefont {Syrwid},
  \citenamefont {{\L}ebek}, \citenamefont {Grochowski},\ and\ \citenamefont
  {Rz{\k a}{\.z}ewski}}]{Syrwid2022}%
  \BibitemOpen
  \bibfield  {author} {\bibinfo {author} {\bibfnamefont {A.}~\bibnamefont
  {Syrwid}}, \bibinfo {author} {\bibfnamefont {M.}~\bibnamefont {{\L}ebek}},
  \bibinfo {author} {\bibfnamefont {P.~T.}\ \bibnamefont {Grochowski}},\ and\
  \bibinfo {author} {\bibfnamefont {K.}~\bibnamefont {Rz{\k a}{\.z}ewski}},\
  }\href {https://doi.org/10.1103/PhysRevA.105.013314} {\bibfield  {journal}
  {\bibinfo  {journal} {Phys. Rev. A}\ }\textbf {\bibinfo {volume} {105}},\
  \bibinfo {pages} {013314} (\bibinfo {year} {2022})}\BibitemShut {NoStop}%
\bibitem [{\citenamefont {{\L}ebek}\ \emph {et~al.}(2022)\citenamefont
  {{\L}ebek}, \citenamefont {Syrwid}, \citenamefont {Grochowski},\ and\
  \citenamefont {Rz{\k a}{\.z}ewski}}]{Lebek2022}%
  \BibitemOpen
  \bibfield  {author} {\bibinfo {author} {\bibfnamefont {M.}~\bibnamefont
  {{\L}ebek}}, \bibinfo {author} {\bibfnamefont {A.}~\bibnamefont {Syrwid}},
  \bibinfo {author} {\bibfnamefont {P.~T.}\ \bibnamefont {Grochowski}},\ and\
  \bibinfo {author} {\bibfnamefont {K.}~\bibnamefont {Rz{\k a}{\.z}ewski}},\
  }\href {https://doi.org/10.1103/PhysRevA.105.L011303} {\bibfield  {journal}
  {\bibinfo  {journal} {Phys. Rev. A}\ }\textbf {\bibinfo {volume} {105}},\
  \bibinfo {pages} {L011303} (\bibinfo {year} {2022})}\BibitemShut {NoStop}%
\bibitem [{\citenamefont {DeMarco}\ and\ \citenamefont
  {Jin}(2002)}]{DeMarco2002}%
  \BibitemOpen
  \bibfield  {author} {\bibinfo {author} {\bibfnamefont {B.}~\bibnamefont
  {DeMarco}}\ and\ \bibinfo {author} {\bibfnamefont {D.~S.}\ \bibnamefont
  {Jin}},\ }\href {https://doi.org/10.1103/PhysRevLett.88.040405} {\bibfield
  {journal} {\bibinfo  {journal} {Phys. Rev. Lett.}\ }\textbf {\bibinfo
  {volume} {88}},\ \bibinfo {pages} {040405} (\bibinfo {year}
  {2002})}\BibitemShut {NoStop}%
\bibitem [{\citenamefont {Du}\ \emph {et~al.}(2008)\citenamefont {Du},
  \citenamefont {Luo}, \citenamefont {Clancy},\ and\ \citenamefont
  {Thomas}}]{Du2008}%
  \BibitemOpen
  \bibfield  {author} {\bibinfo {author} {\bibfnamefont {X.}~\bibnamefont
  {Du}}, \bibinfo {author} {\bibfnamefont {L.}~\bibnamefont {Luo}}, \bibinfo
  {author} {\bibfnamefont {B.}~\bibnamefont {Clancy}},\ and\ \bibinfo {author}
  {\bibfnamefont {J.~E.}\ \bibnamefont {Thomas}},\ }\href
  {https://doi.org/10.1103/PhysRevLett.101.150401} {\bibfield  {journal}
  {\bibinfo  {journal} {Phys. Rev. Lett.}\ }\textbf {\bibinfo {volume} {101}},\
  \bibinfo {pages} {150401} (\bibinfo {year} {2008})}\BibitemShut {NoStop}%
\bibitem [{\citenamefont {Jo}\ \emph {et~al.}(2009)\citenamefont {Jo},
  \citenamefont {Lee}, \citenamefont {Choi}, \citenamefont {Christensen},
  \citenamefont {Kim}, \citenamefont {Thywissen}, \citenamefont {Pritchard},\
  and\ \citenamefont {Ketterle}}]{Jo2009}%
  \BibitemOpen
  \bibfield  {author} {\bibinfo {author} {\bibfnamefont {G.-B.}\ \bibnamefont
  {Jo}}, \bibinfo {author} {\bibfnamefont {Y.-R.}\ \bibnamefont {Lee}},
  \bibinfo {author} {\bibfnamefont {J.-H.}\ \bibnamefont {Choi}}, \bibinfo
  {author} {\bibfnamefont {C.~A.}\ \bibnamefont {Christensen}}, \bibinfo
  {author} {\bibfnamefont {T.~H.}\ \bibnamefont {Kim}}, \bibinfo {author}
  {\bibfnamefont {J.~H.}\ \bibnamefont {Thywissen}}, \bibinfo {author}
  {\bibfnamefont {D.~E.}\ \bibnamefont {Pritchard}},\ and\ \bibinfo {author}
  {\bibfnamefont {W.}~\bibnamefont {Ketterle}},\ }\href
  {https://doi.org/10.1126/science.1177112} {\bibfield  {journal} {\bibinfo
  {journal} {Science}\ }\textbf {\bibinfo {volume} {325}},\ \bibinfo {pages}
  {1521} (\bibinfo {year} {2009})},\ \Eprint {https://arxiv.org/abs/19762638}
  {19762638} \BibitemShut {NoStop}%
\bibitem [{\citenamefont {Sommer}\ \emph {et~al.}(2011)\citenamefont {Sommer},
  \citenamefont {Ku}, \citenamefont {Roati},\ and\ \citenamefont
  {Zwierlein}}]{Sommer2011}%
  \BibitemOpen
  \bibfield  {author} {\bibinfo {author} {\bibfnamefont {A.}~\bibnamefont
  {Sommer}}, \bibinfo {author} {\bibfnamefont {M.}~\bibnamefont {Ku}}, \bibinfo
  {author} {\bibfnamefont {G.}~\bibnamefont {Roati}},\ and\ \bibinfo {author}
  {\bibfnamefont {M.~W.}\ \bibnamefont {Zwierlein}},\ }\href
  {https://doi.org/10.1038/nature09989} {\bibfield  {journal} {\bibinfo
  {journal} {Nature}\ }\textbf {\bibinfo {volume} {472}},\ \bibinfo {pages}
  {201} (\bibinfo {year} {2011})}\BibitemShut {NoStop}%
\bibitem [{\citenamefont {Sanner}\ \emph {et~al.}(2012)\citenamefont {Sanner},
  \citenamefont {Su}, \citenamefont {Huang}, \citenamefont {Keshet},
  \citenamefont {Gillen},\ and\ \citenamefont {Ketterle}}]{Sanner2012}%
  \BibitemOpen
  \bibfield  {author} {\bibinfo {author} {\bibfnamefont {C.}~\bibnamefont
  {Sanner}}, \bibinfo {author} {\bibfnamefont {E.~J.}\ \bibnamefont {Su}},
  \bibinfo {author} {\bibfnamefont {W.}~\bibnamefont {Huang}}, \bibinfo
  {author} {\bibfnamefont {A.}~\bibnamefont {Keshet}}, \bibinfo {author}
  {\bibfnamefont {J.}~\bibnamefont {Gillen}},\ and\ \bibinfo {author}
  {\bibfnamefont {W.}~\bibnamefont {Ketterle}},\ }\href
  {https://doi.org/10.1103/PhysRevLett.108.240404} {\bibfield  {journal}
  {\bibinfo  {journal} {Phys. Rev. Lett.}\ }\textbf {\bibinfo {volume} {108}},\
  \bibinfo {pages} {240404} (\bibinfo {year} {2012})}\BibitemShut {NoStop}%
\bibitem [{\citenamefont {Lee}\ \emph {et~al.}(2012)\citenamefont {Lee},
  \citenamefont {Heo}, \citenamefont {Choi}, \citenamefont {Wang},
  \citenamefont {Christensen}, \citenamefont {Rvachov},\ and\ \citenamefont
  {Ketterle}}]{Lee2012}%
  \BibitemOpen
  \bibfield  {author} {\bibinfo {author} {\bibfnamefont {Y.-R.}\ \bibnamefont
  {Lee}}, \bibinfo {author} {\bibfnamefont {M.-S.}\ \bibnamefont {Heo}},
  \bibinfo {author} {\bibfnamefont {J.-H.}\ \bibnamefont {Choi}}, \bibinfo
  {author} {\bibfnamefont {T.~T.}\ \bibnamefont {Wang}}, \bibinfo {author}
  {\bibfnamefont {C.~A.}\ \bibnamefont {Christensen}}, \bibinfo {author}
  {\bibfnamefont {T.~M.}\ \bibnamefont {Rvachov}},\ and\ \bibinfo {author}
  {\bibfnamefont {W.}~\bibnamefont {Ketterle}},\ }\href
  {https://doi.org/10.1103/PhysRevA.85.063615} {\bibfield  {journal} {\bibinfo
  {journal} {Phys. Rev. A}\ }\textbf {\bibinfo {volume} {85}},\ \bibinfo
  {pages} {063615} (\bibinfo {year} {2012})}\BibitemShut {NoStop}%
\bibitem [{\citenamefont {Valtolina}\ \emph {et~al.}(2017)\citenamefont
  {Valtolina}, \citenamefont {Scazza}, \citenamefont {Amico}, \citenamefont
  {Burchianti}, \citenamefont {Recati}, \citenamefont {Enss}, \citenamefont
  {Inguscio}, \citenamefont {Zaccanti},\ and\ \citenamefont
  {Roati}}]{Valtolina2017}%
  \BibitemOpen
  \bibfield  {author} {\bibinfo {author} {\bibfnamefont {G.}~\bibnamefont
  {Valtolina}}, \bibinfo {author} {\bibfnamefont {F.}~\bibnamefont {Scazza}},
  \bibinfo {author} {\bibfnamefont {A.}~\bibnamefont {Amico}}, \bibinfo
  {author} {\bibfnamefont {A.}~\bibnamefont {Burchianti}}, \bibinfo {author}
  {\bibfnamefont {A.}~\bibnamefont {Recati}}, \bibinfo {author} {\bibfnamefont
  {T.}~\bibnamefont {Enss}}, \bibinfo {author} {\bibfnamefont {M.}~\bibnamefont
  {Inguscio}}, \bibinfo {author} {\bibfnamefont {M.}~\bibnamefont {Zaccanti}},\
  and\ \bibinfo {author} {\bibfnamefont {G.}~\bibnamefont {Roati}},\ }\href
  {http://dx.doi.org/10.1038/nphys4108} {\bibfield  {journal} {\bibinfo
  {journal} {Nat. Phys.}\ }\textbf {\bibinfo {volume} {13}},\ \bibinfo {pages}
  {704} (\bibinfo {year} {2017})}\BibitemShut {NoStop}%
\bibitem [{\citenamefont {Amico}\ \emph {et~al.}(2018)\citenamefont {Amico},
  \citenamefont {Scazza}, \citenamefont {Valtolina}, \citenamefont {Tavares},
  \citenamefont {Ketterle}, \citenamefont {Inguscio}, \citenamefont {Roati},\
  and\ \citenamefont {Zaccanti}}]{Amico2018}%
  \BibitemOpen
  \bibfield  {author} {\bibinfo {author} {\bibfnamefont {A.}~\bibnamefont
  {Amico}}, \bibinfo {author} {\bibfnamefont {F.}~\bibnamefont {Scazza}},
  \bibinfo {author} {\bibfnamefont {G.}~\bibnamefont {Valtolina}}, \bibinfo
  {author} {\bibfnamefont {P.~E.~S.}\ \bibnamefont {Tavares}}, \bibinfo
  {author} {\bibfnamefont {W.}~\bibnamefont {Ketterle}}, \bibinfo {author}
  {\bibfnamefont {M.}~\bibnamefont {Inguscio}}, \bibinfo {author}
  {\bibfnamefont {G.}~\bibnamefont {Roati}},\ and\ \bibinfo {author}
  {\bibfnamefont {M.}~\bibnamefont {Zaccanti}},\ }\href
  {https://doi.org/10.1103/PhysRevLett.121.253602} {\bibfield  {journal}
  {\bibinfo  {journal} {Phys. Rev. Lett.}\ }\textbf {\bibinfo {volume} {121}},\
  \bibinfo {pages} {253602} (\bibinfo {year} {2018})}\BibitemShut {NoStop}%
\bibitem [{\citenamefont {Scazza}\ \emph {et~al.}(2020)\citenamefont {Scazza},
  \citenamefont {Valtolina}, \citenamefont {Amico}, \citenamefont {Tavares},
  \citenamefont {Inguscio}, \citenamefont {Ketterle}, \citenamefont {Roati},\
  and\ \citenamefont {Zaccanti}}]{Scazza2020}%
  \BibitemOpen
  \bibfield  {author} {\bibinfo {author} {\bibfnamefont {F.}~\bibnamefont
  {Scazza}}, \bibinfo {author} {\bibfnamefont {G.}~\bibnamefont {Valtolina}},
  \bibinfo {author} {\bibfnamefont {A.}~\bibnamefont {Amico}}, \bibinfo
  {author} {\bibfnamefont {P.~E.}\ \bibnamefont {Tavares}}, \bibinfo {author}
  {\bibfnamefont {M.}~\bibnamefont {Inguscio}}, \bibinfo {author}
  {\bibfnamefont {W.}~\bibnamefont {Ketterle}}, \bibinfo {author}
  {\bibfnamefont {G.}~\bibnamefont {Roati}},\ and\ \bibinfo {author}
  {\bibfnamefont {M.}~\bibnamefont {Zaccanti}},\ }\href
  {https://doi.org/10.1103/PhysRevA.101.013603} {\bibfield  {journal} {\bibinfo
   {journal} {Phys. Rev. A}\ }\textbf {\bibinfo {volume} {101}},\ \bibinfo
  {pages} {013603} (\bibinfo {year} {2020})}\BibitemShut {NoStop}%
\bibitem [{\citenamefont {Adlong}\ \emph {et~al.}(2020)\citenamefont {Adlong},
  \citenamefont {Liu}, \citenamefont {Scazza}, \citenamefont {Zaccanti},
  \citenamefont {Oppong}, \citenamefont {F{\"o}lling}, \citenamefont {Parish},\
  and\ \citenamefont {Levinsen}}]{Adlong2020}%
  \BibitemOpen
  \bibfield  {author} {\bibinfo {author} {\bibfnamefont {H.~S.}\ \bibnamefont
  {Adlong}}, \bibinfo {author} {\bibfnamefont {W.~E.}\ \bibnamefont {Liu}},
  \bibinfo {author} {\bibfnamefont {F.}~\bibnamefont {Scazza}}, \bibinfo
  {author} {\bibfnamefont {M.}~\bibnamefont {Zaccanti}}, \bibinfo {author}
  {\bibfnamefont {N.~D.}\ \bibnamefont {Oppong}}, \bibinfo {author}
  {\bibfnamefont {S.}~\bibnamefont {F{\"o}lling}}, \bibinfo {author}
  {\bibfnamefont {M.~M.}\ \bibnamefont {Parish}},\ and\ \bibinfo {author}
  {\bibfnamefont {J.}~\bibnamefont {Levinsen}},\ }\href
  {https://doi.org/10.1103/PhysRevLett.125.133401} {\bibfield  {journal}
  {\bibinfo  {journal} {Phys. Rev. Lett.}\ }\textbf {\bibinfo {volume} {125}},\
  \bibinfo {pages} {133401} (\bibinfo {year} {2020})}\BibitemShut {NoStop}%
\bibitem [{\citenamefont {Ji}\ \emph {et~al.}(2022)\citenamefont {Ji},
  \citenamefont {Schumacher}, \citenamefont {Assump{\c c}{\~a}o}, \citenamefont
  {Chen}, \citenamefont {M{\"a}kinen}, \citenamefont {Vivanco},\ and\
  \citenamefont {Navon}}]{Ji2022}%
  \BibitemOpen
  \bibfield  {author} {\bibinfo {author} {\bibfnamefont {Y.}~\bibnamefont
  {Ji}}, \bibinfo {author} {\bibfnamefont {G.~L.}\ \bibnamefont {Schumacher}},
  \bibinfo {author} {\bibfnamefont {G.~G.~T.}\ \bibnamefont {Assump{\c
  c}{\~a}o}}, \bibinfo {author} {\bibfnamefont {J.}~\bibnamefont {Chen}},
  \bibinfo {author} {\bibfnamefont {J.~T.}\ \bibnamefont {M{\"a}kinen}},
  \bibinfo {author} {\bibfnamefont {F.~J.}\ \bibnamefont {Vivanco}},\ and\
  \bibinfo {author} {\bibfnamefont {N.}~\bibnamefont {Navon}},\ }\href
  {https://doi.org/10.1103/PhysRevLett.129.203402} {\bibfield  {journal}
  {\bibinfo  {journal} {Phys. Rev. Lett.}\ }\textbf {\bibinfo {volume} {129}},\
  \bibinfo {pages} {203402} (\bibinfo {year} {2022})}\BibitemShut {NoStop}%
\bibitem [{\citenamefont {Lous}\ \emph {et~al.}(2018)\citenamefont {Lous},
  \citenamefont {Fritsche}, \citenamefont {Jag}, \citenamefont {Lehmann},
  \citenamefont {Kirilov}, \citenamefont {Huang},\ and\ \citenamefont
  {Grimm}}]{Lous2018}%
  \BibitemOpen
  \bibfield  {author} {\bibinfo {author} {\bibfnamefont {R.~S.}\ \bibnamefont
  {Lous}}, \bibinfo {author} {\bibfnamefont {I.}~\bibnamefont {Fritsche}},
  \bibinfo {author} {\bibfnamefont {M.}~\bibnamefont {Jag}}, \bibinfo {author}
  {\bibfnamefont {F.}~\bibnamefont {Lehmann}}, \bibinfo {author} {\bibfnamefont
  {E.}~\bibnamefont {Kirilov}}, \bibinfo {author} {\bibfnamefont
  {B.}~\bibnamefont {Huang}},\ and\ \bibinfo {author} {\bibfnamefont
  {R.}~\bibnamefont {Grimm}},\ }\href
  {https://doi.org/10.1103/PhysRevLett.120.243403} {\bibfield  {journal}
  {\bibinfo  {journal} {Phys. Rev. Lett.}\ }\textbf {\bibinfo {volume} {120}},\
  \bibinfo {pages} {243403} (\bibinfo {year} {2018})}\BibitemShut {NoStop}%
\bibitem [{\citenamefont {Huang}\ \emph {et~al.}(2019)\citenamefont {Huang},
  \citenamefont {Fritsche}, \citenamefont {Lous}, \citenamefont {Baroni},
  \citenamefont {Walraven}, \citenamefont {Kirilov},\ and\ \citenamefont
  {Grimm}}]{Huang2019}%
  \BibitemOpen
  \bibfield  {author} {\bibinfo {author} {\bibfnamefont {B.}~\bibnamefont
  {Huang}}, \bibinfo {author} {\bibfnamefont {I.}~\bibnamefont {Fritsche}},
  \bibinfo {author} {\bibfnamefont {R.~S.}\ \bibnamefont {Lous}}, \bibinfo
  {author} {\bibfnamefont {C.}~\bibnamefont {Baroni}}, \bibinfo {author}
  {\bibfnamefont {J.~T.~M.}\ \bibnamefont {Walraven}}, \bibinfo {author}
  {\bibfnamefont {E.}~\bibnamefont {Kirilov}},\ and\ \bibinfo {author}
  {\bibfnamefont {R.}~\bibnamefont {Grimm}},\ }\href
  {https://doi.org/10.1103/PhysRevA.99.041602} {\bibfield  {journal} {\bibinfo
  {journal} {Phys. Rev. A}\ }\textbf {\bibinfo {volume} {99}},\ \bibinfo
  {pages} {041602(R)} (\bibinfo {year} {2019})}\BibitemShut {NoStop}%
\bibitem [{\citenamefont {Grochowski}\ \emph
  {et~al.}(2020{\natexlab{b}})\citenamefont {Grochowski}, \citenamefont
  {Karpiuk}, \citenamefont {Brewczyk},\ and\ \citenamefont {Rz{\k
  a}{\.z}ewski}}]{Grochowski2020}%
  \BibitemOpen
  \bibfield  {author} {\bibinfo {author} {\bibfnamefont {P.~T.}\ \bibnamefont
  {Grochowski}}, \bibinfo {author} {\bibfnamefont {T.}~\bibnamefont {Karpiuk}},
  \bibinfo {author} {\bibfnamefont {M.}~\bibnamefont {Brewczyk}},\ and\
  \bibinfo {author} {\bibfnamefont {K.}~\bibnamefont {Rz{\k a}{\.z}ewski}},\
  }\href {https://doi.org/10.1103/PhysRevLett.125.103401} {\bibfield  {journal}
  {\bibinfo  {journal} {Phys. Rev. Lett.}\ }\textbf {\bibinfo {volume} {125}},\
  \bibinfo {pages} {103401} (\bibinfo {year} {2020}{\natexlab{b}})}\BibitemShut
  {NoStop}%
\bibitem [{\citenamefont {Englert}\ and\ \citenamefont
  {Schwinger}(1982)}]{Englert1982}%
  \BibitemOpen
  \bibfield  {author} {\bibinfo {author} {\bibfnamefont {B.~G.}\ \bibnamefont
  {Englert}}\ and\ \bibinfo {author} {\bibfnamefont {J.}~\bibnamefont
  {Schwinger}},\ }\href {https://doi.org/10.1103/PhysRevA.26.2322} {\bibfield
  {journal} {\bibinfo  {journal} {Phys. Rev. A}\ }\textbf {\bibinfo {volume}
  {26}},\ \bibinfo {pages} {2322} (\bibinfo {year} {1982})}\BibitemShut
  {NoStop}%
\bibitem [{\citenamefont {Englert}\ and\ \citenamefont
  {Schwinger}(1984)}]{Englert1984}%
  \BibitemOpen
  \bibfield  {author} {\bibinfo {author} {\bibfnamefont {B.~G.}\ \bibnamefont
  {Englert}}\ and\ \bibinfo {author} {\bibfnamefont {J.}~\bibnamefont
  {Schwinger}},\ }\href {https://doi.org/10.1103/PhysRevA.29.2331} {\bibfield
  {journal} {\bibinfo  {journal} {Phys. Rev. A}\ }\textbf {\bibinfo {volume}
  {29}},\ \bibinfo {pages} {2331} (\bibinfo {year} {1984})}\BibitemShut
  {NoStop}%
\bibitem [{\citenamefont {Englert}\ and\ \citenamefont
  {Schwinger}(1985)}]{Englert1985}%
  \BibitemOpen
  \bibfield  {author} {\bibinfo {author} {\bibfnamefont {B.~G.}\ \bibnamefont
  {Englert}}\ and\ \bibinfo {author} {\bibfnamefont {J.}~\bibnamefont
  {Schwinger}},\ }\href {https://doi.org/10.1103/PhysRevA.32.47} {\bibfield
  {journal} {\bibinfo  {journal} {Phys. Rev. A}\ }\textbf {\bibinfo {volume}
  {32}},\ \bibinfo {pages} {47} (\bibinfo {year} {1985})}\BibitemShut {NoStop}%
\bibitem [{\citenamefont {Englert}(1988)}]{Englert1988}%
  \BibitemOpen
  \bibfield  {author} {\bibinfo {author} {\bibfnamefont {B.-G.}\ \bibnamefont
  {Englert}},\ }\href@noop {} {\emph {\bibinfo {title} {Lecture {{Notes}} in
  {{Physics}}: {{Semiclassical Theory}} of {{Atoms}}}}}\ (\bibinfo  {publisher}
  {Springer},\ \bibinfo {address} {Berlin, Heidelberg},\ \bibinfo {year}
  {1988})\BibitemShut {NoStop}%
\bibitem [{\citenamefont {Englert}(1992)}]{Englert1992}%
  \BibitemOpen
  \bibfield  {author} {\bibinfo {author} {\bibfnamefont {B.~G.}\ \bibnamefont
  {Englert}},\ }\href {https://doi.org/10.1103/PhysRevA.45.127} {\bibfield
  {journal} {\bibinfo  {journal} {Phys. Rev. A}\ }\textbf {\bibinfo {volume}
  {45}},\ \bibinfo {pages} {127} (\bibinfo {year} {1992})}\BibitemShut
  {NoStop}%
\bibitem [{\citenamefont {Trappe}\ \emph
  {et~al.}(2016{\natexlab{b}})\citenamefont {Trappe}, \citenamefont {Len},
  \citenamefont {Ng}, \citenamefont {M{\"u}ller},\ and\ \citenamefont
  {Englert}}]{Trappe2016}%
  \BibitemOpen
  \bibfield  {author} {\bibinfo {author} {\bibfnamefont {M.-I.}\ \bibnamefont
  {Trappe}}, \bibinfo {author} {\bibfnamefont {Y.~L.}\ \bibnamefont {Len}},
  \bibinfo {author} {\bibfnamefont {H.~K.}\ \bibnamefont {Ng}}, \bibinfo
  {author} {\bibfnamefont {C.~A.}\ \bibnamefont {M{\"u}ller}},\ and\ \bibinfo
  {author} {\bibfnamefont {B.-G.}\ \bibnamefont {Englert}},\ }\href
  {https://doi.org/10.1103/PhysRevA.93.042510} {\bibfield  {journal} {\bibinfo
  {journal} {Phys. Rev. A}\ }\textbf {\bibinfo {volume} {93}},\ \bibinfo
  {pages} {042510} (\bibinfo {year} {2016}{\natexlab{b}})}\BibitemShut
  {NoStop}%
\bibitem [{\citenamefont {Trappe}\ \emph {et~al.}(2017)\citenamefont {Trappe},
  \citenamefont {Len}, \citenamefont {Ng},\ and\ \citenamefont
  {Englert}}]{Trappe2017}%
  \BibitemOpen
  \bibfield  {author} {\bibinfo {author} {\bibfnamefont {M.-I.}\ \bibnamefont
  {Trappe}}, \bibinfo {author} {\bibfnamefont {Y.~L.~L.}\ \bibnamefont {Len}},
  \bibinfo {author} {\bibfnamefont {H.~K.~K.}\ \bibnamefont {Ng}},\ and\
  \bibinfo {author} {\bibfnamefont {B.~G.}\ \bibnamefont {Englert}},\ }\href
  {https://doi.org/10.1016/j.aop.2017.07.020} {\bibfield  {journal} {\bibinfo
  {journal} {Ann. Phys.}\ }\textbf {\bibinfo {volume} {385}},\ \bibinfo {pages}
  {136} (\bibinfo {year} {2017})}\BibitemShut {NoStop}%
\bibitem [{\citenamefont {Chau}\ \emph {et~al.}(2018)\citenamefont {Chau},
  \citenamefont {Hue}, \citenamefont {Trappe},\ and\ \citenamefont
  {Englert}}]{Chau2018}%
  \BibitemOpen
  \bibfield  {author} {\bibinfo {author} {\bibfnamefont {T.~T.}\ \bibnamefont
  {Chau}}, \bibinfo {author} {\bibfnamefont {J.~H.}\ \bibnamefont {Hue}},
  \bibinfo {author} {\bibfnamefont {M.-I.}\ \bibnamefont {Trappe}},\ and\
  \bibinfo {author} {\bibfnamefont {B.~G.}\ \bibnamefont {Englert}},\ }\href
  {https://doi.org/10.1088/1367-2630/aacde1} {\bibfield  {journal} {\bibinfo
  {journal} {New J. Phys.}\ }\textbf {\bibinfo {volume} {20}},\ \bibinfo
  {pages} {073003} (\bibinfo {year} {2018})}\BibitemShut {NoStop}%
\bibitem [{\citenamefont {Englert}(2019{\natexlab{a}})}]{Englert2019}%
  \BibitemOpen
  \bibfield  {author} {\bibinfo {author} {\bibfnamefont {B.-G.}\ \bibnamefont
  {Englert}},\ }\href {https://doi.org/10.1142/11602} {\emph {\bibinfo {title}
  {Julian {{Schwinger}} and the {{Semiclassical Atom}}}}}\ (\bibinfo
  {publisher} {Proceedings of the Julian Schwinger Centennial Conference; World
  Scientific},\ \bibinfo {year} {2019})\BibitemShut {NoStop}%
\bibitem [{\citenamefont {Ancilotto}(2015)}]{Ancilotto2015}%
  \BibitemOpen
  \bibfield  {author} {\bibinfo {author} {\bibfnamefont {F.}~\bibnamefont
  {Ancilotto}},\ }\href {https://doi.org/10.1103/PhysRevA.92.061602} {\bibfield
   {journal} {\bibinfo  {journal} {Phys. Rev. A}\ }\textbf {\bibinfo {volume}
  {92}},\ \bibinfo {pages} {061602(R)} (\bibinfo {year} {2015})}\BibitemShut
  {NoStop}%
\bibitem [{\citenamefont {Das}\ and\ \citenamefont {Banerjee}(2018)}]{Das2018}%
  \BibitemOpen
  \bibfield  {author} {\bibinfo {author} {\bibfnamefont {A.~K.}\ \bibnamefont
  {Das}}\ and\ \bibinfo {author} {\bibfnamefont {A.}~\bibnamefont {Banerjee}},\
  }\href {https://doi.org/10.1140/epjd/e2018-80483-6} {\bibfield  {journal}
  {\bibinfo  {journal} {Eur. Phys. J. D}\ }\textbf {\bibinfo {volume} {72}},\
  \bibinfo {pages} {111} (\bibinfo {year} {2018})}\BibitemShut {NoStop}%
\bibitem [{\citenamefont {Ma}\ \emph {et~al.}(2012)\citenamefont {Ma},
  \citenamefont {Pilati}, \citenamefont {Troyer},\ and\ \citenamefont
  {Dai}}]{Ma2012}%
  \BibitemOpen
  \bibfield  {author} {\bibinfo {author} {\bibfnamefont {P.~N.}\ \bibnamefont
  {Ma}}, \bibinfo {author} {\bibfnamefont {S.}~\bibnamefont {Pilati}}, \bibinfo
  {author} {\bibfnamefont {M.}~\bibnamefont {Troyer}},\ and\ \bibinfo {author}
  {\bibfnamefont {X.}~\bibnamefont {Dai}},\ }\href
  {https://doi.org/10.1038/nphys2348} {\bibfield  {journal} {\bibinfo
  {journal} {Nat. Phys.}\ }\textbf {\bibinfo {volume} {8}},\ \bibinfo {pages}
  {601} (\bibinfo {year} {2012})}\BibitemShut {NoStop}%
\bibitem [{\citenamefont {Van~Zyl}\ \emph {et~al.}(2013)\citenamefont
  {Van~Zyl}, \citenamefont {Zaremba},\ and\ \citenamefont
  {Pisarski}}]{VanZyl2013}%
  \BibitemOpen
  \bibfield  {author} {\bibinfo {author} {\bibfnamefont {B.~P.}\ \bibnamefont
  {Van~Zyl}}, \bibinfo {author} {\bibfnamefont {E.}~\bibnamefont {Zaremba}},\
  and\ \bibinfo {author} {\bibfnamefont {P.}~\bibnamefont {Pisarski}},\ }\href
  {https://doi.org/10.1103/PhysRevA.87.043614} {\bibfield  {journal} {\bibinfo
  {journal} {Phys. Rev. A}\ }\textbf {\bibinfo {volume} {87}},\ \bibinfo
  {pages} {043614} (\bibinfo {year} {2013})}\BibitemShut {NoStop}%
\bibitem [{\citenamefont {Gangwar}\ \emph {et~al.}(2020)\citenamefont
  {Gangwar}, \citenamefont {Banerjee},\ and\ \citenamefont
  {Das}}]{Gangwar2020}%
  \BibitemOpen
  \bibfield  {author} {\bibinfo {author} {\bibfnamefont {R.}~\bibnamefont
  {Gangwar}}, \bibinfo {author} {\bibfnamefont {A.}~\bibnamefont {Banerjee}},\
  and\ \bibinfo {author} {\bibfnamefont {A.}~\bibnamefont {Das}},\ }\href
  {https://doi.org/10.1088/1361-6455/ab5f76} {\bibfield  {journal} {\bibinfo
  {journal} {J. Phys. B}\ }\textbf {\bibinfo {volume} {53}},\ \bibinfo {pages}
  {035301} (\bibinfo {year} {2020})}\BibitemShut {NoStop}%
\bibitem [{\citenamefont {Vilhena}\ \emph {et~al.}(2014)\citenamefont
  {Vilhena}, \citenamefont {R{\"a}s{\"a}nen}, \citenamefont {Marques},\ and\
  \citenamefont {Pittalis}}]{Vilhena2014}%
  \BibitemOpen
  \bibfield  {author} {\bibinfo {author} {\bibfnamefont {J.~G.}\ \bibnamefont
  {Vilhena}}, \bibinfo {author} {\bibfnamefont {E.}~\bibnamefont
  {R{\"a}s{\"a}nen}}, \bibinfo {author} {\bibfnamefont {M.~A.}\ \bibnamefont
  {Marques}},\ and\ \bibinfo {author} {\bibfnamefont {S.}~\bibnamefont
  {Pittalis}},\ }\href {https://doi.org/10.1021/ct4010728} {\bibfield
  {journal} {\bibinfo  {journal} {J. Chem. Theory Comput.}\ }\textbf {\bibinfo
  {volume} {10}},\ \bibinfo {pages} {1837} (\bibinfo {year}
  {2014})}\BibitemShut {NoStop}%
\bibitem [{\citenamefont {Trappe}\ \emph {et~al.}(2019)\citenamefont {Trappe},
  \citenamefont {Ho},\ and\ \citenamefont {Adam}}]{Trappe2019}%
  \BibitemOpen
  \bibfield  {author} {\bibinfo {author} {\bibfnamefont {M.-I.}\ \bibnamefont
  {Trappe}}, \bibinfo {author} {\bibfnamefont {D.~Y.}\ \bibnamefont {Ho}},\
  and\ \bibinfo {author} {\bibfnamefont {S.}~\bibnamefont {Adam}},\ }\href
  {https://doi.org/10.1103/PhysRevB.99.235415} {\bibfield  {journal} {\bibinfo
  {journal} {Phys. Rev. B}\ }\textbf {\bibinfo {volume} {99}},\ \bibinfo
  {pages} {235415} (\bibinfo {year} {2019})}\BibitemShut {NoStop}%
\bibitem [{\citenamefont {Englert}(2019{\natexlab{b}})}]{Englert2019a}%
  \BibitemOpen
  \bibfield  {author} {\bibinfo {author} {\bibfnamefont {B.-G.}\ \bibnamefont
  {Englert}},\ }\href@noop {} {\bibfield  {journal} {\bibinfo  {journal}
  {arXiv:1907.04751, Chapter 17, pp. 261-269 in: Proceedings of the Julian
  Schwinger Centennial Conference; B.-G. Englert (ed.); World Scientific}\ }
  (\bibinfo {year} {2019}{\natexlab{b}})}\BibitemShut {NoStop}%
\bibitem [{\citenamefont {Trappe}\ \emph
  {et~al.}(2023{\natexlab{a}})\citenamefont {Trappe}, \citenamefont {Hue},\
  and\ \citenamefont {Englert}}]{Trappe2023}%
  \BibitemOpen
  \bibfield  {author} {\bibinfo {author} {\bibfnamefont {M.-I.}\ \bibnamefont
  {Trappe}}, \bibinfo {author} {\bibfnamefont {J.~H.}\ \bibnamefont {Hue}},\
  and\ \bibinfo {author} {\bibfnamefont {B.-G.}\ \bibnamefont {Englert}},\
  }\href@noop {} {\bibfield  {journal} {\bibinfo  {journal} {arXiv:2106.07839,
  pp. 251--267 in: Density Functionals for Many-Particle Systems: Mathematical
  Theory and Physical Applications of Effective Equations; B.-G. Englert, H.
  Siedentop, and M.-I. Trappe (eds.); Lecture Notes Series, IMS, World
  Scientific, Singapore}\ } (\bibinfo {year} {2023}{\natexlab{a}})}\BibitemShut
  {NoStop}%
\bibitem [{\citenamefont {Englert}\ \emph {et~al.}(2023)\citenamefont
  {Englert}, \citenamefont {Hue}, \citenamefont {Huang}, \citenamefont
  {Paraniak},\ and\ \citenamefont {Trappe}}]{Englert2023}%
  \BibitemOpen
  \bibfield  {author} {\bibinfo {author} {\bibfnamefont {B.-G.}\ \bibnamefont
  {Englert}}, \bibinfo {author} {\bibfnamefont {J.~H.}\ \bibnamefont {Hue}},
  \bibinfo {author} {\bibfnamefont {Z.~C.}\ \bibnamefont {Huang}}, \bibinfo
  {author} {\bibfnamefont {M.~M.}\ \bibnamefont {Paraniak}},\ and\ \bibinfo
  {author} {\bibfnamefont {M.-I.}\ \bibnamefont {Trappe}},\ }\href@noop {}
  {\bibfield  {journal} {\bibinfo  {journal} {arXiv:2206.10097, pp. 287--308
  in: Density Functionals for Many-Particle Systems: Mathematical Theory and
  Physical Applications of Effective Equations; B.-G. Englert, H. Siedentop,
  and M.-I. Trappe (eds.); Lecture Notes Series, IMS, World Scientific,
  Singapore}\ } (\bibinfo {year} {2023})}\BibitemShut {NoStop}%
\bibitem [{\citenamefont {Trappe}\ and\ \citenamefont
  {Chisholm}(2023)}]{Trappe2023a}%
  \BibitemOpen
  \bibfield  {author} {\bibinfo {author} {\bibfnamefont {M.-I.}\ \bibnamefont
  {Trappe}}\ and\ \bibinfo {author} {\bibfnamefont {R.~A.}\ \bibnamefont
  {Chisholm}},\ }\href {https://doi.org/10.1038/s41467-023-36628-4} {\bibfield
  {journal} {\bibinfo  {journal} {Nat. Commun.}\ }\textbf {\bibinfo {volume}
  {14}},\ \bibinfo {pages} {1089} (\bibinfo {year} {2023})}\BibitemShut
  {NoStop}%
\bibitem [{\citenamefont {Trappe}\ \emph
  {et~al.}(2023{\natexlab{b}})\citenamefont {Trappe}, \citenamefont {Witt},\
  and\ \citenamefont {Manzhos}}]{Trappe2023b}%
  \BibitemOpen
  \bibfield  {author} {\bibinfo {author} {\bibfnamefont {M.-I.}\ \bibnamefont
  {Trappe}}, \bibinfo {author} {\bibfnamefont {W.~C.}\ \bibnamefont {Witt}},\
  and\ \bibinfo {author} {\bibfnamefont {S.}~\bibnamefont {Manzhos}},\ }\href
  {http://arxiv.org/abs/2304.10059} {\bibfield  {journal} {\bibinfo  {journal}
  {arXiv.2304.10059}\ } (\bibinfo {year} {2023}{\natexlab{b}})}\BibitemShut
  {NoStop}%
\bibitem [{\citenamefont {Aichinger}\ and\ \citenamefont
  {Krotscheck}(2005)}]{Aichinger2005}%
  \BibitemOpen
  \bibfield  {author} {\bibinfo {author} {\bibfnamefont {M.}~\bibnamefont
  {Aichinger}}\ and\ \bibinfo {author} {\bibfnamefont {E.}~\bibnamefont
  {Krotscheck}},\ }\href {https://doi.org/10.1016/j.commatsci.2004.11.002}
  {\bibfield  {journal} {\bibinfo  {journal} {Comput. Mater. Sci.}\ }\textbf
  {\bibinfo {volume} {34}},\ \bibinfo {pages} {188} (\bibinfo {year}
  {2005})}\BibitemShut {NoStop}%
\bibitem [{\citenamefont {Bloom}(1975)}]{Bloom1975}%
  \BibitemOpen
  \bibfield  {author} {\bibinfo {author} {\bibfnamefont {P.}~\bibnamefont
  {Bloom}},\ }\href {https://doi.org/10.1103/PhysRevB.12.125} {\bibfield
  {journal} {\bibinfo  {journal} {Phys. Rev. B}\ }\textbf {\bibinfo {volume}
  {12}},\ \bibinfo {pages} {125} (\bibinfo {year} {1975})}\BibitemShut
  {NoStop}%
\bibitem [{\citenamefont {He}(2014)}]{He2014}%
  \BibitemOpen
  \bibfield  {author} {\bibinfo {author} {\bibfnamefont {L.}~\bibnamefont
  {He}},\ }\href {https://doi.org/10.1103/PhysRevA.90.053633} {\bibfield
  {journal} {\bibinfo  {journal} {Phys. Rev. A}\ }\textbf {\bibinfo {volume}
  {90}},\ \bibinfo {pages} {053633} (\bibinfo {year} {2014})}\BibitemShut
  {NoStop}%
\bibitem [{\citenamefont {Conduit}(2010)}]{Conduit2010}%
  \BibitemOpen
  \bibfield  {author} {\bibinfo {author} {\bibfnamefont {G.~J.}\ \bibnamefont
  {Conduit}},\ }\href {https://doi.org/10.1103/PhysRevA.82.043604} {\bibfield
  {journal} {\bibinfo  {journal} {Phys. Rev. A}\ }\textbf {\bibinfo {volume}
  {82}},\ \bibinfo {pages} {043604} (\bibinfo {year} {2010})}\BibitemShut
  {NoStop}%
\bibitem [{\citenamefont {Conduit}(2013)}]{Conduit2013}%
  \BibitemOpen
  \bibfield  {author} {\bibinfo {author} {\bibfnamefont {G.~J.}\ \bibnamefont
  {Conduit}},\ }\href {https://doi.org/10.1103/PhysRevB.87.184414} {\bibfield
  {journal} {\bibinfo  {journal} {Phys. Rev. B}\ }\textbf {\bibinfo {volume}
  {87}},\ \bibinfo {pages} {184414} (\bibinfo {year} {2013})}\BibitemShut
  {NoStop}%
\bibitem [{\citenamefont {Bertaina}(2013)}]{Bertaina2013}%
  \BibitemOpen
  \bibfield  {author} {\bibinfo {author} {\bibfnamefont {G.}~\bibnamefont
  {Bertaina}},\ }\href {https://doi.org/10.1140/epjst/e2013-01763-9} {\bibfield
   {journal} {\bibinfo  {journal} {Eur. Phys. J. Spec. Top.}\ }\textbf
  {\bibinfo {volume} {217}},\ \bibinfo {pages} {153} (\bibinfo {year}
  {2013})}\BibitemShut {NoStop}%
\bibitem [{\citenamefont {Whitehead}\ \emph {et~al.}(2016)\citenamefont
  {Whitehead}, \citenamefont {Schonenberg}, \citenamefont {Kongsuwan},
  \citenamefont {Needs},\ and\ \citenamefont {Conduit}}]{Whitehead2016}%
  \BibitemOpen
  \bibfield  {author} {\bibinfo {author} {\bibfnamefont {T.~M.}\ \bibnamefont
  {Whitehead}}, \bibinfo {author} {\bibfnamefont {L.~M.}\ \bibnamefont
  {Schonenberg}}, \bibinfo {author} {\bibfnamefont {N.}~\bibnamefont
  {Kongsuwan}}, \bibinfo {author} {\bibfnamefont {R.~J.}\ \bibnamefont
  {Needs}},\ and\ \bibinfo {author} {\bibfnamefont {G.~J.}\ \bibnamefont
  {Conduit}},\ }\href {https://doi.org/10.1103/PhysRevA.93.042702} {\bibfield
  {journal} {\bibinfo  {journal} {Phys. Rev. A}\ }\textbf {\bibinfo {volume}
  {93}},\ \bibinfo {pages} {042702} (\bibinfo {year} {2016})}\BibitemShut
  {NoStop}%
\bibitem [{\citenamefont {Pilati}\ \emph {et~al.}(2021)\citenamefont {Pilati},
  \citenamefont {Orso},\ and\ \citenamefont {Bertaina}}]{Pilati2021}%
  \BibitemOpen
  \bibfield  {author} {\bibinfo {author} {\bibfnamefont {S.}~\bibnamefont
  {Pilati}}, \bibinfo {author} {\bibfnamefont {G.}~\bibnamefont {Orso}},\ and\
  \bibinfo {author} {\bibfnamefont {G.}~\bibnamefont {Bertaina}},\ }\href
  {https://doi.org/10.1103/PhysRevA.103.063314} {\bibfield  {journal} {\bibinfo
   {journal} {Phys. Rev. A}\ }\textbf {\bibinfo {volume} {103}},\ \bibinfo
  {pages} {063314} (\bibinfo {year} {2021})}\BibitemShut {NoStop}%
\bibitem [{\citenamefont {Heiselberg}(2011)}]{Heiselberg2011}%
  \BibitemOpen
  \bibfield  {author} {\bibinfo {author} {\bibfnamefont {H.}~\bibnamefont
  {Heiselberg}},\ }\href {https://doi.org/10.1103/PhysRevA.83.053635}
  {\bibfield  {journal} {\bibinfo  {journal} {Phys. Rev. A}\ }\textbf {\bibinfo
  {volume} {83}},\ \bibinfo {pages} {053635} (\bibinfo {year}
  {2011})}\BibitemShut {NoStop}%
\bibitem [{\citenamefont {He}\ and\ \citenamefont {Huang}(2012)}]{He2012}%
  \BibitemOpen
  \bibfield  {author} {\bibinfo {author} {\bibfnamefont {L.}~\bibnamefont
  {He}}\ and\ \bibinfo {author} {\bibfnamefont {X.-G.}\ \bibnamefont {Huang}},\
  }\href {https://doi.org/10.1103/PhysRevA.85.043624} {\bibfield  {journal}
  {\bibinfo  {journal} {Phys. Rev. A}\ }\textbf {\bibinfo {volume} {85}},\
  \bibinfo {pages} {043624} (\bibinfo {year} {2012})}\BibitemShut {NoStop}%
\bibitem [{\citenamefont {He}\ \emph {et~al.}(2016)\citenamefont {He},
  \citenamefont {Liu}, \citenamefont {Huang},\ and\ \citenamefont
  {Hu}}]{He2016}%
  \BibitemOpen
  \bibfield  {author} {\bibinfo {author} {\bibfnamefont {L.}~\bibnamefont
  {He}}, \bibinfo {author} {\bibfnamefont {X.-J.}\ \bibnamefont {Liu}},
  \bibinfo {author} {\bibfnamefont {X.-G.}\ \bibnamefont {Huang}},\ and\
  \bibinfo {author} {\bibfnamefont {H.}~\bibnamefont {Hu}},\ }\href
  {https://doi.org/10.1103/PhysRevA.93.063629} {\bibfield  {journal} {\bibinfo
  {journal} {Phys. Rev. A}\ }\textbf {\bibinfo {volume} {93}},\ \bibinfo
  {pages} {063629} (\bibinfo {year} {2016})}\BibitemShut {NoStop}%
\bibitem [{\citenamefont {Schmidt}\ \emph {et~al.}(2012)\citenamefont
  {Schmidt}, \citenamefont {Enss}, \citenamefont {Pietil{\"a}},\ and\
  \citenamefont {Demler}}]{Schmidt2012}%
  \BibitemOpen
  \bibfield  {author} {\bibinfo {author} {\bibfnamefont {R.}~\bibnamefont
  {Schmidt}}, \bibinfo {author} {\bibfnamefont {T.}~\bibnamefont {Enss}},
  \bibinfo {author} {\bibfnamefont {V.}~\bibnamefont {Pietil{\"a}}},\ and\
  \bibinfo {author} {\bibfnamefont {E.}~\bibnamefont {Demler}},\ }\href
  {https://doi.org/10.1103/PhysRevA.85.021602} {\bibfield  {journal} {\bibinfo
  {journal} {Phys. Rev. A}\ }\textbf {\bibinfo {volume} {85}},\ \bibinfo
  {pages} {021602} (\bibinfo {year} {2012})}\BibitemShut {NoStop}%
\bibitem [{\citenamefont {Ngampruetikorn}\ \emph {et~al.}(2012)\citenamefont
  {Ngampruetikorn}, \citenamefont {Levinsen},\ and\ \citenamefont
  {Parish}}]{Ngampruetikorn2012}%
  \BibitemOpen
  \bibfield  {author} {\bibinfo {author} {\bibfnamefont {V.}~\bibnamefont
  {Ngampruetikorn}}, \bibinfo {author} {\bibfnamefont {J.}~\bibnamefont
  {Levinsen}},\ and\ \bibinfo {author} {\bibfnamefont {M.~M.}\ \bibnamefont
  {Parish}},\ }\href {https://doi.org/10.1209/0295-5075/98/30005} {\bibfield
  {journal} {\bibinfo  {journal} {Europhys. Lett.}\ }\textbf {\bibinfo {volume}
  {98}},\ \bibinfo {pages} {30005} (\bibinfo {year} {2012})}\BibitemShut
  {NoStop}%
\bibitem [{\citenamefont {Haller}\ \emph {et~al.}(2010)\citenamefont {Haller},
  \citenamefont {Mark}, \citenamefont {Hart}, \citenamefont {Danzl},
  \citenamefont {Reichs{\"o}llner}, \citenamefont {Melezhik}, \citenamefont
  {Schmelcher},\ and\ \citenamefont {N{\"a}gerl}}]{Haller2010a}%
  \BibitemOpen
  \bibfield  {author} {\bibinfo {author} {\bibfnamefont {E.}~\bibnamefont
  {Haller}}, \bibinfo {author} {\bibfnamefont {M.~J.}\ \bibnamefont {Mark}},
  \bibinfo {author} {\bibfnamefont {R.}~\bibnamefont {Hart}}, \bibinfo {author}
  {\bibfnamefont {J.~G.}\ \bibnamefont {Danzl}}, \bibinfo {author}
  {\bibfnamefont {L.}~\bibnamefont {Reichs{\"o}llner}}, \bibinfo {author}
  {\bibfnamefont {V.}~\bibnamefont {Melezhik}}, \bibinfo {author}
  {\bibfnamefont {P.}~\bibnamefont {Schmelcher}},\ and\ \bibinfo {author}
  {\bibfnamefont {H.-C.}\ \bibnamefont {N{\"a}gerl}},\ }\href
  {https://doi.org/10.1103/PhysRevLett.104.153203} {\bibfield  {journal}
  {\bibinfo  {journal} {Phys. Rev. Lett.}\ }\textbf {\bibinfo {volume} {104}},\
  \bibinfo {pages} {153203} (\bibinfo {year} {2010})}\BibitemShut {NoStop}%
\bibitem [{\citenamefont {Massignan}\ \emph {et~al.}(2013)\citenamefont
  {Massignan}, \citenamefont {Yu},\ and\ \citenamefont
  {Bruun}}]{Massignan2013}%
  \BibitemOpen
  \bibfield  {author} {\bibinfo {author} {\bibfnamefont {P.}~\bibnamefont
  {Massignan}}, \bibinfo {author} {\bibfnamefont {Z.}~\bibnamefont {Yu}},\ and\
  \bibinfo {author} {\bibfnamefont {G.~M.}\ \bibnamefont {Bruun}},\ }\href
  {https://doi.org/10.1103/PhysRevLett.110.230401} {\bibfield  {journal}
  {\bibinfo  {journal} {Phys. Rev. Lett.}\ }\textbf {\bibinfo {volume} {110}},\
  \bibinfo {pages} {230401} (\bibinfo {year} {2013})}\BibitemShut {NoStop}%
\bibitem [{\citenamefont {Tajima}\ and\ \citenamefont
  {Uchino}(2018)}]{Tajima2018}%
  \BibitemOpen
  \bibfield  {author} {\bibinfo {author} {\bibfnamefont {H.}~\bibnamefont
  {Tajima}}\ and\ \bibinfo {author} {\bibfnamefont {S.}~\bibnamefont
  {Uchino}},\ }\href {https://doi.org/10.1088/1367-2630/aad1e7} {\bibfield
  {journal} {\bibinfo  {journal} {New J. Phys.}\ }\textbf {\bibinfo {volume}
  {20}},\ \bibinfo {pages} {073048} (\bibinfo {year} {2018})}\BibitemShut
  {NoStop}%
\bibitem [{\citenamefont {Huang}\ and\ \citenamefont
  {Cazalilla}(2023)}]{Huang2023}%
  \BibitemOpen
  \bibfield  {author} {\bibinfo {author} {\bibfnamefont {C.-H.}\ \bibnamefont
  {Huang}}\ and\ \bibinfo {author} {\bibfnamefont {M.~A.}\ \bibnamefont
  {Cazalilla}},\ }\href {https://doi.org/10.1088/1367-2630/acd8e4} {\bibfield
  {journal} {\bibinfo  {journal} {New J. Phys.}\ }\textbf {\bibinfo {volume}
  {25}},\ \bibinfo {pages} {063005} (\bibinfo {year} {2023})}\BibitemShut
  {NoStop}%
\bibitem [{\citenamefont {Feng}\ and\ \citenamefont {Yin}(2020)}]{Feng2020}%
  \BibitemOpen
  \bibfield  {author} {\bibinfo {author} {\bibfnamefont {X.-J.}\ \bibnamefont
  {Feng}}\ and\ \bibinfo {author} {\bibfnamefont {L.}~\bibnamefont {Yin}},\
  }\href {https://doi.org/10.1088/1674-1056/aba9ca} {\bibfield  {journal}
  {\bibinfo  {journal} {Chin. Phys. B}\ }\textbf {\bibinfo {volume} {29}},\
  \bibinfo {pages} {110306} (\bibinfo {year} {2020})}\BibitemShut {NoStop}%
\bibitem [{\citenamefont {Ong}\ \emph {et~al.}(2015)\citenamefont {Ong},
  \citenamefont {Cheng}, \citenamefont {Arakelyan},\ and\ \citenamefont
  {Thomas}}]{Ong2015}%
  \BibitemOpen
  \bibfield  {author} {\bibinfo {author} {\bibfnamefont {W.}~\bibnamefont
  {Ong}}, \bibinfo {author} {\bibfnamefont {C.}~\bibnamefont {Cheng}}, \bibinfo
  {author} {\bibfnamefont {I.}~\bibnamefont {Arakelyan}},\ and\ \bibinfo
  {author} {\bibfnamefont {J.~E.}\ \bibnamefont {Thomas}},\ }\href
  {https://doi.org/10.1103/PhysRevLett.114.110403} {\bibfield  {journal}
  {\bibinfo  {journal} {Phys. Rev. Lett.}\ }\textbf {\bibinfo {volume} {114}},\
  \bibinfo {pages} {110403} (\bibinfo {year} {2015})}\BibitemShut {NoStop}%
\bibitem [{\citenamefont {Fr{\"o}hlich}\ \emph {et~al.}(2011)\citenamefont
  {Fr{\"o}hlich}, \citenamefont {Feld}, \citenamefont {Vogt}, \citenamefont
  {Koschorreck}, \citenamefont {Zwerger},\ and\ \citenamefont
  {K{\"o}hl}}]{Frohlich2011}%
  \BibitemOpen
  \bibfield  {author} {\bibinfo {author} {\bibfnamefont {B.}~\bibnamefont
  {Fr{\"o}hlich}}, \bibinfo {author} {\bibfnamefont {M.}~\bibnamefont {Feld}},
  \bibinfo {author} {\bibfnamefont {E.}~\bibnamefont {Vogt}}, \bibinfo {author}
  {\bibfnamefont {M.}~\bibnamefont {Koschorreck}}, \bibinfo {author}
  {\bibfnamefont {W.}~\bibnamefont {Zwerger}},\ and\ \bibinfo {author}
  {\bibfnamefont {M.}~\bibnamefont {K{\"o}hl}},\ }\href
  {https://doi.org/10.1103/PhysRevLett.106.105301} {\bibfield  {journal}
  {\bibinfo  {journal} {Phys. Rev. Lett.}\ }\textbf {\bibinfo {volume} {106}},\
  \bibinfo {pages} {105301} (\bibinfo {year} {2011})}\BibitemShut {NoStop}%
\bibitem [{\citenamefont {Kohstall}\ \emph {et~al.}(2012)\citenamefont
  {Kohstall}, \citenamefont {Zaccanti}, \citenamefont {Jag}, \citenamefont
  {Trenkwalder}, \citenamefont {Massignan}, \citenamefont {Bruun},
  \citenamefont {Schreck},\ and\ \citenamefont {Grimm}}]{Kohstall2012}%
  \BibitemOpen
  \bibfield  {author} {\bibinfo {author} {\bibfnamefont {C.}~\bibnamefont
  {Kohstall}}, \bibinfo {author} {\bibfnamefont {M.}~\bibnamefont {Zaccanti}},
  \bibinfo {author} {\bibfnamefont {M.}~\bibnamefont {Jag}}, \bibinfo {author}
  {\bibfnamefont {A.}~\bibnamefont {Trenkwalder}}, \bibinfo {author}
  {\bibfnamefont {P.}~\bibnamefont {Massignan}}, \bibinfo {author}
  {\bibfnamefont {G.~M.}\ \bibnamefont {Bruun}}, \bibinfo {author}
  {\bibfnamefont {F.}~\bibnamefont {Schreck}},\ and\ \bibinfo {author}
  {\bibfnamefont {R.}~\bibnamefont {Grimm}},\ }\href
  {https://doi.org/10.1038/nature11065} {\bibfield  {journal} {\bibinfo
  {journal} {Nature}\ }\textbf {\bibinfo {volume} {485}},\ \bibinfo {pages}
  {615} (\bibinfo {year} {2012})}\BibitemShut {NoStop}%
\bibitem [{\citenamefont {Koschorreck}\ \emph {et~al.}(2012)\citenamefont
  {Koschorreck}, \citenamefont {Pertot}, \citenamefont {Vogt}, \citenamefont
  {Fr{\"o}hlich}, \citenamefont {Feld},\ and\ \citenamefont
  {K{\"o}hl}}]{Koschorreck2012}%
  \BibitemOpen
  \bibfield  {author} {\bibinfo {author} {\bibfnamefont {M.}~\bibnamefont
  {Koschorreck}}, \bibinfo {author} {\bibfnamefont {D.}~\bibnamefont {Pertot}},
  \bibinfo {author} {\bibfnamefont {E.}~\bibnamefont {Vogt}}, \bibinfo {author}
  {\bibfnamefont {B.}~\bibnamefont {Fr{\"o}hlich}}, \bibinfo {author}
  {\bibfnamefont {M.}~\bibnamefont {Feld}},\ and\ \bibinfo {author}
  {\bibfnamefont {M.}~\bibnamefont {K{\"o}hl}},\ }\href
  {https://doi.org/10.1038/nature11151} {\bibfield  {journal} {\bibinfo
  {journal} {Nature}\ }\textbf {\bibinfo {volume} {485}},\ \bibinfo {pages}
  {619} (\bibinfo {year} {2012})}\BibitemShut {NoStop}%
\bibitem [{\citenamefont {Ciamei}\ \emph {et~al.}(2022)\citenamefont {Ciamei},
  \citenamefont {Finelli}, \citenamefont {Cosco}, \citenamefont {Inguscio},
  \citenamefont {Trenkwalder},\ and\ \citenamefont {Zaccanti}}]{Ciamei2022}%
  \BibitemOpen
  \bibfield  {author} {\bibinfo {author} {\bibfnamefont {A.}~\bibnamefont
  {Ciamei}}, \bibinfo {author} {\bibfnamefont {S.}~\bibnamefont {Finelli}},
  \bibinfo {author} {\bibfnamefont {A.}~\bibnamefont {Cosco}}, \bibinfo
  {author} {\bibfnamefont {M.}~\bibnamefont {Inguscio}}, \bibinfo {author}
  {\bibfnamefont {A.}~\bibnamefont {Trenkwalder}},\ and\ \bibinfo {author}
  {\bibfnamefont {M.}~\bibnamefont {Zaccanti}},\ }\href
  {https://doi.org/10.1103/PhysRevA.106.053318} {\bibfield  {journal} {\bibinfo
   {journal} {Phys. Rev. A}\ }\textbf {\bibinfo {volume} {106}},\ \bibinfo
  {pages} {053318} (\bibinfo {year} {2022})}\BibitemShut {NoStop}%
\bibitem [{\citenamefont {Darkwah~Oppong}\ \emph {et~al.}(2019)\citenamefont
  {Darkwah~Oppong}, \citenamefont {Riegger}, \citenamefont {Bettermann},
  \citenamefont {H{\"o}fer}, \citenamefont {Levinsen}, \citenamefont {Parish},
  \citenamefont {Bloch},\ and\ \citenamefont
  {F{\"o}lling}}]{DarkwahOppong2019}%
  \BibitemOpen
  \bibfield  {author} {\bibinfo {author} {\bibfnamefont {N.}~\bibnamefont
  {Darkwah~Oppong}}, \bibinfo {author} {\bibfnamefont {L.}~\bibnamefont
  {Riegger}}, \bibinfo {author} {\bibfnamefont {O.}~\bibnamefont {Bettermann}},
  \bibinfo {author} {\bibfnamefont {M.}~\bibnamefont {H{\"o}fer}}, \bibinfo
  {author} {\bibfnamefont {J.}~\bibnamefont {Levinsen}}, \bibinfo {author}
  {\bibfnamefont {M.~M.}\ \bibnamefont {Parish}}, \bibinfo {author}
  {\bibfnamefont {I.}~\bibnamefont {Bloch}},\ and\ \bibinfo {author}
  {\bibfnamefont {S.}~\bibnamefont {F{\"o}lling}},\ }\href
  {https://doi.org/10.1103/PhysRevLett.122.193604} {\bibfield  {journal}
  {\bibinfo  {journal} {Phys. Rev. Lett.}\ }\textbf {\bibinfo {volume} {122}},\
  \bibinfo {pages} {193604} (\bibinfo {year} {2019})}\BibitemShut {NoStop}%
\bibitem [{\citenamefont {Ravensbergen}\ \emph {et~al.}(2020)\citenamefont
  {Ravensbergen}, \citenamefont {Soave}, \citenamefont {Corre}, \citenamefont
  {Kreyer}, \citenamefont {Huang}, \citenamefont {Kirilov},\ and\ \citenamefont
  {Grimm}}]{Ravensbergen2020}%
  \BibitemOpen
  \bibfield  {author} {\bibinfo {author} {\bibfnamefont {C.}~\bibnamefont
  {Ravensbergen}}, \bibinfo {author} {\bibfnamefont {E.}~\bibnamefont {Soave}},
  \bibinfo {author} {\bibfnamefont {V.}~\bibnamefont {Corre}}, \bibinfo
  {author} {\bibfnamefont {M.}~\bibnamefont {Kreyer}}, \bibinfo {author}
  {\bibfnamefont {B.}~\bibnamefont {Huang}}, \bibinfo {author} {\bibfnamefont
  {E.}~\bibnamefont {Kirilov}},\ and\ \bibinfo {author} {\bibfnamefont
  {R.}~\bibnamefont {Grimm}},\ }\href
  {https://doi.org/10.1103/PhysRevLett.124.203402} {\bibfield  {journal}
  {\bibinfo  {journal} {Phys. Rev. Lett.}\ }\textbf {\bibinfo {volume} {124}},\
  \bibinfo {pages} {203402} (\bibinfo {year} {2020})}\BibitemShut {NoStop}%
\bibitem [{\citenamefont {{von Milczewski}}\ \emph {et~al.}(2022)\citenamefont
  {{von Milczewski}}, \citenamefont {Rose},\ and\ \citenamefont
  {Schmidt}}]{vonMilczewski2022}%
  \BibitemOpen
  \bibfield  {author} {\bibinfo {author} {\bibfnamefont {J.}~\bibnamefont {{von
  Milczewski}}}, \bibinfo {author} {\bibfnamefont {F.}~\bibnamefont {Rose}},\
  and\ \bibinfo {author} {\bibfnamefont {R.}~\bibnamefont {Schmidt}},\ }\href
  {https://doi.org/10.1103/PhysRevA.105.013317} {\bibfield  {journal} {\bibinfo
   {journal} {Phys. Rev. A}\ }\textbf {\bibinfo {volume} {105}},\ \bibinfo
  {pages} {013317} (\bibinfo {year} {2022})}\BibitemShut {NoStop}%
\bibitem [{\citenamefont {D'Alberto}\ \emph {et~al.}(2024)\citenamefont
  {D'Alberto}, \citenamefont {Cardarelli}, \citenamefont {Galli}, \citenamefont
  {Bertaina},\ and\ \citenamefont {Pieri}}]{DAlberto2024}%
  \BibitemOpen
  \bibfield  {author} {\bibinfo {author} {\bibfnamefont {J.}~\bibnamefont
  {D'Alberto}}, \bibinfo {author} {\bibfnamefont {L.}~\bibnamefont
  {Cardarelli}}, \bibinfo {author} {\bibfnamefont {D.~E.}\ \bibnamefont
  {Galli}}, \bibinfo {author} {\bibfnamefont {G.}~\bibnamefont {Bertaina}},\
  and\ \bibinfo {author} {\bibfnamefont {P.}~\bibnamefont {Pieri}},\ }\href
  {https://arxiv.org/abs/2402.13664v1} {\bibfield  {journal} {\bibinfo
  {journal} {arXiv:2402.13664}\ } (\bibinfo {year} {2024})}\BibitemShut
  {NoStop}%
\bibitem [{\citenamefont {DeSalvo}\ \emph {et~al.}(2019)\citenamefont
  {DeSalvo}, \citenamefont {Patel}, \citenamefont {Cai},\ and\ \citenamefont
  {Chin}}]{DeSalvo2019}%
  \BibitemOpen
  \bibfield  {author} {\bibinfo {author} {\bibfnamefont {B.~J.}\ \bibnamefont
  {DeSalvo}}, \bibinfo {author} {\bibfnamefont {K.}~\bibnamefont {Patel}},
  \bibinfo {author} {\bibfnamefont {G.}~\bibnamefont {Cai}},\ and\ \bibinfo
  {author} {\bibfnamefont {C.}~\bibnamefont {Chin}},\ }\href
  {https://doi.org/10.1038/s41586-019-1055-0} {\bibfield  {journal} {\bibinfo
  {journal} {Nature}\ }\textbf {\bibinfo {volume} {568}},\ \bibinfo {pages}
  {61} (\bibinfo {year} {2019})}\BibitemShut {NoStop}%
\bibitem [{\citenamefont {Perdew}\ \emph {et~al.}(2021)\citenamefont {Perdew},
  \citenamefont {Ruzsinszky}, \citenamefont {Sun}, \citenamefont {Nepal},\ and\
  \citenamefont {Kaplan}}]{Perdew2021a}%
  \BibitemOpen
  \bibfield  {author} {\bibinfo {author} {\bibfnamefont {J.~P.}\ \bibnamefont
  {Perdew}}, \bibinfo {author} {\bibfnamefont {A.}~\bibnamefont {Ruzsinszky}},
  \bibinfo {author} {\bibfnamefont {J.}~\bibnamefont {Sun}}, \bibinfo {author}
  {\bibfnamefont {N.~K.}\ \bibnamefont {Nepal}},\ and\ \bibinfo {author}
  {\bibfnamefont {A.~D.}\ \bibnamefont {Kaplan}},\ }\href
  {https://doi.org/10.1073/pnas.2017850118} {\bibfield  {journal} {\bibinfo
  {journal} {Proc. Natl. Acad. Sci.}\ }\textbf {\bibinfo {volume} {118}},\
  \bibinfo {pages} {e2017850118} (\bibinfo {year} {2021})}\BibitemShut
  {NoStop}%
\end{thebibliography}%
